%Paper: hep-th/9307010
%From: Hugh Osborn <H.Osborn@damtp.cambridge.ac.uk>
%Date: Fri, 2 Jul 93 09:16 BST
%Date (revised): Wed, 4 Aug 93 15:17 BST

%Plain TeX file

\magnification\magstep1
%For smaller font footnotes
\newdimen\itemindent \itemindent=32pt
\def\textindent#1{\parindent=\itemindent\let\par=\resetpar%
\indent\llap{#1\enspace}\ignorespaces}

\let\oldpar=\par
\def\resetpar{\oldpar\parindent=20pt\let\par=\oldpar}

\font\ninerm=cmr9 \font\ninesy=cmsy9
\font\eightrm=cmr8 \font\sixrm=cmr6
\font\eighti=cmmi8 \font\sixi=cmmi6
\font\eightsy=cmsy8 \font\sixsy=cmsy6
\font\eightbf=cmbx8 \font\sixbf=cmbx6
\font\eightit=cmti8
\def\eightpoint{\def\rm{\fam0\eightrm}
  \textfont0=\eightrm \scriptfont0=\sixrm \scriptscriptfont0=\fiverm
  \textfont1=\eighti  \scriptfont1=\sixi  \scriptscriptfont1=\fivei
  \textfont2=\eightsy \scriptfont2=\sixsy \scriptscriptfont2=\fivesy
  \textfont3=\tenex   \scriptfont3=\tenex \scriptscriptfont3=\tenex
  \textfont\itfam=\eightit  \def\it{\fam\itfam\eightit}%
  \textfont\bffam=\eightbf  \scriptfont\bffam=\sixbf
  \scriptscriptfont\bffam=\fivebf  \def\bf{\fam\bffam\eightbf}%
  \normalbaselineskip=9pt
  \setbox\strutbox=\hbox{\vrule height7pt depth2pt width0pt}%
  \let\big=\eightbig  \normalbaselines\rm}
\catcode`@=11 %
\def\eightbig#1{{\hbox{$\textfont0=\ninerm\textfont2=\ninesy
  \left#1\vbox to6.5pt{}\right.\n@space$}}}
\def\vfootnote#1{\insert\footins\bgroup\eightpoint
  \interlinepenalty=\interfootnotelinepenalty
  \splittopskip=\ht\strutbox %
  \splitmaxdepth=\dp\strutbox %
  \leftskip=0pt \rightskip=0pt \spaceskip=0pt \xspaceskip=0pt
  \textindent{#1}\footstrut\futurelet\next\fo@t}
\catcode`@=12 %

%If AMS fonts are absent remove the following two lines and uncomment the
%next two
\input mssymb.tex
\def \bR{{\Bbb R}}
\font \bigbf=cmbx10 scaled \magstep1
\def \olr{{\raise7pt\hbox{$\leftrightarrow  \! \! \! \! \!$}}}
\def\pmb#1{\setbox0=\hbox{#1}%
 \kern-.025em\copy0\kern-\wd0
 \kern.05em\copy0\kern-\wd0
 \kern-.025em\raise.0433em\box0 }
\def \bxi{\pmb{$\xi$}}
\def \ollr{{\raise7pt\hbox{$\leftrightarrow  \! \! \! \! \! \!$}}}
\def \l{\langle}
\def \r{\rangle}
\def \de{\delta}
\def \si{\sigma}

\def \nab{\nabla}
\def \hA{{\hat A}}

\def \pr{\partial}
\def \hX{{\hat X}}
\def \bx{{\bf x}}
\def \hx{{\hat x}}
\def \hy{{\hat y}}
\def \hz{{\hat z}}
\def \hs{{\hat s}}

\def \tr{{\rm tr \, }}

\def \vep{\varepsilon}
\def \half{{\textstyle {1 \over 2}}}
\def \thir{{\textstyle {1 \over 3}}}
\def \quar{{\textstyle {1 \over 4}}}

\def \hh{{1\over 2}}
\def \ts{ \textstyle}

\def \A{{\cal A}}
\def \B{{\cal B}}
\def \E{{\cal E}}

\def \H{{\cal H}}
\def \I{{\cal I}}
\def \O{{\cal O}}

\def \R{{\cal R}}
\def \L{{\cal L}}

\def \ep{\epsilon}
\parskip 5pt
{\nopagenumbers
\rightline{DAMTP/93-31}
\vskip 2truecm
\centerline {\bigbf Implications of Conformal Invariance in Field
Theories}
\vskip 5pt
\centerline {\bigbf for General Dimensions}
\vskip 2.0 true cm
\centerline {H. Osborn and A. Petkos}
\vskip 10pt
\centerline {Department of Applied Mathematics and Theoretical Physics,}
\centerline {Silver Street, Cambridge, UK}
\vskip 2.0 true cm
The requirements of conformal invariance for two and three point functions
for general dimension $d$ on flat space are investigated. A compact group
theoretic construction of the three point function for arbitrary spin fields
is presented and it is applied to various cases involving conserved vector
operators and the energy momentum tensor. The restrictions arising from the
associated conservation equations are investigated. It is shown that there
are, for general $d$, three linearly independent conformal invariant forms for
the three point function of the energy momentum tensor, although for $d=3$
there are two and for $d=2$ only one. The form of the three point function
is also demonstrated to simplify considerably when all three points lie on
a straight line. Using this the coefficients of the conformal invariant
three point functions are calculated for free scalar and fermion theories in
general dimensions and for abelian vector fields when $d=4$. Ward identities
relating three and two point functions are also discussed. This requires
careful analysis of the singularities in the short distance expansion and
the method of differential regularisation is found convenient. For $d=4$
the coefficients appearing in the energy momentum tensor three point function
are related to the coefficients of the two possible terms in the trace
anomaly for a conformal theory on a curved space background.
\vfill\eject}
\pageno=1

\leftline{\bigbf 1 Introduction}
\medskip
There has been an enormous literature in recent years  devoted to
conformal field theories in two dimensions. The starting point of such
discussions [1,2] is almost invariably the operator product expansion for
$T(z)=T_{zz}(x)$, where $T_{\mu\nu}(x)$ is the traceless two dimensional
energy momentum tensor and $z=x_1+i x_2$, which has the form
$$ T(z) T(w) \sim {\half c\over (z-w)^4} + {2\over (z-w)^2} T(w)
+ {1\over z-w} T'(w) \, .
\eqno (1.1)$$
An alternative and equivalent result for this operator product expansion
is given by the conformal invariant forms for the two and three point
functions of $T(z)$,
$$ \eqalign {
\l T(z) \, T(w) \r = {}& {\half c\over (z-w)^4} \, , \cr
\l T(z_1) \, T(z_2) \, T(z_3) \r = {}& {c \over (z_1-z_2)^2 (z_2-z_3)^2
(z_3 - z_1)^2} \, . \cr}
\eqno (1.2) $$
Writing $T(z)=\sum_n L_n z^{-n-2}$ then (1.1) is also equivalent to the
Virasoro algebra where where $c$ is the central charge. When $c \leq 1$,
there is a  complete classification of unitary conformal field theories
but also for $c>1$ there is a extremely large set of explicitly
constructed examples of conformal
field theories with very rich mathematical structure.
Besides string theory these are of relevance to two
dimensional statistical physics systems at critical points defined by
vanishing $\beta$ functions where the trace of the energy momentum
tensor vanishes and the systems enjoy conformal invariance.  The work
on conformal field theories in two dimensions has also led to further
understanding of non  conformally invariant quantum field theories.
Motivated by the definition of $c$ in terms of the two point function
of the energy momentum tensor as in (1.2) in conformal field theories
Zamolodchikov [3] showed how to define for non conformal theories a quantity
$C(t)$, where $t = \ln a + {\rm const.}$ for $a$
some distance scale, which is monotonically decreasing under renormalisation
flow for $t$ increasing and is stationary only at critical points where
the $\beta$ function vanishes. As $t\to \infty$ and the critical point is
approached $C(\infty)$ becomes equal to the Virasoro central charge of
the conformal field theory associated with the critical point.

Although the conformal group in two  dimensions is infinite
dimensional, and additional powerful constraints such as modular
invariance are present, there remains a clear motivation for studying
conformal field theories in three and four dimensions which are
respectively relevant for realistic statistical physics systems at
their critical points and for physical quantum field theories.
However conformal field theories in dimensions $d > 2$ are relatively
far less well explored although much work was undertaken twenty or so years
ago [4,5,6,7,8,9,10,11,12] under the influence of the discovery of scaling
in deep inelastic scattering and more recently there have been discussions
motivated in part by the wish to extend two dimensional results [13,14,15,16].
In three dimensions there should be many
non-trivial examples corresponding to the different universality
classes of statistical physics systems at their various critical points.
Nevertheless, it is very difficult to construct explicitly solvable
examples except in some approximation based on free field  theories.
In four dimensions it is likely that conformal field theories are
rather rare, based on the strong evidence of triviality for most quantum field
theories when the cut off is removed.  Possible examples are $N = 4$
supersymmetric gauge theories when the $\beta$ function is identically
zero and also $SU(N)$ gauge theories in the $N \rightarrow \infty$
limit with large numbers of fermions in the fundamental representation.
If the number of fermion representations $f = {11 \over 2} N-k$
with $k \ll N$ then there is an infra-red stable fixed point where
$\beta(g_*)=0$ for $g_*^2 N = {\rm O}(N^{-1})$ which should be
perturbatively accessible [17,18].  For $d=4 - \vep$, and also $d = 2 + \vep$,
there are of course examples of conformal field theories whose critical
exponents may be determined as an expansion in $\vep$.

In general for $d > 2$ it is not clear what are the crucial parameters
specifying a conformal field theory analogous to $c$ when $ d=2$.  The
most obvious generalisation of $c$ is in terms of the overall scale of
the two point function of the energy momentum tensor which has a unique
form in the conformal limit. Indeed, Cappelli {\it et al.} [19]
and also Shore [20] considered this in an attempt to
generalise the Zamolodchikov $c$-theorem to $d > 2$.  This endeavour
is unfortunately not fully successful in that the supposed candidate
for $C(t)$ is not monotonic under renormalisation flow.
An alternative approach considered by Cardy [21], and pursued subsequently
by Jack and Osborn [18,22], is through the
trace of the energy momentum tensor $T_{\mu \nu}$ on curved space which
is a $c$-number in the conformal limit. In two dimensions for conformal
theories $g^{\mu\nu} T_{\mu\nu} \propto R$, the scalar curvature, where
the coefficient of proportionality provides an alternative
definition of the Virasoro central charge $c$.  In four dimensions,
assuming conformal invariance, $g^{\mu\nu} T_{\mu\nu}$ contains two terms
proportional to $F,\, G$ where $F,\, G$ are dimension 4 scalars
constructed from the Riemann tensor.  The coefficient of $F$ is
related to the two point function of the energy momentum tensor while
the coefficient of $G$, which is the Euler density, was suggested by Cardy,
and subsequently analysed in perturbation theory by Jack and Osborn,
as a candidate for a $c$-theorem.  Although the perturbative evidence
is encouraging [18] there are no general results for the desired
positivity conditions which ensure a monotonic renormalisation flow.

In this paper the intention is to initiate an analysis of three point
functions of the energy momentum tensor and other operators in dimensions
$d>2$, extending the results in (1.1) and (1.2). In
conformal field theories three point functions are essentially unique
since the conformal group is transitive on three points.  We will show that
there are three linearly independent forms for the
conformally invariant three point function for the energy
momentum tensor in arbitrary dimension.
However in three dimensions this is restricted to two and for $d=2$ to one
(and the coefficient is just the central charge $c$ again).  In four
dimensions the three coefficients present in the general conformally invariant
expression may be related to the coefficients of $F,\, G$ in
the expansion of the trace of the energy momentum tensor on curved space [23].

The analysis of the consequences of conformal invariances rapidly becomes
tedious with the proliferation of indices.  In the  next section we describe
a group theoretic approach for conformal invariant three point
functions which is based on a paper of Mack [10] some time ago.  This gives
a relatively compact construction and allows a simple analysis of the
conditions for conservation of the energy momentum tensor $\partial_\mu
T_{\mu \nu} = 0$ and also for vector currents $V_\mu$ when we require
$\partial_\mu V_\mu = 0$. The construction is shown to be both necessary and
sufficient for any conformal invariant three point function by showing how
the complete three point function is determined by the the leading term in
the operator product expansion for two operators.
In section 3 the general construction is applied to various cases
involving $T_{\mu \nu}$ and $V_\mu$ and also scalar operators of arbitrary
dimension.  In section 4 an alternative approach based on requiring
the three points to lie on a straight line is described.  In this case
the conformal invariant form simplifies and a straightforward
algebraic approach is feasible.  It is also possible to connect
specific cases to the more general analysis of the previous section.
The discussion of this collinear configuration is further advantageous
since it
is then feasible to derive fairly simply the forms of the three point
function in the examples of conformal field theories in general
dimensions provided by free scalars and fermions.  This is undertaken
in section 5 where we calculate the coefficients defining the three
point function of the energy momentum tensor in these cases as well
as for other examples discussed earlier.  For $d = 4$ we  also
consider the conformal field theory defined by free vector fields.
These three field theories give rise to three linearly independent
forms for the energy momentum tensor three point function so that
the general form may be realised as a linear combination of the
expressions for free scalars, fermions and vectors.

In section 6 we discuss the Ward identities relating two and three
point functions.  These are derived by considering formulae for
diffeomorphism invariance and scale invariance on curved space and
then reducing to flat space.  To verify these identities using our
general results it is necessary to analyse carefully the form of the
three point function as two points become close together (equivalent
to an operator product expansion) and to ensure that the singular
short distance terms are well defined distributions. This is achieved
by adapting the technique of differential regularisation [24]. As a result
of the Ward identity there is a linear relation between the three
coefficients defining the general three point function of the energy
momentum tensor and the scale of the unique two point function in
a conformal theory. Explicit forms for the leading and next to leading
terms involving scalar and conserved vector fields and also the energy
momentum tensor
in the operator product expansion of the energy momentum tensor with the
same operators are found. In section 7 we also construct a Hamiltonian
operator $H$ from the energy momentum tensor. Knowing the action of
$H$ enables us to relate any three point function containing $T_{\mu\nu}$
and the corresponding two point function without $T_{\mu\nu}$. The
same formulae as in section 6 are obtained but without the need for
careful regularisation of the coefficient functions in the
short distance operator product expansion. In section 8 the effects of
$c$-number contributions, depending on external fields and the metric in a
curved space background, to the trace of the energy momentum tensor are
considered when $d=4$. These correspond to anomalies in the simple conformal
Ward identities and reflect the existence of singularities in the three point
functions when all points are coincident. Using dimensional regularisation
the anomalous terms are considered for the three point function of the energy
momentum tensor with two scalar fields with dimension $\eta \approx 2$ and with
two conserved currents and for the three point function of the energy
momentum tensor itself. In the last case the coefficients in the general three
point function are related to the coefficients of the $F$ and $G$ terms in the
trace of the energy momentum tensor on curved space. In section 9 we show
that in some cases an alternative form for three point functions involving
vector currents and the energy momentum tensor, in which their conservation is
manifest, is possible. This is achieved by pulling out derivatives from the
expression for the three point function. In the final section 10
some general remarks about our results and possible directions for future
investigations are presented. In appendix A some formulae used in the
discussion
of the energy momentum tensor three point function are collected.
%\vfill\eject
\bigskip
\leftline{\bigbf 2 Conformal Invariance for Two and Three Point
Functions}
\medskip
Conformal transformations may be defined as coordinate transformations
preserving the infinitesimal euclidean length element up to a local
scale factor,
$$ x_\mu \to x'{}_{\! \mu} (x) = (gx)_\mu \, ,
\quad dx'{}_{\! \mu} dx'{}_{\! \mu} = \Omega^g(x)^{-2} dx_\mu dx_\mu \ .
\eqno (2.1) $$
For any such conformal transformation $g$ we may define a local orthogonal
transformation by
$$\R^g_{\mu \alpha}(x) = \Omega^g (x)
{\pr x'{}_{\! \mu} \over \pr x_\alpha} \ , \quad \R^g_{\mu \alpha} (x)
\R^g_{\nu \alpha} (x) = \delta_{\mu \nu} \, ,
\eqno (2.2) $$
which in $d$ dimensions is an element of $O(d)$, $\R^{g'}(gx) \R^g (x) =
\R^{g'g}(x)$,  $\R^g(x)^{-1} = \R^{g^{-1}}(gx)$.

Besides conventional constant rotations and translations forming the group
$O(d)\ltimes T_d$
$$ x'{}_{\!\mu} = R_{\mu \nu} x_\nu +  a_\mu \, , \quad
\quad R_{\mu \alpha} R_{\nu \alpha}= \delta_{\mu \nu} \, ,
\eqno (2.3) $$
for which $\Omega^g (x) =1$, there are also constant scale
transformations forming the dilatation group $D$
$$ x'{}_{\! \mu} = \lambda x_\mu \, , \quad \Omega^g (x) = \lambda^{-1}
\eqno (2.4) $$
and special conformal transformations
$$ x'{}_{\! \mu} =  { x_\mu + b_\mu x^2 \over \Omega^g (x)} \, , \quad
\Omega^g (x) = 1 + 2 b{\cdot x} + b^2 x^2 \ .
\eqno (2.5) $$
The set of conformal transformations $\{g\}$ defined by (2.1)
forms the conformal group which is isomorphic to $O(d+1,1)$. It is crucial
to note for our subsequent calculations that the full conformal group may be
generated just by combining rotations and translations, as in (2.3), with
an inversion through the origin represented by the discrete element $i$,
$i^2=1$,
$$ x'{}_{\!\mu} = (ix)_\mu = {x_\mu \over x^2} \, , \quad \R^i_{\mu \nu} (x) =
I_{\mu \nu}(x) \equiv \de_{\mu \nu} - 2 \, {x_\mu x_\nu \over x^2} \, ,
\quad \Omega^i (x) = x^2 \ .
\eqno (2.6) $$
Inversions are not elements of the component of the conformal group
connected to the identity, since $\det I = -1$, but
special conformal transformations in (2.3) are formed by considering
an inversion, a translation and then another inversion.

For a quasi-primary quantum field $\O(x)$ of scale dimension $\eta$
then a finite dimensional representation under conformal transformations
(2.1) is induced by a representation of the little group
$\bigl(O(d)\otimes D\bigl) {} \ltimes T_d$ [4] if $\O \to T(g)\O$ where
$$ \bigl ( T(g) \O \bigl ){}^i (x') =
\Omega^g (x)^{\eta}  D^i {}_{\! j} ( \R^g(x) ) \O^j (x) \, ,
\eqno (2.7) $$
with $\R^g(x)$ as given by (2.2). The index $i$ here denotes the components in
some representation of the rotation group $O(d)$ so that for $R_{\mu \nu}$ any
orthogonal rotation matrix $D^i {}_{\! j} (R)$ is the corresponding element
in this representation acting on the fields $\O^i$\footnote{*}{We assume that
if $\O^i$ forms a representation of $SO(d)$ it may be extended to $O(d)$. In
this case ${\bar D}(R) \equiv D(I^{-1}RI) \simeq D(R)$ if $\det I = -1$. For
some
representations this is not possible but such complications [9] are not
important
here.}.
For two such fields $\O_1^{i_1}(x_1)$ and $\O_2^{i_2} (x_2)$ of equal scale
dimension, $\eta_1 = \eta_{2} = \eta $,
we may define a conformally invariant two point function by
$$ \l \O_1^{i_1}(x_1)\, \O_2^{i_2} (x_2) \r =
{1\over (x_{12}^{\, 2})^\eta_{\vphantom d}} \, P^{i_1 i_2} ( x_{12} ) \, ,
\quad x_{12} = x_1 - x_2 \ ,
\eqno (2.8) $$
if $P^{i_1 i_2} (x) $ is required to satisfy
$$ D_1^{\, i_1} {}_{\! j_1}(\R(x_1) )  D_2^{\, i_2} {}_{\! j_2}(\R(x_2))
P^{j_1 j_2} (x_{12}) = P^{i_1 i_2} (x'_{12} ) \, , \quad
P^{i_1 i_2} (\lambda x) = P^{i_1 i_2} (x) \ ,
\eqno (2.9) $$
using $x_{12}'^{\, 2} = x_{12}^{\, 2} /\bigl ( \Omega^g(x_1)\Omega^g(x_2)\bigl
)$.
A solution of this condition is provided by
$$ P^{i_1 i_2} (x_{12} ) =
D_1^{\, i_1} {}_{\! j_1} ( I(x_{12})) \, g^{j_1 i_2} \, ,
\eqno (2.10) $$
where $g^{i_1 i_2}$ is an invariant tensor for the representations
$D_1$ and $D_2$, i.e.
$$
D_1^{\, i_1} {}_{\! j_1} ( R) D_2^{\, i_2} {}_{\! j_2} (R)  g^{j_1 j_2} =
g^{i_1 i_2} \ \hbox{for all}\ R \, .
\eqno (2.11) $$
Since $D(R) D(I(x)) D(R)^{-1} = D(I(Rx))$ to verify that (2.10)
satisfies (2.9) it is sufficient to demonstrate that $D(I(x_1))
D (I(x_{12})) D(I(x_2)) = D(I(x'_{12}))$ which follows from, by
direct calculation,
$$ I_{\mu \alpha} (x_1) I_{\alpha \beta} (x_{12} ) I_{\beta \nu}
(x_2) = I_{\mu \nu} (x'_{12}) \, , \quad x'_{12} = {x_1 \over x_1^{\, 2}}
- {x_2 \over x_2^{\, 2}} \ .
\eqno (2.12) $$
For fields of differing spins then of course there is no tensor
$g^{i_1 i_2}$ satisfying (2.11) and the two point function is zero.

The above results show that $D(I(x_{12}))$ acts effectively as a
parallel transport matrix between $x_1$ and $x_2$
for local conformal rotations. This is crucial in constructing
an analogous formula for three point functions, adapting some
results of Mack [10] in the context of operator product expansions.
Since conformal transformations map any three points into any other
three points the three point function is also essentially unique
in general dimension $d$. Our discussion for arbitrary representations
for the fields $\O_1 , \O_2 , \O_3$ is based on writing
$$ \eqalign {
\l \O_1^{i_1}(x_1)\, \O_2^{i_2} (x_2) \, \O_3^{i_3} (x_3) \r = {}&
{1\over (x_{12}^{\, 2})^{\de_{12}}\,(x_{23}^{\, 2})^{\de_{23}}\,
(x_{31}^{\, 2})^{\de_{31}}}\cr
& \times D_1^{\, i_1} {}_{\! j_1} (I(x_{13}))
D_2^{\, i_2} {}_{\! j_2} (I(x_{23})) \, t^{j_1 j_2 i_3} (X_{12}) \, ,
\cr }
\eqno (2.13) $$
where $t^{i_1 i_2 i_3}(X)$ is a homogeneous tensor satisfying
$$ \eqalign {
 D_1^{\, i_1} {}_{\! j_1} (R) D_2^{\, i_2} {}_{\! j_2} (R)
D_3^{\, i_3} {}_{\! j_3} (R) & \, t^{j_1 j_2 j_3} (X) =
t^{i_1 i_2 i_3}(RX) \ \hbox {for all}\ R \ ,\cr
t^{i_1 i_2 i_3}(\lambda X) = {}& \lambda^q t^{i_1 i_2 i_3}(X) \cr}
\eqno (2.14) $$
and
$$ X_{12} = -X_{21} = {x_{13} \over x_{13}^{\, 2}} -
{x_{23} \over x_{23}^{\, 2}} \, ,
\quad X_{12}^{\, 2} = {x_{12}^{\, 2}\over x_{13}^{\, 2} x_{23}^{\, 2}}
\ . \eqno (2.15) $$
By virtue of (2.13,14,15) the scaling properties of the fields are
satisfied if
$$ \eqalign {
\de_{12} = {}& \half (\eta_1 + \eta_2 - \eta_3 + q ) \, , \cr
\de_{23} = {}& \half (\eta_2 + \eta_3 - \eta_1 - q ) \, , \cr
\de_{31} = {}& \half (\eta_3 + \eta_1 - \eta_2 - q ) \, . \cr}
\eqno (2.16) $$
By a rescaling $t^{i_1 i_2 i_3}(X) \to (X^2)^{-p}t^{i_1 i_2 i_3}(X)$
then $q\to q-2p$ so that given the form of $X_{12}^{\, 2}$ in (2.15)
the expression (2.13) is unchanged. Clearly it is possible to set
$q=0$ or $q=1$ which is sometimes convenient later. The verification that
(2.13) is in accord with the required transformation properties of
the fields, as given by (2.7), depends on (2.14) together with the
essential result that for an inversion
$$
I_{\mu \nu} (x_3) X_{12\, \nu} = {1\over x_3^{\, 2}} \, X'_{12\, \mu}
\ , \eqno (2.17) $$
where $X'_{12}$ is formed from $x'_1, x'_2 , x'_3$ as in (2.15).

The expression (2.13) for the three point function is asymmetric
in its treatment of the external fields. Nevertheless this is only
apparent. If we use
$$
I_{\mu \alpha}(x_{13}) I_{\alpha \nu} ( x_{23} ) = I_{\mu \alpha}
(x_{12}) I_{\alpha \nu} (X_{13}) \, , \quad
I_{\mu \alpha}(x_{23}) I_{\alpha \nu} ( x_{13} ) = I_{\mu \alpha}
(x_{21}) I_{\alpha \nu} (X_{23}) \, ,
\eqno (2.18) $$
as well as
$$
I_{\mu \alpha}(x_{23}) X_{12\, \alpha} = {x_{12}^{\, 2} \over
x_{13}^{\, 2}}\, X_{13 \, \mu} \, , \quad
I_{\mu \alpha}(x_{13}) X_{12\, \alpha} = {x_{12}^{\, 2} \over
x_{23}^{\, 2}}\, X_{32 \, \mu} \, ,
\eqno (2.19) $$
then we may show, by virtue of (2.14),
$$ \eqalign {
D_1^{\, i_1} {}_{\! j_1} & (I(x_{13}))
D_2^{\, i_2} {}_{\! j_2} (I(x_{23})) \, t^{j_1 j_2 i_3} (X_{12}) \cr
& = \Bigl ( {x_{12}^{\, 2} \over x_{13}^{\, 2}} \Bigl )^q \,
D_1^{\, i_1} {}_{\! j_1} (I(x_{12}))
D_3^{\, i_3} {}_{\! j_3} (I(x_{32})) \, {\tilde t}^{\, j_1 i_2 j_3}
(X_{13}) \, , \cr
& = \Bigl ( {x_{12}^{\, 2} \over x_{23}^{\, 2}} \Bigl )^q \,
D_2^{\, i_2} {}_{\! j_2} (I(x_{21}))
D_3^{\, i_3} {}_{\! j_3} (I(x_{31})) \, {\hat t}^{\, i_1 j_2 j_3}
(X_{32}) \, , \cr
& {\tilde t}^{\, i_1 i_2 i_3} (X) =  D_1^{\, i_1} {}_{\! j_1}
(I(X)) \, t^{j_1 i_2 i_3} (X) \, , \quad {\hat t}^{\, i_1 i_2 i_3} (X)
=  D_2^{\, i_2} {}_{\! j_2} (I(X)) \, t^{i_1 j_2 i_3} (X) \, . \cr}
\eqno (2.20) $$
Hence we may equivalently write an analogous representation for
$ \l \O_1^{i_1}(x_1)\, \O_2^{i_2} (x_2) \, \O_3^{i_3} (x_3) \r$ to that
given by (2.13) and (2.16) with on the r.h.s $2 \leftrightarrow 3$ and
$t^{j_1 j_2 i_3} \to {\tilde t}^{\, j_1 i_2 j_3}$ or $1 \leftrightarrow 3$ and
$t^{j_1 j_2 i_3} \to {\hat t}^{\, i_1 j_2 j_3}$.
If the three point function is symmetric, for all fields $\O_1 , \O_2 ,
\O_3$ belonging to the same representation, then it is necessary that
$$
t^{i_2 i_1 i_3}(X) = t^{i_1 i_2 i_3} (-X) \, , \quad
 D^{\, i_1} {}_{\! j_1} (I(X)) \, {t}^{j_1 i_2 i_3} (X) =
t^{i_3 i_1 i_2} (-X) \ .
\eqno (2.21) $$

We are here mostly interested in applying this formalism to cases
involving vector fields $V_\mu (x)$, of dimension $d-1$, and the
energy momentum tensor $T_{\mu \nu}(x)$, which is symmetric and
traceless and of dimension $d$. These satisfy the conservation
equations
$$ \pr_\mu V_\mu = 0 \, , \qquad \pr_\mu T_{\mu \nu} = 0 \ .
\eqno (2.22) $$
The two point functions, in accord with the general result in
(2.8) and (2.10), are
$$ \eqalign {
\l V_{\mu} (x) \, V_\nu (0) \r = {}&{C_V\over x^{2(d-1)}} \, I_{\mu \nu}
(x) \, , \quad
\l T_{\mu \nu}(x) \, T_{\si \rho} (0)\r = {C_T\over x^{2d}} \,
\I_{\mu \nu,\si \rho} (x) \, , \cr
\I_{\mu \nu,\si \rho} & (x) = \half \bigl ( I_{\mu \si}(x) I_{\nu \rho}
(x) + I_{\mu \rho}(x) I_{\nu \si} (x) \bigl ){} - {1\over d}\, \de_{\mu
\nu} \de_{\si \rho} \ . \cr}
\eqno (2.23) $$
$C_V,\, C_T$ are constants determining the overall scale of these
two point functions.
$\I_{\mu\nu,\si\rho}$ represents the inversion operator on symmetric
traceless tensors. Corresponding to $I(x)^2=1$ we have
$$ \I_{\mu \nu,\alpha \beta}(x) \, \I_{\alpha \beta, \si \rho} (x)
= \half \bigl (\de_{\mu \si} \de_{\nu \rho} + \de_{\mu \rho}
\de_{\nu \si} \bigl){} - {1\over d} \, \de_{\mu \nu} \de_{\si \rho}
\equiv \E_{\mu\nu,\si\rho} \, ,
\eqno (2.24) $$
with $\E$ representing the projection operator onto the space of symmetric
traceless tensors.
It is easy to verify that (2.23) is consistent with the conservation
equations (2.22) (see also below).

For a three point function involving a vector field we may write
from (2.13), for $\eta_\pm = \eta_2 \pm \eta_3$,
$$ \eqalign {
\l V_\mu (x_1)\, \O_2^{i_2} (x_2) \, \O_3^{i_3} (x_3) \r = {}&
{1\over x_{12}^{\, d-1+\eta_- + q}\, x_{23}^{\, \eta_+ - d+1-q}\,
x_{31}^{\, d-1-\eta_- - q}}\cr
& \times I_{\mu \nu} (x_{13}) D_2^{\, i_2} {}_{\! j_2} (I(x_{23})) \,
t_\nu {}^{\! j_2 i_3} (X_{12}) \, .
\cr }
\eqno (2.25) $$
To verify current conservation we may use
$$ \eqalign {
&\pr_{\mu} \Bigl ( {1\over (x^{2})^\lambda } \, I_{\mu \nu}
(x) \Bigl ){} = 2(\lambda - d+1)\, {x_{\nu}\over
(x^{2})^{\lambda + 1} } \, ,\cr
& \pr_{1\mu} \Bigl ( {1\over (x_{13}^{\, 2})^\lambda\, (x_{12}^{\, 2})
^{d-1-\lambda}} \, I_{\mu \nu}
(x_{13}) \Bigl ){} = 2(\lambda - d+1)\,
{1\over (x_{13}^{\, 2})^{\lambda + 1}\, (x_{12}^{\, 2})^{d-1-\lambda}}
\, {X_{12\, \nu} \over X_{12}^{\, 2}} \, , \cr
& ~~~~~~~~~~ \pr_{1\mu} \, X_{12\, \nu} =
{1\over x_{13}^{\, 2}} \,  I_{\mu \nu}(x_{13}) \ .\cr}
\eqno (2.26) $$
Hence current conservation in (2.25) requires
$$ \Bigl ( \pr_\mu  - (d-1+\eta_- +q)
{X_\mu \over X^2}\Bigl ) t_\mu {}^{\! i_2 i_3} (X) = 0 \ .
\eqno (2.27) $$
It is easy to see that this is invariant under rescaling of
$t_\mu {}^{\! i_2 i_3} (X)$ by some power of $X^2$ with the
appropriate change in $q$.

For the energy momentum tensor we may also write
$$ \eqalign {
\l T_{\mu \nu} (x_1)\, \O_2^{i_2} (x_2) \, \O_3^{i_3} (x_3) \r = {}&
{1\over x_{12}^{\, d+\eta_- + q}\, x_{23}^{\, \eta_+ - d -q}\,
x_{31}^{\, d-\eta_- -q}} \cr
& \times\I_{\mu \nu,\si\rho}(x_{13}) D_2^{\, i_2} {}_{\! j_2}
(I(x_{23})) \, t_{\si \rho} {}^{\! j_2 i_3} (X_{12}) \, .
\cr }
\eqno (2.28) $$
In this case for determining the corresponding conservation equation
$$ \eqalign {
&\pr_{\mu} \Bigl ( {1\over (x^{2})^\lambda } \, \I_{\mu \nu,\si \rho}
(x) \Bigl ){} = 2(\lambda - d)\, {x_{\nu}\over
(x^{2})^{\lambda + 1} } \,\Bigl ( \half \bigl (x_\si I_{\nu \rho}(x)
+ x_\rho I_{\nu \si} (x) \bigl ) {} + {1\over d} \, x_\nu
\de_{\si \rho} \Bigl ) \, , \cr
& \pr_{1\mu} \Bigl ( {1\over (x_{13}^{\, 2})^\lambda\, (x_{12}^{\, 2})
^{d-\lambda}} \, \I_{\mu \nu,\si \rho}
(x_{13}) \Bigl ){} = 2(\lambda - d)\,
{1\over (x_{13}^{\, 2})^{\lambda + 1}\, (x_{12}^{\, 2})^{d-\lambda}}
\, {1 \over X_{12}^{\, 2}} \cr
& ~~~~~~~~~~~~~~~~~~~~~~~~~\times\Bigl ( \half \bigl ( I_{\nu \rho}(
x_{13}) X_{12\, \si} + I_{\nu \si}(x_{13}) X_{12\, \rho}\bigl){}
- {1\over d}\, I_{\nu \alpha}(x_{13})X_{12 \, \alpha} \, \de_{\si
\rho} \Bigl ) \, . \cr}
\eqno (2.29) $$
Assuming $t_{\mu \nu}{}^{\! i_2 i_3}$ is symmetric and traceless
with respect to $\mu,\nu$ then it is easy to obtain the condition
$$ \Bigl (\pr_\mu  - (d+\eta_- +q)
{X_\mu \over X^2}\Bigl ) t_{\mu\nu} {}^{\! i_2 i_3} (X) = 0 \ .
\eqno (2.30) $$
For both (2.27) and (2.30) the solutions may be restricted to
homogeneous polynomials in $X$ if $q$ is a sufficiently large integer.

The construction described above is manifestly sufficient for forming
conformally invariant three point functions. It is also necessary.
To show this we assume the existence of an operator product expansion where
for $x_1 \to x_2$ the contributions arising from the
leading singular term for each quasi-primary operator
$\O^i$ appearing in the expansion are of the form
$$ \O_1^{i_1}(x_1)\, \O_2^{i_2} (x_2) \sim A^{i_1 i_2}{}_{\! i}(x_{12}) \O^{i}
(x_2) \, .
\eqno (2.31) $$
For conformal invariance, if $\O^i$ on the r.h.s. of (2.31) has
dimension $\eta$ and transforms as in (2.7), we require
$$ \eqalign {
D_1^{\, i_1} {}_{\! j_1} (R) D_2^{\, i_2} {}_{\! j_2} (R)
A^{j_1 j_2}{}_{\! j}(x) D^{j} {}_{\! i} (R) = {}&
A^{i_1 i_2}{}_{\! i}(Rx) \ \hbox {for all}\ R \ ,\cr
A^{i_1 i_2}{}_{\! i}(\lambda x) = {}& \lambda^{\eta - \eta_1 - \eta_2}
A^{i_1 i_2}{}_{\! i}(x) \, . \cr}
\eqno (2.32) $$
To use this result we consider the conformal transformation on the general
three point function arising from an inversion through the point $z$
$$ x \to x' = {x-z \over (x-z)^2} \, .
\eqno (2.33) $$
Using the transformation properties given by (2.6) and (2.7)
$$ \eqalign {
\l & \O_1^{i_1}(x_1)\, \O_2^{i_2} (x_2) \, \O_3^{i_3} (x_3) \r \cr & =
x_1^{\prime \, 2\eta_1} x_2^{\prime \, 2\eta_2} x_3^{\prime \, 2\eta_3}
\, D_1^{\, i_1} {}_{\! j_1} (I(x'_1)) D_2^{\, i_2} {}_{\! j_2} (I(x'_2))
D_3^{\, i_3} {}_{\! j_3} (I(x'_3))
\l \O_1^{j_1}(x'_1)\, \O_2^{j_2} (x'_2) \, \O_3^{j_3} (x'_3) \r \, . \cr }
\eqno (2.34) $$
Now, following similar arguments of Cardy [13], if we let $z\to x_3$ then
$x'_{13} \sim x'_{23} \sim -x'_3$ and hence $|x'_{12}|\ll |x'_{13}|,
\, |x'_{23}|$ so that we may use the operator product expansion (2.31).
Assuming that the operator $\O^i$ has a non zero two point function with
$\O^{i_3}$, and hence $\eta=\eta_3$, then applying the results (2.8,9,10) for
the two point function to this case it is easy to see that
$$ \eqalign {
\l \O_1^{j_1}(x'_1)\, \O_2^{j_2} (x'_2) \, \O_3^{j_3} (x'_3) \r \sim {}&
A^{j_1 j_2}{}_{\! j}(x'_{12})\, \l \O^{j} (x'_2) \, \O_3^{j_3} (x'_3) \r \,
,\cr
\l \O^{j} (x) \, \O_3^{k} (0) \r = {}& {1\over x^{2\eta_3}} \,
D_3^{\, k} {}_{\! j'} (I(x)) g^{jj'} \, . \cr}
\eqno (2.35) $$
Inserting (2.35) in (2.34) and taking the limit, when
$x'_1 \to x_{13}^{\vphantom d}/ x_{13}^{\, 2}, \  x'_2 \to
x_{23}^{\vphantom d}/x_{23}^{\, 2}$ and also
$x'_{12}\to X_{12}$ as defined by (2.15), gives finally
$$
\l \O_1^{i_1}(x_1)\, \O_2^{i_2} (x_2) \, \O_3^{i_3} (x_3) \r =
{1\over x_{13}^{\, 2\eta_1}\,x_{23}^{\, 2\eta_2}} \,
D_1^{\, i_1} {}_{\! j_1} (I(x_{13}))
D_2^{\, i_2} {}_{\! j_2} (I(x_{23})) \,
A^{j_1 j_2}{}_{\! j} (X_{12}) g^{ji_3} \, ,
\eqno (2.36) $$
so that the whole three point function is determined by the leading singular
operator product coefficient.
Comparing with (2.13) shows that the two expressions are equivalent if
$$ t^{i_1i_2i_3} (X) = (X^2)^{\hh(\eta_1 + \eta_2 - \eta_3 + q)} \,
A^{i_1 i_2}{}_{\! i} (X) g^{i i_3} \, .
\eqno (2.37) $$
By virtue of (2.32) $t^{i_1i_2i_3} (X)$ satisfies the conditions required in
(2.14). For a conserved vector field $V_\mu$ or the energy
momentum tensor $T_{\mu\nu}$ it is not difficult to see that (2.27) and (2.30)
are the same as
$$ \pr_\mu A_\mu{}^{i}{}_{j}(X) = 0 \, , \qquad
\pr_\mu A_{\mu\nu}{}^{i}{}_{j}(X) = 0 \, ,
\eqno (2.38) $$
for the corresponding coefficients of the leading singular term in the
operator product expansion.

If the operator product expansion (2.31) is extended to general set of
quasi-primary operators, labelled by $a,b,\dots$, so that
$$ \O_a^i (x) \O_b^j(0) \sim A_{ab,c\, k}^{~~ij} (x) \O_c^k (0) \, ,
\quad A_{ab,c\, k}^{~~ij} (x) = A_{ba,c\, k}^{~~ji} (-x) =
{\rm O} ( x^{\eta_c -\eta_a -\eta_b} ) \, ,
\eqno (2.39) $$
and if the two point functions in this operator basis are diagonal,
$$ \l \O_a^i (x) \, \O_b^j(0) \r = {1\over x^{2\eta_a}} \, D_{a\, k}^{\, i}
(I(x)) g_a^{kj} \, \de_{ab}^{\vphantom g} \, ,
\eqno (2.40) $$
then using (2.20) consistency of different short distance limits
determining the three point function requires
$$
A_{ac,b\, \ell}^{~~ik} (x) g_b^{\ell j} = (x^2)^{\eta_b-\eta_c}_{\vphantom d}
D_{a\, \ell}^{\, i} (I(x)) A_{ab,c\, m}^{~~\ell j} (x) g_c^{m k}  \, .
\eqno (2.41) $$
This condition is essentially equivalent to associativity of the operator
product expansion in this case\footnote{*}{For two vector currents and an axial
current similar conditions were obtained by Crewther [25].}.
\bigskip
\leftline{\bigbf 3 Applications of General Formalism}
\medskip
We here apply the previous general results to various specific
cases involving a scalar field $\O$ of dimension $\eta$, the conserved
vector current $V_\mu$ of dimension $d-1$ and the energy momentum
tensor $T_{\mu \nu}$ of dimension $d$.

For the three point function for energy momentum tensor and two
scalar fields it is easy to see that we may write
$$ \eqalign {
\l T_{\mu \nu} (x_1)\, \O (x_2) \, \O (x_3) \r = {}&
{1\over x_{12}^{\, d \vphantom \eta}\, x_{23}^{\, 2\eta - d}\,
x_{31}^{\, d \vphantom \eta}} \, {t}_{\mu \nu} (X_{23}) \cr
= {}& {1\over x_{12}^{\, d\vphantom \eta}\, x_{23}^{\, 2\eta - d}\,
x_{31}^{\, d\vphantom \eta}} \, \I_{\mu \nu,\si \rho}(x_{13})\,
t_{\si \rho} (X_{12}) \, ,\cr}
\eqno (3.1) $$
where we have taken $t_{\mu \nu}$ to be homogeneous of degree zero.
It is easy to see that the conditions for tracelessness and the
conservation equation coincide to give
$$ t_{\mu \nu}(X) = a\, h^1_{\mu \nu}(\hX) \, , \quad
 h^1_{\mu \nu}(\hX) = \hX_\mu \hX_\nu - {1\over d} \, \de_{\mu \nu}
\, , \quad \hX_\mu = {X_\mu \over \sqrt {X^2}} \ .
\eqno (3.2) $$
For two energy momentum tensors it is natural to write from (2.13)
$$
\l T_{\mu \nu} (x_1)\, T_{\si \rho} (x_2) \, \O (x_3) \r =
{1\over x_{12}^{\, 2d-\eta}\, x_{23}^{\, \eta}\,
x_{31}^{\, \eta}} \,  \I_{\mu \nu,\alpha \beta}(x_{13})
\I_{\si \rho,\gamma\delta}(x_{23})
{t}_{\alpha\beta\gamma\delta} (X_{12}) \, ,
\eqno (3.3) $$
with ${t}_{\alpha\beta\gamma\delta} (X)$ again homogeneous of
degree zero. A general expansion consistent with the required
symmetries and tracelessness is given by
$$ \eqalign {
{t}_{\alpha\beta\gamma\delta} (X) = {}&
a \,h^1_{\alpha \beta}(\hX) h^1_{\gamma \delta}(\hX) +
b \,h^2_{\alpha \beta\gamma \delta}(\hX) +
c \,h^3_{\alpha \beta\gamma \delta} \, ,\cr
h^2_{\mu \nu \si \rho}(\hX) = {}&\hX_\mu \hX_{\si}\de_{\nu \rho}
+ (\mu\leftrightarrow\nu,\si\leftrightarrow\rho) -{4\over d}
\hX_\mu \hX_\nu\de_{\si \rho} -{4\over d} \hX_\si \hX_\rho \de_{\mu\nu}
+ {4\over d^2} \de_{\mu \nu} \de_{\si\rho} \, , \cr
h^3_{\mu \nu \si \rho} = {}& \de_{\mu \si} \de_{\nu\rho}
+\de_{\mu \rho} \de_{\nu\si} -
{2\over d}\, \de_{\mu \nu} \de_{\si\rho} \, , \cr}
\eqno (3.4) $$
where $h^1_{\mu\nu}(\hX)$ is defined in (3.2). The conservation
equation from (2.30) becomes
$$ \Bigl (\pr_\mu  - (2d-\eta)
{X_\mu \over X^2}\Bigl ) t_{\mu\nu\si\rho} (X) = 0 \, ,
\eqno (3.5) $$
which gives two linear relations between $a,b,c$
$$ \eqalign {
& a+4b - \half(d-\eta)(d-1)(a+4b) - d\eta \, b =0 \, , \cr
& a+4b + d(d-\eta)b + d(2d-\eta)c = 0 \, . \cr}
\eqno (3.6) $$
In consequence the conformal invariant form for the $\l TT\O \r$
three point function is unique up to an overall constant.

For vector currents we write
$$ \l V^a_\mu (x_1) \, V^b_\nu (x_2) \, V^c_\omega (x_3) \r
= {f^{abc}\over x_{12}^{\, d \vphantom \eta}\, x_{23}^{\, d-2}\,x_{31}^{\,
d-2}} \,
I_{\mu \alpha}(x_{13}) I_{\nu\beta}(x_{23}) t_{\alpha \beta \omega}
(X_{12}) \, ,
\eqno (3.7) $$
where $a,b,c$ are to be regarded here as
group indices with $f^{abc}$ a corresponding totally antisymmetric
structure constant and we have set $q=1$. From (2.21) and (2.27) we
require
$$ \eqalign {
I_{\mu \alpha}(X) t_{\alpha \nu \omega} (X) =  - t_{\omega \mu
\nu} (X) \, , \quad &t_{\mu \nu \omega}(X) = t_{\nu \mu \omega}(X) \, , \cr
\Bigl ( \pr_\mu - d\, {X_\mu \over X^2} \Bigl ) t_{\mu \nu \omega} (X)
= {}& 0 \, . \cr}
\eqno (3.8) $$
It is easy to see that the general solution is
$$ t_{\mu \nu \omega}(X) = a\, {X_\mu X_\nu X_\omega\over X^2}
+ b \bigl ( X_\mu \de_{\nu\omega} + X_\nu \de_{\mu \omega} -
X_{\omega} \de_{\mu \nu} \bigl ) \, ,
\eqno (3.9) $$
for independent constants $a,b$. By using results such as (2.18,19)
the $\l VVV\r$ three point function may then be expressed
more symmetrically as
$$ \eqalign {
\l V^a_\mu (x_1) \, V^b_\nu (x_2) \, V^c_\omega (x_3) \r
= {}&{f^{abc}\over x_{12}^{\, d-2}\, x_{23}^{\, d-2}\,x_{31}^{\, d-2}}
\biggl \{(a-2b) X_{23\, \mu}\, X_{31\, \nu}\, X_{12\, \omega} \cr
{} - b \Bigl( &
{1\over x_{23}^{\, 2}}X_{23\, \mu} I_{\nu \omega}(x_{23}) +
{1\over x_{13}^{\, 2}}X_{31\, \nu} I_{\mu \omega}(x_{13}) +
{1\over x_{12}^{\, 2}}X_{12\, \omega} I_{\mu \nu}(x_{12})\Bigl )
\biggl \} \, ,\cr}
\eqno (3.10) $$
which is in accord with previous results found over twenty years
ago [7].

For the three point function involving two vector currents and
the energy momentum tensor we may write
$$ \eqalign {
\l T_{\mu \nu}^{\vphantom v}(x_1) \,V^a_\si (x_2) \, V^b_\rho (x_3) \r
& = {\de^{ab}\over x_{12}^{\, d \vphantom \eta}\,
x_{13}^{\, d \vphantom \eta}\,x_{23}^{\, d-2}} \,
I_{\si \alpha}(x_{21}) I_{\rho\beta}(x_{31}) t_{\mu\nu\alpha \beta}
(X_{23}) \cr
& = {\de^{ab}\over x_{12}^{\, d \vphantom \eta}\,
x_{13}^{\, d \vphantom \eta}\,x_{23}^{\, d-2}} \,
\I_{\mu\nu,\gamma\delta}(x_{13})
I_{\si \alpha}(x_{23}) {\tilde t}_{\gamma\delta\alpha \rho}
(X_{12}) \, ,\cr
{\tilde t}_{\mu\nu\si\rho}(X)=I_{\si\alpha}(X)&
t_{\mu\nu\alpha\rho}(X) \, ,
\quad t_{\mu\nu\si\rho}=t_{\mu\nu\rho\si}=t_{\nu\mu\si\rho}\, ,
\quad t_{\mu\mu\si\rho}=0 \, ,\cr}
\eqno (3.11) $$
with $t_{\mu\nu\si\rho}(X)$ homogeneous of degree zero in $X$. The
conservation equations for the vector currents and the energy
momentum tensor require
$$ \eqalign{
\Bigl (\pr_\si - (d-2) {X_\si\over X^2}\Bigl ) & t_{\mu\nu\si\rho}
(X) = 0 \, ,\cr
\Bigl (\pr_\mu - d\, {X_\mu \over X^2}\Bigl ) {\tilde t}_{\mu\nu\si\rho}
(X) = 0 \, , & \quad
\Bigl (\pr_\si - d\, {X_\si \over X^2}\Bigl ) {\tilde t}_{\mu\nu\si\rho}
(X) = 0 \, ,\cr}
\eqno (3.12) $$
where the second equation for ${\tilde t}$ follows from that for $t$ given
its definition in (3.11). A general expression for $t_{\mu\nu\si\rho}(X)$
satisfying the constraints
in (3.11), with the definitions in (3.2) and (3.4), and the corresponding
form for ${\tilde t}_{\mu\nu\si\rho}(X)$ are given by
$$ \eqalign {
t_{\mu\nu\si\rho}^{\vphantom h}(X) = {}& a\, h^1_{\mu\nu}(\hX)
\de_{\si\rho}^{\vphantom b} + b \,h^1_{\mu\nu}(\hX)h^1_{\si\rho}(\hX) +
c\, h^2_{\mu\nu\si\rho}(\hX) + e \, h^3_{\mu\nu\si\rho} \, , \cr
{\tilde t}_{\mu\nu\si\rho}^{\vphantom h}(X) = {}& {1\over d^2}
\bigl ( d(d-2) a -2(d-1)
(b+4c) -4de \bigl ) h^1_{\mu\nu}(\hX)\de_{\si\rho}^{\vphantom h} \cr
& - {1\over d}\bigl( 2da + (d-2)(b+4c) \bigl )
h^1_{\mu\nu}(\hX)h^1_{\si\rho}(\hX) \cr
& + e \bigl( h^3_{\mu\nu\si\rho} - h^2_{\mu\nu\si\rho}(\hX) \bigl ) {} -
(e+c) {\tilde h}_{\mu\nu\si\rho}(\hX) \, , \cr
{\tilde h}_{\mu\nu\si\rho}(\hX) = {}& \hX_\mu \hX_\si \de_{\nu\rho}
+ \hX_\nu \hX_\si \de_{\mu\rho} - \hX_\mu \hX_\rho \de_{\nu\si}
- \hX_\nu \hX_\rho \de_{\mu\si} \, . \cr}
\eqno (3.13) $$
Using results such as (A.2) and (A.4) (3.12) gives the relations
$$ da -2b +2(d-2)c=0 \, , \quad b-d(d-2) e = 0 \, .
\eqno (3.14) $$
Hence $a,b$ may be found in terms of $c,e$ and
there are two linearly independent amplitudes in this case.

With the experience gained in the above simpler cases we may now
turn to consider the three point function for three energy
momentum tensors which is the main aim of this paper. According to
the general formalism we may write
$$ \eqalign {
\l T_{\mu \nu}(x_1) \,T_{\si\rho} (x_2) \, T_{\alpha \beta} (x_3) \r
= {}&{1\over x_{12}^{\, d \vphantom \eta}\,
x_{13}^{\, d \vphantom \eta}\,x_{23}^{\, d \vphantom \eta}} \,
\I_{\mu\nu,\mu'\nu'}(x_{13}) \I_{\si\rho,\si'\rho'}(x_{23}) \,
t_{\mu'\nu'\si'\rho'\alpha \beta} (X_{12}) \, ,\cr}
\eqno (3.15) $$
with $t_{\mu\nu\si\rho\alpha\beta}(X)$ homogeneous of degree zero
in $X$, symmetric and traceless on
each pair of indices $\mu\nu, \ \si\rho$ and $\alpha\beta$ and from
(2.21) satisfying
$$ \eqalignno {
t_{\mu\nu\si\rho\alpha\beta}(X) = {}&
t_{\si\rho\mu\nu\alpha\beta}(X) \, . &(3.16a) \cr
\I_{\mu\nu,\mu'\nu'}(X)t_{\mu'\nu'\si\rho\alpha\beta}(X)={}&
t_{\alpha\beta\mu\nu\si\rho}(X) \, , &(3.16b) \cr}
$$
The conservation equations require just
$$ \Bigl ( \pr_\mu - d\, {X_\mu \over X^2} \Bigl )
t_{\mu\nu\si\rho\alpha\beta}(X) = 0 \ .
\eqno (3.17) $$
Defining
$$ \eqalign {
h^4_{\mu\nu\si\rho\alpha\beta}(\hX) = {}& h^3_{\mu\nu\si\alpha}
\hX_\rho \hX_\beta + (\si\leftrightarrow\rho,\alpha\leftrightarrow
\beta) \cr
& - {2\over d} \, \de_{\si \rho}^{\vphantom h}h^2_{\mu\nu\alpha\beta}(\hX)
- {2\over d} \, \de_{\alpha\beta}^{\vphantom h}h^2_{\mu\nu\si\rho}(\hX)
-{8\over d^2}\, \de_{\si\rho}^{\vphantom h}\de_{\alpha\beta}^{\vphantom h}
h^1_{\mu\nu}(\hX) \, ,\cr
h^5_{\mu\nu\si\rho\alpha\beta} = {}& \de_{\mu\si}^{\vphantom h}
\de_{\nu\alpha}^{\vphantom h}\de_{\rho\beta}^{\vphantom h}
+ (\mu\leftrightarrow\nu,\si\leftrightarrow\rho,
\alpha\leftrightarrow\beta)\cr
&-{4\over d}\,\de_{\mu\nu}^{\vphantom h} h^3_{\si\rho\alpha\beta}
-{4\over d}\,\de_{\si\rho}^{\vphantom h} h^3_{\mu\nu\alpha\beta}
-{4\over d}\,\de_{\alpha\beta}^{\vphantom h} h^3_{\mu\nu\si\rho}
-{8\over d^2} \, \de_{\mu\nu}^{\vphantom h}\de_{\si\rho}^{\vphantom h}
\de_{\alpha\beta}^{\vphantom h} \, , \cr}
\eqno (3.18) $$
a general expansion for $t_{\mu\nu\si\rho\alpha\beta}(X)$
compatible with (3.16a) has the form
$$ \eqalign {
t_{\mu\nu\si\rho\alpha\beta}(X) = {}& a\,
h^5_{\mu\nu\si\rho\alpha\beta} + b\,
h^4_{\alpha\beta\mu\nu\si\rho}(\hX) + b'\bigl (
h^4_{\mu\nu\si\rho\alpha\beta}(\hX) +
h^4_{\si\rho\mu\nu\alpha\beta}(\hX)\bigl) \cr
& + c\, h^3_{\mu\nu\si\rho}h^1_{\alpha\beta}(\hX) + c'\bigl (
h^3_{\si\rho\alpha\beta}h^1_{\mu\nu}(\hX) +
h^3_{\mu\nu\alpha\beta}h^1_{\si\rho}(\hX)\bigl) \cr
& + e\, h^2_{\mu\nu\si\rho}(\hX)h^1_{\alpha\beta}(\hX) + e'\bigl (
h^2_{\si\rho\alpha\beta}(\hX)h^1_{\mu\nu}(\hX) +
h^2_{\mu\nu\alpha\beta}(\hX)h^1_{\si\rho}(\hX)\bigl) \cr
& +f\, h^1_{\mu\nu}(\hX)h^1_{\si\rho}(\hX)h^1_{\alpha\beta}(\hX) \,.
\cr}
\eqno (3.19)$$
In this case it is necessary to be more systematic in solving (3.16b)
and (3.17). From the results in (A.1) (3.16b) gives
$$ b+b'=-2a\, , \quad c'=c \, , \quad e+e'=-4b'-2c \, ,
\eqno (3.20) $$
so that $a,b,c,e,f$ may be regarded as independent. Then using (A.2)
imposition of (3.17) gives in addition
$$ \eqalign{
d^2 a + 2(b+b') -(d-2)b'-dc +e'={}& 0 \, , \cr
d(d+2)(2b'+c)+4(e+e')+f={}& 0 \, . \cr}
\eqno (3.21)$$
Hence there remain three undetermined coefficients which may be
taken as $a,b,c$ ($f=(d+4)(d-2)(4a+2b-c),\ e'=-(d+4)(d-2)a-(d-2)b
+dc,\ e=(d+2)(da+b-c)$).
\bigskip
\leftline{\bigbf 4 Collinear Frame}
\medskip
The form of the conformally invariant three point function
simplifies considerably if the three points are constrained to
lie on a straight line. In this configuration the invariance
group is restricted to $O(d-1)$ rotations in the perpendicular
plane, scale transformations and also translations and special
conformal transformations, as in (2.5), along the line. The latter
allow any three points to be mapped to any other three points
preserving cyclic order. With these constraints it is not difficult
to see in general that, for $n_\mu$ a unit vector,
$$\eqalign{
\l \O^{i_1}(x)\, \O^{i_2} (y)& \, \O^{i_3} (z) \r \bigl |_{x_\mu=\hx
n_\mu,y_\nu=\hy n_\nu, z_\omega=\hz n_\omega}\cr
& = {1\over (\hx-\hy)^{\eta_1+\eta_2-\eta_3}\,
(\hx-\hz)^{\eta_1+\eta_3-\eta_2}\,
(\hy-\hz)^{\eta_2+\eta_3-\eta_2}}\, \A^{i_1 i_2 i_3} \, ,\cr}
\eqno (4.1) $$
where $\A^{i_1 i_2 i_3}$ is a constant $O(d-1)$ invariant tensor and
we suppose $\hx>\hy>\hz$. An
alternative construction of  conformally invariant three point
functions is then to analyse the constraints in a collinear frame
where the form is particularly simple and then to obtain the results
for a general configuration by conformal transformation. The
calculations are very straightforward except that in order to impose
conservation equations, such as (2.22) it is necessary to consider an
infinitesimal conformal transformation away from
the collinear configuration. We illustrate
this procedure for some of the cases considered in the previous
section here since it is useful to consider the collinear form in
making the connection with specific conformally invariant models
later.

The first non trivial case to consider is that for  $\l TT\O \r$
where for $x_\mu = (\hx,{\bf 0}), \, y_\nu = (\hy,{\bf 0}), \,
z_\omega =(\hz,{\bf 0})$ we may write
$$ \eqalign{
\l T_{\mu \nu} (x)\, T_{\si \rho} (y) \, \O (z) \r
\equiv {}& T^{TT\O}_{\mu\nu\si\rho}(x,y,z) =
{1\over (\hx-\hy)^{2d-\eta}\, (\hx-\hz)^{\eta}\,
(\hy-\hz)^{\eta}} \, \A^{TT\O}_{\mu\nu\si\rho} \, , \cr
&  \A^{TT\O}_{\mu\nu\si\rho}= \A^{TT\O}_{\si\rho\mu\nu}=
\A^{TT\O}_{\nu\mu\si\rho} \, . \cr}
\eqno (4.2) $$
$\A^{TT\O}$ may be decomposed as
$$ \eqalign {
\A^{TT\O}_{1111}={}& \alpha \, , \quad \A^{TT\O}_{ij11}= \beta \,
\de_{ij}^{\vphantom \eta}
,\, \quad \A^{TT\O}_{i1k1}= \gamma \,\de_{ik}^{\vphantom \eta} , \cr
\quad \A^{TT\O}_{ijk\ell} = {}&\delta \,\de_{ij}^{\vphantom \eta}
\de_{k\ell}^{\vphantom \eta} + \ep
(\de_{ik}^{\vphantom \eta}\de_{j\ell}^{\vphantom \eta}
+ \de_{i\ell}^{\vphantom \eta} \de_{jk}^{\vphantom \eta} ) \, , \cr}
\eqno (4.3) $$
where $i,j\dots$ denote components orthogonal to the $1$ direction.
It is easy to see that the tracelessness condition
$\A^{TT\O}_{\mu\mu\si\rho}=0$ requires
$$
\alpha+(d-1)\beta=0 \, ,\quad \beta+(d-1)\delta + 2\ep = 0 \, .
\eqno (4.4)$$
We now consider an infinitesimal conformal transformation $x_\mu\to x_\mu+\de
x_\mu$,
$$ \eqalign {
\de x_i = {}& \xi_i \bigl( (\hy + \hz)x_1 - x^2 - \hy \hz \bigl ){}
+ 2 x_i \, \xi_j x_j \, , \cr
\de x_1 = {} & - \xi_i x_i \bigl (\hy + \hz - 2 x_1 \bigl ) \, ,\cr}
\eqno (4.5) $$
such that
$$
(\hy,{\bf 0})\to (\hy,{\bf 0}) \, , \quad
(\hz,{\bf 0})\to (\hz,{\bf 0}) \, , \quad
(\hx,{\bf 0})\to (\hx,\de{\bf x}) \, , \quad \de {\bf x} =
(\hx-\hy)(\hx-\hz) \bxi \, .
\eqno (4.6) $$
In this case from (2.2)
$$ \R_{\mu\nu}(x) \approx \pmatrix { 1 & -\xi_j(\hy+\hz-2x_1)\cr
\xi_i(\hy+\hz-2x_1) & \de_{ij} \cr} \, ,\quad \Omega(x) \approx
1-2\xi_i x_i \, .
\eqno (4.7) $$
Using this transformation, with $x'_\mu = (\hx,\de {\bf x})$ and
evaluating $\R_{\mu\nu}(x)$ at $(\hx,{\bf 0})$ and $ (\hy,{\bf 0})$,
it is straightforward to find to first order in $\de {\bf x}$
$$ \eqalign {
T^{TT\O}_{i111}(x',y,z) = {}& \Bigl ( {1\over \hx-\hy} +
{1\over \hx-\hz}\Bigl ) \bigl ( \de x_i T^{TT\O}_{1111}(x,y,z) -
\de x_j T^{TT\O}_{ij11}(x,y,z)\bigl ) \cr
{}& -\Bigl ( {1\over \hx-\hy} - {1\over \hx-\hz}\Bigl )
\bigl ( \de x_k T^{TT\O}_{i1k1}(x,y,z) +\de x_\ell T^{TT\O}_{i11\ell}(x,y,z)
\bigl ) \, ,\cr
T^{TT\O}_{i1k\ell}(x',y,z) = {}&
\Bigl ( {1\over \hx-\hy} + {1\over \hx-\hz}\Bigl ) \bigl (
\de x_i T^{TT\O}_{11k\ell}(x,y,z) -\de x_j T^{TT\O}_{ijk\ell}(x,y,z)
\bigl ) \cr
{}& +\Bigl ( {1\over \hx-\hy} - {1\over \hx-\hz}\Bigl ) \bigl (
\de x_k T^{TT\O}_{i11\ell}(x,y,z) +\de x_\ell T^{TT\O}_{i1k1}(x,y,z)
\bigl ) \, ,\cr
T^{TT\O}_{ijk1}(x',y,z) = {}& \Bigl ( {1\over \hx-\hy} +
{1\over \hx-\hz}\Bigl ) \bigl (
\de x_i T^{TT\O}_{1jk1}(x,y,z) +\de x_j T^{TT\O}_{ijk\ell}(x,y,z)
\bigl ) \cr
{}& +\Bigl ( {1\over \hx-\hy} - {1\over \hx-\hz}\Bigl ) \bigl (
\de x_k T^{TT\O}_{ij11}(x,y,z) -\de x_\ell T^{TT\O}_{i1k1}(x,y,z)
\bigl ) \, .\cr}
\eqno (4.8) $$
Hence we may now impose current conservation, $\pr_i T_{i1} + \pr_1
T_{11} =0$ and also $\pr_i T_{ij} + \pr_1 T_{1j} =0$, to give in
addition to (4.4)
$$ 2\gamma = (d-\eta) \beta \, , \quad \beta -\de -d\ep =
(d-\eta) \gamma \, .
\eqno (4.9) $$
With four independent relations given by (4.4) and (4.9) there
is just one possible conformal invariant form up to an overall
constant. It is easy to relate this approach to our previous results.
{}From (3.4) in the collinear frame then $\A_{\mu \nu\si\rho}$ in (4.2)
becomes
$$  \A^{TT\O}_{\mu \nu\si\rho} = a\, h^1_{\mu \nu}(\hX)h^1_{\si\rho}(\hX) +
b\,h^2_{\mu \nu\si\rho}(\hX) + c\, h^3_{\mu \nu\si\rho} \, ,
\eqno (4.10) $$
where $h^1,h^2,h^3$ may be decomposed as in (4.3), with now $\hX_\mu
= (1,{\bf 0})$, to give
$$ \gamma = b+c \, \quad d^2 \, \de = a+4b -2d\, c \, , \quad
\ep = c \, ,
\eqno (4.11) $$
with $\alpha,\beta$ determined by (4.4). The relations (4.9) are then
equivalent to  (3.6).

In a similar fashion we may discuss the $\l TVV\r$ three point function
which may be written in the collinear frame as
$$ \eqalign{
\l T_{\mu \nu}^{\vphantom v} (x)\, V^a_{\si} (y) \, V^b_\rho (z) \r
= {}& \de^{ab} T^{TVV}_{\mu\nu\si\rho}(x,y,z) =
{\de^{ab} \over (\hx-\hy)^{d}\, (\hx-\hz)^{d}\,
(\hy-\hz)^{d-2}} \, \A^{TVV}_{\mu\nu\si\rho} \, , \cr
&  \A^{TVV}_{\mu\nu\si\rho}= \A^{TVV}_{\mu\nu\rho\si}=
\A^{TVV}_{\nu\mu\si\rho} \, , \cr}
\eqno (4.12) $$
where
$$ \eqalign {
\A^{TVV}_{1111}={}& \alpha \, , \quad \A^{TVV}_{ij11}= \beta \,
\de_{ij}^{\vphantom \eta}
\, , \quad \A^{TVV}_{11k\ell}= \gamma \,\de_{k\ell}^{\vphantom \eta} \, , \quad
\A^{TVV}_{i1k1}= \de \,\de_{ik}^{\vphantom \eta}\, , \cr
\quad \A^{TVV}_{ijk\ell} = {}&\rho \,\de_{ij}^{\vphantom \eta}
\de_{k\ell}^{\vphantom \eta} + \tau
(\de_{ik}^{\vphantom \eta}\de_{j\ell}^{\vphantom \eta} +
\de_{i\ell}^{\vphantom \eta} \de_{jk}^{\vphantom \eta} ) \, . \cr}
\eqno (4.13) $$
The tracelessness conditions are
$$
\alpha+(d-1)\beta=0 \, ,\quad \gamma +(d-1)\rho + 2\tau = 0 \, .
\eqno (4.14)$$
As before we may impose the conservation equation $\pr_\mu T_{\mu \nu}
=0$ by considering an infinitesimal conformal transformation so that
$x\to x'$ which is no longer on the straight line defined by $y$ and
$z$
$$ \eqalign {
T^{TVV}_{i111}(x',y,z) = {}& \Bigl ( {1\over \hx-\hy} +
{1\over \hx-\hz}\Bigl ) \bigl ( \de x_i T^{TVV}_{1111}(x,y,z) -
\de x_j T^{TVV}_{ij11}(x,y,z)\bigl ) \cr
{}& -\Bigl ( {1\over \hx-\hy} - {1\over \hx-\hz}\Bigl )
\bigl ( \de x_k T^{TVV}_{i1k1}(x,y,z) -\de x_\ell T^{TVV}_{i11\ell}(x,y,z)
\bigl ) \, ,\cr
T^{TVV}_{i1k\ell}(x',y,z) = {}&
\Bigl ( {1\over \hx-\hy} + {1\over \hx-\hz}\Bigl ) \bigl (
\de x_i T^{TVV}_{11k\ell}(x,y,z) -\de x_j T^{TVV}_{ijk\ell}(x,y,z)
\bigl ) \cr
{}& +\Bigl ( {1\over \hx-\hy} - {1\over \hx-\hz}\Bigl ) \bigl (
\de x_k T^{TVV}_{i11\ell}(x,y,z) -\de x_\ell T^{TVV}_{i1k1}(x,y,z)
\bigl ) \, ,\cr
T^{TVV}_{ijk1}(x',y,z) = {}& \Bigl ( {1\over \hx-\hy} +
{1\over \hx-\hz}\Bigl ) \bigl (
\de x_i T^{TVV}_{1jk1}(x,y,z) +\de x_j T^{TVV}_{ijk\ell}(x,y,z)
\bigl ) \cr
{}& +\Bigl ( {1\over \hx-\hy} - {1\over \hx-\hz}\Bigl ) \bigl (
\de x_k T^{TVV}_{ij11}(x,y,z) +\de x_\ell T^{TVV}_{i1k1}(x,y,z)
\bigl ) \, .\cr}
\eqno (4.15) $$
It is then easy to obtain the additional equation to (4.14)
$$ \beta + \rho + d\tau = 0 \, .
\eqno (4.16) $$
To impose $\pr_\si V_\si =0$ we may shift $y\to y'$ and then find
$$ \beta + \gamma - 2\de =0 \, .
\eqno (4.17) $$
With four relations amongst six variables it is clear that there are
two linearly independent conformal invariant expressions in this case
which may be specified by $\rho,\tau$. To see
the connection with our earlier treatment we may use (3.13) to
find in the collinear frame
$$
\A^{TVV}_{\mu\nu\si\rho} = a\, h^1_{\mu\nu}(\hX)\de_{\si\rho}^{\vphantom h}
+ (b+8c+8e)\, h^1_{\mu\nu}(\hX) h^1_{\si\rho}(\hX)
- (c+2e)\, h^2_{\mu\nu\si\rho}(\hX) + e \, h^3_{\mu\nu\si\rho} \, ,
\eqno (4.18) $$
for $\hX_\mu= (1,{\bf 0})$. Comparing with (4.13) we find
$$ \eqalign {
\beta  = {}& -{1\over d}( a+2e) - {1\over d^2}(d-1) (b+4c) \, , \quad
\de = -b -e \, , \cr
\rho ={}& -{1\over d}( a+2e) + {1\over d^2} (b+4c)\, , \qquad
\tau = e \, , \cr}
\eqno (4.19) $$
with $\alpha,\gamma$ given by (4.14). It is easy to see then that the
constraints (4.16,17) are equivalent to (3.14) (note that $\rho,\tau$
or $c=\half d(\rho+\tau),e=\tau$ may be regarded as independent variables).

Finally in this section we consider again the three point function
for the energy momentum tensor. In the collinear frame as above we
may write
$$ \eqalign {
\l T_{\mu \nu}(x) \,T_{\si\rho} (y) \, T_{\alpha \beta} (z) \r
= {}&{1\over (\hx-\hy)^{d}\, (\hx-\hz)^{d}\,
(\hy-\hz)^{d}} \, \A^{TTT}_{\mu\nu\si\rho\alpha\beta} \, , \cr
\A^{TTT}_{\mu\nu\si\rho\alpha\beta} = {}&
\A^{TTT}_{\si\rho\mu\nu\alpha\beta} =
\A^{TTT}_{\alpha\beta\mu\nu\si\rho} \, , \cr}
\eqno (4.20) $$
with $\A^{TTT}_{\mu\nu\si\rho\alpha\beta}$ also symmetric and traceless
on each pair of indices $\mu\nu,\ \si\rho$ and $\alpha\beta$. This
may therefore be decomposed as
$$ \eqalignno {
\A^{TTT}_{111111}={}& \alpha \, , \qquad \A^{TTT}_{ij1111}= \beta \,\de_{ij}
\, , \qquad \A^{TTT}_{i1k11}= \gamma \,\de_{ik} \, , \cr
\quad \A^{TTT}_{ijk\ell 11} = {}&\de \,\de_{ij}\de_{k\ell} + \ep
(\de_{ik}\de_{j\ell} + \de_{i\ell} \de_{jk} ) \, ,\quad
\A^{TTT}_{ijk1m1} = \rho \,\de_{ij}\de_{k\ell} + \tau
(\de_{ik}\de_{j\ell} + \de_{i\ell} \de_{jk} ) \, ,\cr
\quad \A^{TTT}_{ijk\ell mn} = {}&r \,\de_{ij}\de_{k\ell}\de_{mn}\cr
{} +& s\bigl ( \de_{ij}(\de_{km}\de_{\ell n} + \de_{kn} \de_{\ell m} )
+ \de_{kl}(\de_{im}\de_{jn} + \de_{in} \de_{jm} )
+ \de_{mn}(\de_{ik}\de_{j\ell} + \de_{i\ell} \de_{jk} )\bigl) \cr
{} +& t (\de_{ik} \de_{jm} \de_{\ell n} + \de_{jk} \de_{im} \de_{\ell n}
+ \de_{i\ell} \de_{jm} \de_{kn} + \de_{j\ell} \de_{im} \de_{kn}\cr
&~~~+ \de_{ik} \de_{jn} \de_{\ell m} + \de_{jk} \de_{in} \de_{\ell m}
+ \de_{i\ell} \de_{jn} \de_{km} + \de_{j\ell} \de_{in} \de_{km})\, ,
& (4.21)\cr} $$
where tracelessness requires
$$ \eqalign{
\alpha + (d-1) \beta ={}& 0 \,, \quad \beta+(d-1) \de + 2\ep = 0 \, ,
\quad \gamma + (d-1) \rho + 2\tau = 0 \, , \cr
\de + (d-1)r + 4s = {} & 0 \, ,\quad
\ep + (d-1) s + 4t = 0 \, .\cr}
\eqno (4.22) $$
Following similar arguments to the above the requirements following
from the conservation equation imply two further relations so that
there are three linearly independent coefficients which may be taken as
$r,s,t$. In addition to (4.22) the conservation equations then
determine $\rho,\tau$ by
$$ \eqalign {
2\tau + dr + (d+4)s + 2t = {}& 0 \, , \cr
2\rho - dr + (d-2)s + 2(d+4) t = {}& 0 \, . \cr}
\eqno (4.23) $$
To compare with our previous results we may use (3.15) and (3.19)
to obtain in the collinear frame the explicitly symmetric form
$$ \eqalignno {
\A^{TTT}_{\mu\nu\si\rho\alpha\beta} = {}& a\,
h^5_{\mu\nu\si\rho\alpha\beta} + b\bigl (
h^4_{\alpha\beta\mu\nu\si\rho}(\hX) +
h^4_{\mu\nu\si\rho\alpha\beta}(\hX) +
h^4_{\si\rho\mu\nu\alpha\beta}(\hX)\bigl) \cr
& + c\bigl ( h^3_{\mu\nu\si\rho}h^1_{\alpha\beta}(\hX) +
h^3_{\si\rho\alpha\beta}h^1_{\mu\nu}(\hX) +
h^3_{\mu\nu\alpha\beta}h^1_{\si\rho}(\hX)\bigl) \cr
& + (e-8a-8b)\big( h^2_{\mu\nu\si\rho}(\hX)h^1_{\alpha\beta}(\hX) +
h^2_{\si\rho\alpha\beta}(\hX)h^1_{\mu\nu}(\hX) +
h^2_{\mu\nu\alpha\beta}(\hX)h^1_{\si\rho}(\hX)\bigl) \cr
& +(f+128a+96b-16c-16e)\, h^1_{\mu\nu}(\hX)h^1_{\si\rho}(\hX)
h^1_{\alpha\beta}(\hX) \, , & (4.24)\cr}
$$
after using (A.1) and (3.20). Comparing this with (4.21) we find
$$ \eqalign {
d\rho = {}& 4a+2b-c-e \, , \qquad \quad \tau = a+b \, ,\cr
d^3 r = {}& 16(d-2)a-24b+6dc +16c+4e-f \, , \quad ds=-4a-c \, , \quad
t=a \, . \cr}
\eqno (4.25) $$
It is then straightforward to check that (4.23) is equivalent to
(3.21) with (3.20) (note $d^2r=16a-2db+(d+4)c$).

Although our discussion is concerned with general dimensions $d$
it is important to recognise that $d=2,3$ are special cases. When
$d=2$ there is only one transverse dimension. Hence in (4.3) only
$\de + 2\ep$ has any significance, similarly in (4.13) for $\rho +
2\tau$ while in (4.21) the relevant quantities are $\de+2\ep$,
$\rho + 2\tau$ and $r + 6s+8t$. These restrictions reduce the
number of conformal invariant forms to one for both the $\l TVV\r$
and $\l TTT\r$ three point functions. Less obviously when $d=3$
then in (4.21) $\quad \A^{TTT}_{ijk\ell mn}$ depends only on $r-4t$
and $s+2t$, rather than $r,s,t$ being independent variables.
Consequently the number of conformal invariants for $\l TTT\r$
when $d=3$ is just two.
\bigskip
\leftline{\bigbf 5 Free Field Theories}
\medskip
For general dimension $d$ the only completely explicit conformal
field theories are those provided by free scalar or free fermion
fields. In the scalar case we may write
$$\eqalign {
T_{\mu \nu} = {}& \pr_\mu \phi \pr_\nu \phi - \quar \, {1\over d-1} \bigl (
(d-2) \pr_\mu \pr_\nu + \de_{\mu \nu} \pr^2 \bigl ) \phi^2 \ ,\cr
V^a_\mu = {}& \phi t_\phi^a \pr_\mu \phi \, , \quad (t_\phi^a)^T = -
t_\phi^a \, , \quad [t_\phi^a , \, t_\phi^b] = f^{abc} t_\phi^c \, .
\cr}
\eqno (5.1) $$
In the fermion case
$$ \eqalign{
T_{\mu \nu} = {}& \half {\bar \psi} ( \gamma_\mu {\olr \pr}_\nu +
\gamma_\nu{\olr \pr}_\mu ) \psi \, ,\qquad {\olr \pr}_\mu =
\half (\pr_\mu - {\overleftarrow \pr}_{\! \mu} ) \, , \cr
V^a_\mu = {}& {\bar \psi} t_\psi^a \gamma_\mu \psi \, , \quad
(t_\psi^a)^\dagger = -t_\psi^a \, ,
\quad [t_\psi^a , \, t_\psi^b] = f^{abc} t_\psi^c \, ,\cr}
\eqno (5.2) $$
for $(\gamma_\mu)^\dagger = \gamma_\mu$ euclidean gamma matrices,
$\{\gamma_\mu , \gamma_\nu \} = 2\de_{\mu\nu}$. The
basic two point functions for the massless scalar, fermion fields
are
$$ \l \phi(x)\, \phi(0) \r = {1\over (d-2)S_d}\, {1\over x^{d-2}} \, ,
\quad \l \psi (x) {\bar \psi} (0) \r = {1\over S_d}\, {\gamma{\cdot x}
\over x^d} \, ,
\eqno (5.3) $$
for $S_d = 2\pi^{\hh d}/\Gamma(\hh d)$.

It is easy to use these results to determine the form of the two
point functions of $V_\mu^a$ and $T_{\mu\nu}$ which are compatible
with (2.23) or
$$ \l V^a_{\mu} (x) \, V^b_\nu (0) \r = \de^{ab}\,{C_V\over x^{2(d-1)}}
\, I_{\mu \nu}(x) \, .
\eqno (5.4) $$
In the scalar case, if $\phi$ has $n_\phi$ components and $\tr(t_\phi^a
t_\phi^b) = - N_\phi \de^{ab}$, then
$$ C_V = {N_\phi \over d-2}\, {1\over S_d^{\, 2}}\, , \qquad
C_T = n_\phi\, {d\over d-1}\, {1\over S_d^{\, 2}}\, ,
\eqno(5.5) $$
while in the fermion case, if there are $n_\psi$ Dirac fields and
$\tr(t_\psi^a t_\psi^b) = - N_\psi \de^{ab}$, then
$$ C_V = N_\psi \, 2^{\hh d}\, {1\over S_d^{\, 2}}\, , \qquad
C_T = n_\psi\, \half d \, 2^{\hh d} \, {1\over S_d^{\, 2}}\, .
\eqno(5.6) $$

For determining three point functions it is considerably easier
to evaluate them in the collinear frame where, as discussed in the
previous section, the functional form is very much simpler than in
the general case. Thus for the vector currents, if $x,y,z$ lie on
the 1 axis with $x_1=\hx,y_1=\hy,z_1=\hz$ as earlier, then
from (3.7) and (3.9)
$$ \eqalign {
\l V^a_\mu (x) \, V^b_\nu (y) \, V^c_\omega (z) \r
= {}& {f^{abc}\over (\hx-\hy)^{d-1}\, (\hx-\hz)^{d-1}\,
(\hy-\hz)^{d-1}} \, \A^{VVV}_{\mu\nu\omega} \, , \cr
\A^{VVV}_{\mu\nu\omega} ={}&I_{\mu\alpha}(\hX) I_{\nu\beta}(\hX)
t_{\alpha\beta\omega}(\hX) \cr
={}& (a+4b)\hX_\mu\hX_\nu\hX_\omega -
b(\hX_\mu \de_{\nu\omega} + \hX_\nu \de_{\mu\omega} + \hX_\omega
\de_{\mu\nu} ) \, , \cr \quad \hX_\mu = (1,{\bf 0}) \, , \qquad
& \A^{VVV}_{111} = a+b \, , \quad \A^{VVV}_{ij1} = - b\, \de_{ij} \, .
\cr}
\eqno (5.7) $$
Hence we may determine $a,b$ and then obtain the general expression
for $\l VVV\r$ from (3.7) and (3.9). For the scalar case
$$ S_d^{\, 3} a = \half N_\phi \, {d\over d-2} \, ,\qquad
S_d^{\, 3} b = \half N_\phi \, {1\over d-2} \, ,
\eqno (5.8) $$
while in the fermion case
$$ a = 0\, ,\qquad\quad
S_d^{\, 3} b = N_\psi \, 2^{\hh d} \, .
\eqno (5.9) $$

For the determination of $\l TVV\r$ for free scalar or fermion fields
we again follow the prescription of restricting calculations
of the three point function for these theories to the collinear
configuration. It is sufficient to calculate only $\l T_{ij} V_k
V_\ell \r$, involving only components orthogonal to the 1 axis
on which $x,y,z$ lie, although results for
components along the 1 direction provide a useful check through
the general relations derived in the previous section.
For the scalar case in the notation of (4.13) or (3.13)
$$ \tau = e = {d\over 2(d-1)(d-2)} \,N_\phi\, {1\over S_d^{\, 3}}\, , \quad
\rho = {d-4\over 2(d-1)(d-2)}\, N_\phi\,{1\over S_d^{\, 3}} \, , \quad
c = {d\over2(d-1)}\,N_\phi\, {1\over S_d^{\, 3}}\, .
\eqno (5.10) $$
In the fermion case we find
$$ \tau = e = 0\, , \quad
\rho = 2^{\hh d} \,N_\psi\, {1\over S_d^{\, 3}} \, , \quad
c = \half d\, 2^{\hh d} \,N_\psi \, {1\over S_d^{\, 3}}\, .
\eqno (5.11) $$

For the $\l TTT\r$ three point function we may follow similar procedures with
rather more labour, especially in the scalar case. It is convenient to
calculate also $\l T\phi^2 \phi^2 \r$ and also $\l TT\phi^2\r$, where $\phi^2$
is a scalar operator of dimension $d-2$, and check that these are of the
required form for conformal invariance such as prescribed by (4.2,3,4) and
(4.9) in the latter case. The results for scalar fields, given the form (5.1)
for $T_{\mu\nu}$, are
$$ \eqalign {
r = {}& n_\phi\, {1\over 8}\, {1\over (d-1)^3}(d^3 + 28d-16)\, {1\over
S_d^{\, 3}} \, , \qquad
s = n_\phi\, {1\over 8}\, {d\over (d-1)^3}(d^2-8d+4)\, {1\over
S_d^{\, 3}} \, , \cr
t = {}& a = n_\phi\, {1\over 8}\, {d^3\over (d-1)^3}\, {1\over S_d^{\, 3}}\,,
\quad b = - n_\phi\, {1\over 8}\, {d^4\over (d-1)^3}\, {1\over S_d^{\, 3}}\,,
\quad c = - n_\phi\, {1\over 8}\, {d^2(d-2)^2\over (d-1)^3}\,
{1\over S_d^{\, 3}} \, . \cr}
\eqno (5.12) $$
For fermion fields we obtain
$$
r = - n_\psi \, \hh\, 2^{\hh d}\, {1\over S_d^{\, 3}} \, , \quad
s = n_\psi \, {1\over 8}d \, 2^{\hh d}\, {1\over S_d^{\, 3}} \, , \quad
t= a =0 \, , \quad 2b=c
= - n_\psi \, {1\over 8}d^2\, 2^{\hh d}\, {1\over S_d^{\, 3}} \, .
\eqno (5.13) $$

For $d=4$ there is an additional conformal field theory described in terms
of free abelian vector fields. The energy momentum tensor is
$$
T_{\mu \nu} = F_{\mu\lambda} F_{\nu\lambda} - \quar\, \de_{\mu\nu}
F_{\alpha\beta} F_{\alpha \beta} + s\, X_{\mu\nu} \, ,
\eqno (5.14) $$
where $s$ is the nilpotent BRS operator, $s^2=0$, and $X_{\mu\nu}$ contains
all contributions depending on the gauge fixing term and ghost fields. This
term does not contribute for correlation functions involving gauge invariant
operators and physical states and for our purposes can be neglected. The
basic two point function is then
$$ \l F_{\mu\nu}(x)\, F_{\si\rho}(0) \r = {2\over S_4}\, {1\over x^4}\,
\bigl ( I_{\mu\si}(x)I_{\nu\rho}(x) - I_{\mu\rho}(x)I_{\nu\si}(x) \bigl ) \, .
\eqno (5.15) $$
For the energy momentum tensor two point function, comparing with (2.23), we
obtain
$$ C_T = 16 \, {1\over S_4^{\, 2}} \, .
\eqno (5.16) $$
For the three point function (note that $\l TT\O\r =0$ for $\O= \quar
F_{\alpha\beta}F_{\alpha\beta}$) the results are
$$ r = - 48\, {1\over S_4^{\, 3}} \, , \quad
s = 32\, {1\over S_4^{\, 3}} \, , \quad
t = a= - 16\, {1\over S_4^{\, 3}} \, , \quad b= 0 \, , \quad
c = - 64 \, {1\over S_4^{\, 3}} \, .
\eqno (5.17) $$
It is not difficult to see that the solutions provided by (5.12,13), for
$d=4$, and (5.17) are linearly independent thereby realising the full range
of possibilities for $\l TTT\r$ in this case.

Calculations for three point functions of the energy momentum tensor and
vector currents were also undertaken by Stanev [26] for massless free scalar,
fermion and vector fields in four dimensions.
Due to the complexity of the resulting formulae it
is difficult to compare results in detail but we agree on the number of
independent amplitudes.
\bigskip
\leftline{\bigbf 6 Ward Identities and Short Distance Expansions}
\medskip
To derive Ward identities relating two and three point functions we
assume that we may define a functional $W(g,A,J)$ depending on a
background metric $g_{\mu\nu}$ and gauge field $A^a{}_{\! \mu}$
and a scalar source $J$ such that the energy momentum tensor
$T_{\mu\nu}$, vector current $V^{a\mu}$ and scalar operator $\O$
may be defined through functional derivatives by
$$ \eqalign {
\l T_{\mu\nu}(x)\r = {}& -{2\over \sqrt {g(x)}}\,
{\de\over\de g^{\mu\nu}(x)} \, W \, ,\cr
\l V^{a\mu}(x)\r = {}& - {1 \over \sqrt {g(x)}}\,{\de\over\de A^a{}_{\!
\mu}(x)}
\, W \, ,\quad \l \O(x)\r = - {1 \over \sqrt {g(x)}}\,{\de \over \de J(x)} \,
W \, . \cr}
\eqno (6.1) $$
Assuming $W$ is a scalar and invariant under diffeomorphisms requires
$$ \int \! d^d x \, \Bigl ( -(\nab^\mu v^\nu + \nab^\nu v^\mu)
{\de\over\de g^{\mu\nu}} + (v^\nu \pr_\nu A^a{}_{\! \mu} +\pr_\mu
v^\nu A^a{}_{\! \nu} ){\de\over\de A^a{}_{\! \mu}} + v^\mu \pr_\mu J
\, {\de \over \de J} \Bigl ) W = 0 \, ,
\eqno (6.2) $$
for arbitrary $v^\mu (x)$. Under local scale variations of the metric we
require for arbitrary $\si(x)$
$$  \int \! d^d x \, \si \Bigl ( 2 g^{\mu\nu}
{\de\over\de g^{\mu\nu}} + (d-\eta) J
\, {\de \over \de J} \Bigl ) W = 0 \, ,
\eqno (6.3) $$
since $J$ has dimension $d-\eta$. In general this identity has anomalies
reflecting the introduction of a mass scale in quantum field theories
through renormalisation. In conformal field theories where the $\beta$
functions vanish these anomalous
effects are restricted to additional local terms in (6.3) involving
scalars constructed from $g_{\mu \nu},A^a{}_{\! \mu}$ and $J$ of the
appropriate dimension. The relevant terms for $d=4$ will be considered
later. In addition we assume invariance under local gauge
transformations which implies
$$ \int \! d^dx \, \bigl (\pr_\mu \Lambda^a + f^{abc}A^b{}_{\! \mu}
\Lambda^c ) {\de\over\de A^a{}_{\! \mu}} \, W = 0 \, ,
\eqno (6.4) $$
where for simplicity we suppose that $\O$ is a singlet under the gauge
group. (6.2,3) and (6.4) are equivalent to
$$\eqalign {
&\nab^\mu \l T_{\mu\nu} \r +\nab_\nu A^a {}_{\! \mu} \, \l V^{a\mu}\r
+ \nab_\mu \bigl ( A^a {}_{\! \nu} \, \l V^{a\mu}\r\bigl ){} +
\pr_\nu J \, \l \O\r = 0 \, , \cr
& g^{\mu\nu} \l T_{\mu\nu} \r + (d-\eta) J \l \O \r = 0 \, ,\quad
\nab_\mu \l V^{a \mu} \r + f^{abc}A^b{}_{\! \mu} \l V^{c\mu}\r = 0
\, .\cr}
\eqno (6.5) $$

The Ward identities that we use are then obtained by functional
differentiation of (6.5) and restricting to flat space and also $A^a{}_{\!
\mu}$ and $J$ to be zero. The completely symmetric three
point function of the energy momentum tensor is therefore defined by
$$ \eqalignno { \!\!\!\!\!\!\!\!
\l T_{\mu \nu}&(x) \,T_{\si\rho} (y) \, T_{\alpha \beta}  (z) \r
=  - 8\, {\de\over\de g^{\alpha\beta}(z)} {\de\over\de g^{\si\rho}(y)}
{\de\over\de g^{\mu\nu}(x)} \, W \Bigl |_{g_{\mu\nu}=A^a{}_{\!\mu}=J=0}\cr
= {}& - 8\, {\de\over\de g^{\alpha\beta}(z)} \, {1\over \sqrt{g(y)}}
{\de\over\de g^{\si\rho}(y)} \, {1\over \sqrt{g(x)}}
{\de\over\de g^{\mu\nu}(x)} \, W \Bigl |_{g_{\mu\nu}=A^a{}_{\!\mu}=J=0}&(6.6)
\cr
& + g_{\alpha\beta}(z) \bigl ( \de^d(z-y) + \de^d(z-x) \bigl )
\l T_{\mu\nu}(x)\, T_{\si\rho}(y)\r %\cr
+ g_{\si\rho}(y) \de^d(y-x) \l T_{\mu\nu}(x)\, T_{\alpha\beta}(z)\r \, , \cr}
$$
and similarly for other cases.

Firstly we consider the purely vector identity
$$
\pr_\mu^{\vphantom d} \l V^a_\mu (x) \, V^b_\nu (y) \, V^c_\omega (z) \r
= f^{abd}\, \de^d(x-y)\, \l V^d_\nu (x) \, V^c_\omega (z) \r{}
- f^{acd}\, \de^d(x-z)\, \l V^d_\omega (x) \, V^b_\nu (y) \r \, .
\eqno (6.7) $$
{}From the result (3.7) we may easily find
$$
 \l V^a_\mu (x) \, V^b_\nu (y) \, V^c_\omega (z) \r =
{f^{abc}\over s^{d-2}\, (x-z)^{d-2}(y-z)^d}\, I_{\nu\si}(s)
I_{\omega\rho}(x-z)\, I_{\si\beta}(X) t_{\mu\beta\rho}(X) \, ,
\eqno (6.8) $$
where
$$ s_\mu = x_\mu-y_\mu \, , \qquad
X_\mu = {s_\mu\over s^2} - {(x-z)_\mu \over (x-z)^2} \, .
\eqno (6.9) $$
Hence we may easily find the short distance limit
$$
\l V^a_\mu (x) \, V^b_\nu (y) \, V^c_\omega (z) \r
\sim f^{abc} \, {1\over s^{d}}t_{\mu\nu\rho} (s)
\, {I_{\rho\omega}(y-z) \over (y-z)^{2(d-1)}} \ \ \hbox{as} \ x \to y \, .
\eqno (6.10) $$
Using the explicit form of $t_{\mu\nu\rho}(s)$, as given by (3.9) and
obtained by solving (3.8), we now find taking care over the differentiation
of singular functions of $x$
$$ \pr_\mu \Bigl ( {1\over s^{d}}t_{\mu\nu\rho} (s)\Bigl ){}
= \Bigl ( {1\over d}\,  a + b \Bigl ) \de_{\nu\rho} \, S_d \de^d (s) \, ,
\quad S_d = {2\pi^{\hh d}\over \Gamma (\half d)} \, .
\eqno (6.11) $$
Hence the short distance singular term for $s=x-y\to 0$ in the representation
(6.8) is responsible for the first term on the right hand side of the Ward
identity (6.7) (the other term of course arises from the singularities as
$x\to z$) if
$$ S_d \Bigl ( {1\over d}\, a + b \Bigl ){} = C_V \, ,
\eqno (6.12) $$
given the expression (5.4) for the two point amplitude. We may trivially see
that this is compatible with the results of calculations in (5.5) and (5.8,9).

For three point functions containing the energy momentum tensor we begin with
that involving the scalar fields of dimension $\eta$. This has been discussed
before [13] but is considered again here as a simple warm up exercise. From
(6.5)
it is easy to derive the identities
$$ \eqalignno{ \!\!\!\!\!\!\!\!\!\!\!\!
\pr_\mu^{\vphantom d} \l T_{\mu\nu} (x) \, \O (y) \, \O (z) \r
= {}& \pr_\nu \de^d(x-y)\, \l \O (x) \, \O (z) \r
+ \pr_\nu \de^d(x-z)\, \l \O (x) \, \O (y) \r \, , &(6.13a)\cr
\!\!\!\!\!\!\!\!\!\!\!\! \l T_{\mu\mu} (x) \, \O (y) \, \O (z) \r
= {}& (d-\eta) \bigl ( \de^d(x-y)\, \l \O (x) \, \O (z) \r
+ \de^d(x-z)\, \l \O (x) \, \O (y) \r \bigl ) \, . &(6.13b)\cr}
$$
{}From (3.1) and (3.2), with $s,X$ as in (6.9), the short distance limit is
$$ \eqalignno {
\l T_{\mu\nu} (x) \, \O (y) \, \O (z) \r = {}&
{1\over s^d \, (x-z)^d \, (y-z)^{2\eta-d}} \, t_{\mu\nu} (X) \cr
\sim {}& A_{\mu\nu}(s) \, {N\over (y-z)^{2\eta}} +
B_{\mu\nu\lambda} (s) \, {\pr\over \pr y_\lambda}
{N \over (y-z)^{2\eta}} \ \ \hbox{as} \ x\to y \, , \cr
A_{\mu\nu}(s)N = {1\over s^d}\, t_{\mu\nu} (s)
= {}& {a\over s^d} \, \Bigl ( {s_\mu s_\nu \over s^2} - {1\over d} \,
\de_{\mu  \nu} \Bigl ) \, , \quad B_{\mu\nu\lambda}(s) = {1\over \eta }
\bigl ( \half s^2\pr_\lambda A_{\mu\nu} (s) + d s_\lambda A_{\mu\nu}(s)\bigl )
\, , \cr
B_{\mu\nu\lambda}(s)N = {}& {a\over2\eta s^d} \Bigl ( s_\mu \de_{\nu\lambda} +
s_\nu \de_{\mu\lambda} - s_\lambda \de_{\mu\nu} + (d-2) {s_\mu s_\nu
s_\lambda \over s^2} \Bigl ) \, , & (6.14) \cr}
$$
where the coefficient $N$ is defined by the scale of the two point function
$$ \l \O (x) \, \O (z) \r = {N\over (x-z)^{2\eta}} \, .
\eqno (6.15) $$
The result (6.14) for $s\to 0$ is then equivalent to the operator product
expansion
$$
T_{\mu\nu}(x) \O(y) \sim A_{\mu\nu}(s) \, \O(y)
+ B_{\mu\nu\lambda}(s) \, \pr_\lambda \O(y) \, .
\eqno (6.16) $$

Since the leading singular coefficient $A_{\mu\nu}(s) = {\rm O}(s^{-d})$
it is not well defined as a distribution on ${\bR}^d$ and requires
a regularisation prescription to ensure an unambiguous
evaluation of $\de$-function contributions to $\pr_\mu A_{\mu\nu}(s)$
and $A_{\mu\mu}(s)$ and verify the identities (6.13a,b). It is convenient here
to adopt the methods of differential regularisation [24] and write a
regularised
expression, identical with $A_{\mu\nu}(s)$ for $s\ne 0$, as
$$
\hA_{\mu\nu}(s) =  {a\over Nd} \Bigl ({1\over d-2} \, \pr_\mu\pr_\nu
{1\over s^{d-2}} + C \,\de_{\mu\nu}\, S_d \de^d (s)\Bigl )  \, .
\eqno (6.17) $$
Expressing $\hA_{\mu\nu}(s)$ in terms of derivatives acting on $1/s^{d-2}$ then
enables integration over smooth test functions on ${\bR}^d$ to be defined
by integration by parts. The term containing the arbitrary coefficient $C$
is a reflection of the freedom in any regularisation of $A_{\mu\nu}(s)$.
Given the form (6.17) we then calculate, using standard results such as
$-\pr^2 1/s^{d-2} = (d-2)S_d \de^d (s)$,
$$ \pr_\mu \hA_{\mu\nu}(s) = {a\over Nd}(C- 1) S_d \, \pr_\nu \de (s) \, ,
\qquad \hA_{\mu\mu}(s) = {a\over Nd}(dC-1) \, S_d \de^d (s) \, .
\eqno (6.18) $$
On the other hand $B_{\mu\nu\lambda}(s) = {\rm O}(s^{-d+1})$ does not require
any such regularisation as in (6.17) and unambiguously we find, either
directly or from $ \pr_\mu \hA_{\mu\nu}(s)$,
$$ \pr_\mu B_{\mu\nu\lambda} (s) = {a\over N\eta d}(d-1) \, \de_{\nu\lambda}
\, S_d \de^d (s) \, , \quad B_{\mu\mu\lambda} (s) = 0 \, .
\eqno (6.19) $$
Compatibility with the identities (6.13a,b), using (6.18) and (6.19) in
(6.14) with the replacement $A_{\mu\nu}(s)\to \hA_{\mu\nu}(s)$, then requires
$$
S_d ( C - 1)\, a = d N \, , \quad
S_d ( dC -1 )\, a = d (d-\eta) N \, , \quad
S_d (d-1) \,a = - d\eta N \, ,
\eqno (6.20) $$
which consistently determine $a,C$ giving $\eta C = - (d-1-\eta)$.

The next case to consider involves the $\l TVV\r$ three point function for
which (6.5) leads to the identities
$$ \eqalign{
\pr_\mu^{\vphantom d} \l T_{\mu\nu} (x)& \,
V^a{}_{\! \si} (y) \, V^b{}_{\! \rho}(z) \r \cr
{}& = \pr_\nu \de^d(x-y)\, \l V^a{}_{\! \si} (x) \, V^b{}_{\! \rho} (z) \r
- \pr_\mu \bigl ( \de^d(x-y)\de_{\nu\si} \l V^a{}_{\! \mu} (x) \,
V^b{}_{\! \rho} (z) \r \bigl ) \cr
{}& \, + \pr_\nu \de^d(x-z) \, \l V^a{}_{\! \si} (y) \,
V^b{}_{\! \rho} (x) \r
- \pr_\mu \bigl ( \de^d(x-z)\de_{\nu\rho} \l V^a{}_{\! \si} (y) \,
V^b{}_{\! \mu} (x) \r \bigl ) \, , \cr
\l & T_{\mu\mu} (x) \,
V^a{}_{\! \si} (y) \, V^b{}_{\! \rho}(z) \r = 0 \, ,\qquad
\pr^y {}_{\!\! \si} \l T_{\mu\nu} (x) \,
V^a{}_{\! \si} (y) \, V^b{}_{\! \rho}(z) \r = 0 \, . \cr}
\eqno (6.21) $$
{}From (3.11), with notation as in (6.9) and neglecting group indices which are
unimportant in this context,
we find for the leading term in the short distance limit $s=x-y \to 0$
$$ \eqalign{ \!\!\!\!\!\!\!
\l T_{\mu\nu} (x) \, V_{\si} (y) \, V_{\rho}(z) \r
= {}& {1\over s^d}\, I_{\si \alpha}(s)\, t_{\mu\nu\alpha\beta}
(X) \, {I_{\beta\rho}(x-z) \over (x-z)^d(y-z)^{d-2}} \cr
\sim {}&
A_{\mu\nu\si\beta} (s) \, C_V{I_{\beta\rho}(y-z) \over (y-z)^{2d-2}}
\, , \quad
A_{\mu\nu\si\rho} (s)C_V = {1\over s^d} \, {\tilde t}_{\mu\nu\si\rho}(s) \, .
\cr}
\eqno (6.22) $$
In order to be in accord with (6.21) it is necessary to require
that $A_{\mu\nu\si\rho} (s)\to \hA_{\mu\nu\si\rho} (s)$ representing a
distribution with the properties
$$ \eqalignno {
\pr_\mu \hA_{\mu\nu\si\rho}(s) = {}& \bigl ( \de_{\si\rho}\pr_\nu -
\de_{\nu\si}
\pr_\rho \bigl ) \de^d (s)  \, , & (6.23a) \cr
\hA_{\mu\mu\si\rho}(s) = 0 \, , \qquad & \pr_\si \hA_{\mu\nu\si\rho}(s) = 0 \,
,
& (6.23b) \cr} $$
including $\de$ function contributions. For $s\ne 0$ $\hA_{\mu\nu\si\rho}(s)$
is determined by (3.13) and (3.14). To obtain a well defined
distribution on $\bR^d$ we follow the previous case and pull out derivatives
in the spirit of differential regularisation so that it may be expressed in the
form
$$ \eqalign {
\hA_{\mu\nu\si\rho}(s)C_V = {}& - {2c\over d(d-2)} \, \de_{\si\rho} \, \pr_\mu
\pr_\nu {1\over s^{d-2}}
+ {2e\over d(d-2)} \, \de_{\mu\nu} \, \pr_\si \pr_\rho
{1\over s^{d-2}} \cr
& - {e\over d(d-4)}\, \pr_\mu\pr_\nu\pr_\si\pr_\rho {1\over s^{d-4}}
- {c+ de\over d(d-2)} \, \bigl ( \de_{\nu\rho} \, \pr_\mu \pr_\si
+ \de_{\mu\rho} \, \pr_\nu \pr_\si \bigl ) {1\over s^{d-2}} \cr
& + {c-(d-2)e\over d(d-2)} \, \bigl ( \de_{\nu\si} \, \pr_\mu \pr_\rho
+ \de_{\mu\si} \, \pr_\nu \pr_\rho \bigl ) {1\over s^{d-2}} \cr
&
+ \bigl ( C\, \de_{\mu\nu} \de_{\si\rho} + D \, \de_{\mu\si} \de_{\nu\rho}
+ E \, \de_{\mu\rho}\de_{\nu\si} \bigl ) S_d \de^d(s) \, , \cr}
\eqno (6.24) $$
where $C,D,E$ represent the potential arbitrariness in the distribution. With
this form there is no ambiguity in determining the $\de$ function contributions
on differentiation or taking the trace and imposing (6.23b) gives
$$
C = {2\over d}\, e \, , \qquad D=E = - {1\over d}(c+de)
\eqno (6.25) $$
and then it is easy to verify (6.23a) if
$$ 2S_d (c+e) = d\, C_V \, .
\eqno (6.26) $$
This agrees with the results of our calculations for free fields (5.5) and
(5.10) or (5.6) and (5.11).

$A_{\mu\nu\si\rho}(s)$ clearly represents
the coefficient of the leading singular term in the operator product
expansion of $T_{\mu \nu}(x)$ and $V^a{}_{\! \si}(y)$. The next
leading singular coefficient, which is ${\rm O}(s^{-d+1})$, also contributes
$\de$ function terms to the Ward identity (6.21) on taking the
divergence although a regularisation as in $A_{\mu\nu\si\rho}(s) \to
\hA_{\mu\nu\si\rho}(s)$ is unnecessary in this case. This
coefficient may be obtained by expanding (6.22) where the ${\rm O}(s^{-d+1})$
terms in the short distance limit are
$$ \eqalign {
& A_{\mu\nu\si\beta} (s) s_\lambda {\pr\over \pr y_\lambda}
C_V{I_{\beta\rho}(y-z) \over (y-z)^{2d-2}} -
\B_{\mu\nu\si\beta\lambda} (s) \, 2 (y-z)_{\lambda}
C_V{I_{\beta\rho}(y-z) \over (y-z)^{2d}} \, , \cr
& \B_{\mu\nu\si\beta\lambda} (s) = \half s^2 \pr_\lambda A_{\mu\nu\si\beta} (s)
+  s_\lambda A_{\mu\nu\si\beta} (s) + s_\si A_{\mu\nu\lambda\beta} (s)
-\de_{\si\lambda}\, s_\alpha A_{\mu\nu\alpha\beta} (s) \, . \cr}
\eqno (6.27) $$
It is crucial that $\B_{\mu\nu\si\rho\lambda} (s)$ satisfies
$$ \B_{\mu\nu\si\lambda\lambda} (s) = 0 \, ,
\eqno (6.28) $$
which may be obtained from $s^2 \pr_\si A_{\mu\nu\si\rho} (s) = 0$ and
$A_{\mu\nu\si\rho} (s) = I_{\si\alpha}(s) I_{\rho\beta}(s)
A_{\mu\nu\beta\alpha} (s)$. The relation (6.28) is necessary in order for the
terms in (6.27) to be written in accord with an operator product expansion,
$$ \eqalign {
& T_{\mu\nu} (x) V_\si (y) \sim \hA_{\mu\nu\si\rho} (s) \, V_\rho (y)
+ B_{\mu\nu\si\rho\lambda} (s) \, \pr_\lambda V_{\rho} (y) \, , \cr
B_{\mu\nu\si\rho\lambda} (s) = {}& A_{\mu\nu\si\rho}(s) s_\lambda
- {1\over d}\de_{\rho\lambda} A_{\mu\nu\si\beta}(s) s_\beta
+ {d-1 \over d(d-2)} \B_{\mu\nu\si\rho\lambda} (s) - {1\over d(d-2)}
\B_{\mu\nu\si\lambda\rho} (s) \, , \cr}
\eqno (6.29) $$
where we have required $B_{\mu\nu\si\lambda\lambda} (s) = 0$.
The Ward identities contained in (6.20) are satisfied since
$$ \eqalign {
\pr_\mu B_{\mu\nu\si\rho\lambda} (s) = {}& - \bigl ( \de_{\si\rho}
\de_{\nu\lambda} - \de_{\nu\si} \de_{\rho\lambda} \bigl ) \de^d(s) \, , \cr
\pr_\si B_{\mu\nu\si\rho\lambda} (s) = {}&  \hA_{\mu\nu\lambda\rho}(s) -
{1\over d} \de_{\rho\lambda}\hA_{\mu\nu\beta\beta}(s) \, . \cr}
\eqno (6.30) $$
To show these we use (6.22a,b) (note $\pr_\si (s_\si
\hA_{\mu\nu\lambda\rho}(s))
=0$) to obtain
$$ \pr_\si \B_{\mu\nu\si\rho\lambda} = 0 \, , \quad
\pr_\mu \B_{\mu\nu\si\rho\lambda} = -(d-1) \hA_{\lambda\nu\si\rho} -
\hA_{\rho\nu\si\lambda} + \de_{\rho\lambda} \hA_{\mu\nu\si\mu} \, ,
\eqno (6.31) $$
including possible $\de$ function contributions. The second equation depends on
$$ \eqalign {
(s_\mu & \pr_\lambda - s_\lambda \pr_\mu ) \hA_{\mu\nu\si\rho}(s) \cr
& = - d \hA_{\lambda\nu\si\rho}(s) - \hA_{\si\nu\lambda\rho}(s) +
\de_{\si\lambda}
\hA_{\mu\nu\mu\rho}(s) - \hA_{\rho\nu\si\lambda}(s) + \de_{\rho\lambda}
\hA_{\mu\nu\si\mu}(s) \, , \cr}
\eqno (6.32) $$
which is a consequence of the infinitesimal version of the rotational
covariance equation as in (2.32).

It remains to analyse similarly the three point function of the
energy momentum tensor for which from (6.5), given the definition
(6.6), the crucial identities are
$$ \eqalign{
\pr_\mu^{\vphantom d} \l & T_{\mu\nu} (x) \, T_{\si\rho} (y) \,
T_{\alpha \beta} (z) \r\cr
& =  \pr_\nu \de^d(x-y)\, \l T_{\si\rho} (x) \, T_{\alpha \beta} (z)\r
+\bigl \{ \pr_\si \bigl ( \de^d(x-y) \l T_{\rho\nu} (x) \,
T_{\alpha \beta} (z)\r \bigl ){} + \si \leftrightarrow \rho \bigl\}\cr
&{}\, + \pr_\nu \de^d(x-z)\, \l T_{\si\rho} (y) \, T_{\alpha \beta} (x)\r
+ \bigl \{ \pr_\alpha \bigl ( \de^d(x-z) \l T_{\beta\nu} (x) \,
T_{\si\rho} (y)\r \bigl ){} + \alpha \leftrightarrow \beta \bigl \}\, , \cr
\l & T_{\mu\mu} (x)  \, T_{\si\rho} (y) \, T_{\alpha \beta} (z) \r
=  2  \bigl (\de^d(x-y)+\de^d(x-z)\bigl)\,
\l T_{\si\rho} (y) \, T_{\alpha \beta} (z)   \r  \, . \cr}
\eqno (6.33) $$
As before the $\de$ function contributions arise from the leading short
distance singularities at coincident points. From (3.15) for $x\to y$
$$ \eqalignno{
\l T_{\mu\nu} (x) \, T_{\si\rho} (y) \, T_{\alpha \beta} (z) \r
={}& {1\over s^d}\, \I_{\si\rho,\si'\rho'}(s)\,
t_{\si'\rho'\alpha'\beta'\mu\nu}
(X) \, {\I_{\alpha'\beta'\alpha\beta}(x-z) \over (x-z)^d(y-z)^{d}} \cr
\sim {}&
A_{\mu\nu\si\rho\alpha'\beta'} (s) \, C_T
{\I_{\alpha'\beta'\alpha\beta}(y-z) \over (y-z)^{2d}} \, , & (6.34a) \cr
A_{\mu\nu\si\rho\alpha\beta} (s)C_T = {}& {1\over s^d} \,
t_{\mu\nu\si\rho\alpha\beta}(s) \, , & (6.34b) \cr}
$$
where $C_T$ is the scale of the two point function as in (2.23).
In order to satisfy (6.33) we require the associated regularised distribution
to satisfy
$$ \eqalignno {
\pr_\mu \hA_{\mu\nu\si\rho\alpha\beta} (s) = {}&  \bigl (
\E_{\si\rho,\alpha\beta} \, \pr_\nu + \E_{\nu\rho,\alpha\beta}\, \pr_\si
+ \E_{\nu\si,\alpha\beta}\, \pr_\rho \bigl ) \de^d (s) \, , &(6.35a)\cr
\hA_{\mu\mu\si\rho\alpha\beta} (s) = {}& 2
\E_{\si\rho,\alpha\beta} \, \de^d (s) \, , &(6.35b)\cr}
$$
where $\E$, defined in (2.24), represents the identity on the space of
symmetric traceless tensors. From the original definition (6.34b) also
$$ \hA_{\mu\nu\si\rho\alpha\beta} (s) = \hA_{\si\rho\mu\nu\alpha\beta} (s) \, ,
\quad \hA_{\mu\nu\si\rho\alpha\alpha} (s) = 0 \, .
\eqno (6.36) $$
As before we represent $\hA_{\mu\nu\si\rho\alpha\beta}(s)= {\rm O}(s^{-d})$ in
a
differentially regularised form in terms of derivatives acting on functions of
${\rm O}(s^{-d+2})$ which are unambiguous as distributions. It is convenient
for this purpose then to define the following set of tensors
$H^i_{\mu\nu\si\rho\alpha\beta}(s)$ which are traceless and symmetric
on $\mu\nu, \ \si\rho$ and $\alpha\beta$,
$$ \eqalign {\! \! \! \! \! \!
H^1_{\mu\nu\si\rho\alpha\beta} (s) = {}&\Bigl ( \pr_\mu^{\vphantom h}
\pr_\nu^{\vphantom h} - {1\over d}\, \de_{\mu \nu}^{\vphantom h}
\pr^2 \Bigl ) \, {1\over s^{d-2}} h^1_{\si\rho}(\hs)h^1_{\alpha\beta}(\hs)
\, ,\cr
H^2_{\mu\nu\si\rho\alpha\beta} (s) = {}&\Bigl ( \de_{\si\alpha}^{\vphantom h}
\pr_\rho^{\vphantom h} \pr_\beta^{\vphantom h} + (\si\leftrightarrow \rho,
\alpha \leftrightarrow \beta) \cr
& ~~~~~{} - {4\over d}\, \de_{\si \rho}^{\vphantom h}
\pr_\alpha^{\vphantom h}\pr_\beta^{\vphantom h}
- {4\over d}\, \de_{\alpha \beta}^{\vphantom h}
\pr_\si^{\vphantom h}\pr_\rho^{\vphantom h}
+ {4\over d^2}\, \de_{\si \rho}^{\vphantom h}\de_{\alpha \beta}^{\vphantom h}
\pr^2 \Bigl ) \, {1\over s^{d-2}} h^1_{\mu\nu}(\hs) \, ,\cr
H^3_{\mu\nu\si\rho\alpha\beta} (s) = {}& h^3_{\si\rho\alpha\beta} \,
\Bigl ( \pr_\mu^{\vphantom h}
\pr_\nu^{\vphantom h} - {1\over d}\, \de_{\mu \nu}^{\vphantom h} \Bigl ){}
{1\over s^{d-2}} \, , \cr
H^4_{\mu\nu\si\rho\alpha\beta} (s) = {}&\Bigl ( h^3_{\mu\nu\si\alpha}
\pr_\rho^{\vphantom h} \pr_\beta^{\vphantom h} + (\si\leftrightarrow \rho,
\alpha \leftrightarrow \beta)
- {2\over d}\, \de_{\si \rho}^{\vphantom h}
\bigl ( h^3_{\mu\nu\lambda\alpha}\pr_\lambda^{\vphantom h}
\pr_\beta^{\vphantom h} + (\alpha\leftrightarrow \beta) \bigl ) {}\cr
- & {2\over d}\, \de_{\alpha \beta}^{\vphantom h}
\bigl ( h^3_{\mu\nu\lambda\si}\pr_\lambda^{\vphantom h}
\pr_\rho^{\vphantom h} + (\si\leftrightarrow \rho) \bigl ) {}
+ {8\over d^2}\, \de_{\si\rho}\de_{\alpha \beta}^{\vphantom h}
\Bigl ( \pr_\mu^{\vphantom h}
\pr_\nu^{\vphantom h} - {1\over d}\, \de_{\mu \nu}^{\vphantom h} \Bigl ){}
\Bigl ) {} {1\over s^{d-2}} \, , \cr}
\eqno (6.37) $$
where $\hs = s/\sqrt{s^2}$. With this basis we can express after some labour
$A_{\mu\nu\si\rho\alpha\beta}(s)$, which is given by (3.19) for $ s\ne 0$
with the coefficients written in terms of $a,b,c$ by virtue of (3.20,21), as
a distribution consistent with (6.36) in the form
$$ \eqalign {
\hA_{\mu\nu\si\rho\alpha\beta}^{\vphantom h} (s)C_T &=  {d-2\over d+2}(4a+2b-c)
H^1_{\alpha\beta\mu\nu\si\rho}(s) + {1\over d}(da+b-c)
H^2_{\alpha\beta\mu\nu\si\rho}(s)\cr
&\, - {d(d-2)a-(d-2)b-2c\over d(d+2)}\bigl(
H^2_{\mu\nu\si\rho\alpha\beta}(s) + H^2_{\si\rho\mu\nu\alpha\beta}(s)\bigl )\cr
&\, + {2da+2b-c\over d(d-2)}H^3_{\alpha\beta\mu\nu\si\rho}(s)
 - {2(d-2)a-b-c\over d(d-2)}H^4_{\alpha\beta\mu\nu\si\rho}(s) \cr
&\, - 2\,{(d-2)a-c\over d(d-2)}\bigl(
H^3_{\mu\nu\si\rho\alpha\beta}(s) + H^3_{\si\rho\mu\nu\alpha\beta}(s)\bigl )\cr
&\, +{(d-2)(2a+b)-dc\over d(d^2-4)}\bigl(
H^4_{\mu\nu\si\rho\alpha\beta}(s) + H^4_{\si\rho\mu\nu\alpha\beta}(s)\bigl )\cr
& + \bigl (C\, h^5_{\mu\nu\si\rho\alpha\beta} + D (\de_{\mu\nu}^{\vphantom h}
h^3_{\si\rho\alpha\beta}+\de_{\si\rho}^{\vphantom h} h^3_{\mu\nu\alpha\beta})
\bigl ) S_d \de^d(s) \, . \cr}
\eqno (6.38) $$
$C,D$ are arbitrary coefficients until the imposition of the Ward identities.
{}From the trace identity (6.35b) it is easy to see that (note
$h^3_{\si\rho\alpha\beta} = 2\E_{\si\rho,\alpha\beta}^{\vphantom h}$)
$$ d\, D = C_T \, .
\eqno (6.39) $$
With the results in the appendix in (A.6) we now find
$$ \eqalign {
\pr_\mu^{\vphantom h} \hA_{\mu\nu\si\rho\alpha\beta}^{\vphantom h} (s)C_T &
= \Bigl ( {2\over d^2}(d-1)\bigl ( (d-2)a-c\bigl ){} - {4\over d}\, C + D\Bigl)
h^3_{\si\rho\alpha\beta}\pr_\nu^{\vphantom h} \, S_d\de^d(s)\cr
&\ + \Bigl ( {1\over d} \bigl ( 2(d-2)a-b-c\bigl ){}  + C \Bigl )
\bigl ( h^3_{\nu\si\alpha\beta}\pr_\rho^{\vphantom h} +
 h^3_{\nu\rho\alpha\beta}\pr_\si^{\vphantom h} \bigl ) \, S_d\de^d(s)\cr
&\  + \Bigl ( -{2\over d^2} \bigl ( 2(d-2)a-b-c\bigl ){} -{2\over d}C
+ D \Bigl ) \de_{\si\rho}^{\vphantom h}h^3_{\mu\nu\alpha\beta}
\pr_\mu^{\vphantom h}\,  S_d\de^d(s)\cr
&\  + \Bigl ( - {(d-2)(2a+b)-dc\over d(d+2)} + C \Bigl )\cr
& ~~~~~~~~~\times \Bigl (
h^3_{\si\rho\nu\alpha}\pr_\beta^{\vphantom h}
+ h^3_{\si\rho\nu\beta}\pr_\alpha^{\vphantom h} - {2\over d}\,
\de_{\alpha\beta}^{\vphantom h}h^3_{\si\rho\nu\mu}\pr_\mu^{\vphantom h}\Bigl )
S_d\de^d(s) \, . \cr}
\eqno (6.40) $$
Comparison with (6.35a) then gives the equations
$$ \eqalign {
- {(d-2)(2a+b)-dc\over d(d+2)} + C = {}& 0 \, ,\cr
S_d\Bigl ( {2\over d^2}(d-1)\bigl ( (d-2)a-c\bigl ){} - {4\over d}\, C + D
\Bigl ) {} = {} & \half C_T \, , \cr
S_d \Bigl ( {1\over d} \bigl ( 2(d-2)a-b-c\bigl ){}  + C \Bigl ){} ={}& \half
C_T \, ,\cr }
\eqno (6.41) $$
apart from one which follows trivially given (6.39). It is then a simple
exercise to check their consistency and to finally obtain
$$ 4 S_d \, {(d-2)(d+3)a -2b - (d+1)c \over d(d+2) } = C_T \, .
\eqno (6.42) $$
It may readily be verified that this is in agreement with the results for free
field theories in (5.5) and (5.12), (5.6) and (5.13) or for $d=4$ (5.16)
and (5.17).

Using our results we may write an operator product expansion for the energy
momentum tensor, generalising (1.1) to arbitrary $d$, which takes the form
$$ T_{\mu\nu}(x) T_{\si\rho}(y) \sim C_T {\I_{\mu\nu,\si\rho}(s)\over s^{2d}}
+ \hA_{\mu\nu\si\rho\alpha\beta}(s) T_{\alpha\beta}(y) +
B_{\mu\nu\si\rho\alpha\beta\lambda}(s) \pr_\lambda T_{\alpha\beta}(y) \, ,
\eqno (6.43) $$
where the $c$-number term corresponds to the two point function as in (2.23).
The next to leading operator term may also be found as in our discussion of
the similar $T_{\mu\nu}(x) V_{\si}(y)$ case. By expanding (6.34a) we find
$$ \eqalign { \!\!\!\!\!\!
B_{\mu\nu\si\rho\alpha\beta\lambda}(s) &{} =
A_{\mu\nu\si\rho\alpha\beta}(s) s_\lambda \cr
& \, + {1\over (d+2)(d-1)} \Bigl (
(d+1)\B_{\mu\nu\si\rho\alpha\beta\lambda}(s) -
\B_{\mu\nu\si\rho\lambda\beta\alpha}(s) -
\B_{\mu\nu\si\rho\alpha\lambda\beta}(s) \cr
{} & ~~~~~~~ -  d \, \de_{\alpha\lambda}
A_{\mu\nu\si\rho\gamma\beta}(s) s_\gamma -  d \,\de_{\beta\lambda}
A_{\mu\nu\si\rho\alpha\gamma}(s) s_\gamma + 2\de_{\alpha\beta}
A_{\mu\nu\si\rho\lambda\gamma}(s) s_\gamma \Bigl ) \, , \cr}
\eqno (6.44) $$
where
$$ \eqalign {
\B_{\mu\nu\si\rho\alpha\beta\lambda}(s) = {}& \half s^2 \pr_\lambda
A_{\mu\nu\si\rho\alpha\beta}(s) + s_\si A_{\mu\nu\lambda\rho\alpha\beta}(s)
- \de_{\si\lambda}\,  s_{\si'} A_{\mu\nu\si'\rho\alpha\beta}(s) \cr
& + s_\rho A_{\mu\nu\si\lambda\alpha\beta}(s)
- \de_{\rho\lambda}\,  s_{\rho'} A_{\mu\nu\si\rho'\alpha\beta}(s) \, . \cr}
\eqno (6.45) $$
As before it is crucial that $\B_{\mu\nu\si\rho\lambda\beta\lambda}(s)=0$
and the corresponding condition has been imposed on
$B_{\mu\nu\si\rho\alpha\beta\lambda}(s)$ in (6.44). The Ward identity in
(6.33) is satisfied due to
$$ \eqalign { \!\!\!\!
\pr_\mu B_{\mu\nu\si\rho\alpha\beta\lambda}(s) = {}& -
\de_{\nu\lambda} \E_{\si\rho,\alpha\beta} \, \de^d(s) \cr
& + {1\over (d+2)(d-1)}\Bigl (d\, \de_{\alpha\lambda} \E_{\si\rho,\nu\beta}
+ d\, \de_{\beta\lambda} \E_{\si\rho,\nu\alpha} - 2\de_{\alpha\beta}
\E_{\si\rho,\nu\lambda} \Bigl )\de^d (s) \, , \cr}
\eqno (6.46) $$
where the second line on the r.h.s. of (6.46) does not contribute in the
operator product expansion (6.43). To verify (6.46) we use (6.35a,b)
along with the definition (6.45) and an analogous equation to (6.32)
to obtain
$$ \eqalign {
\pr_\mu \B_{\mu\nu\si\rho\alpha\beta\lambda}(s) = {}& -
d \, \hA_{\lambda\nu\si\rho\alpha\beta}(s) -
\hA_{\alpha\nu\si\rho\lambda\beta}(s)
+ \de_{\alpha\lambda}\,  \hA_{\mu\nu\si\rho\mu\beta}(s) \cr
& - \hA_{\beta\nu\si\rho\alpha\lambda}(s)
+ \de_{\beta\lambda}\,  \hA_{\mu\nu\si\rho\alpha\mu}(s) \cr
& + \bigl (2 \de_{\nu\lambda} \E_{\si\rho,\alpha\beta} +
(d+1) (\de_{\si\lambda} \E_{\nu\rho,\alpha\beta}
+  \de_{\rho\lambda} \E_{\nu\si,\alpha\beta} ) \cr
& ~~~~~ - \de_{\nu\si} \E_{\lambda\rho,\alpha\beta}
- \de_{\nu\rho} \E_{\lambda\si,\alpha\beta} - 2\de_{\si\rho}
\E_{\nu\lambda,\alpha\beta} \bigl ) \de^d(s)  \, . \cr}
\eqno (6.47) $$

Of course in any conformal theory other operators may appear in the
expansion (6.43). In particular any scalar operator $\O$ with dimension
$\eta<d$ will be more significant in the short distance limit than the
contribution of the energy momentum tensor itself shown in (6.43)
if the $\l TT\O\r$ three point function, whose general form was found
in section 3 in (3.3,4) and (3.6), is non zero.
\bigskip
\leftline{\bigbf 7 Hamiltonian}
\medskip
The discussion of Ward identities in the previous section relating two and
three point functions requires a careful treatment of the singularities of
the three point function at coincident points. An alternative derivation
which does not require an analysis of the short distance limit may be given
by constructing the Hamiltonian $H$ operator from the energy momentum tensor.
If $H$ is assumed to generate translations along the 1 direction then it can
be written as
$$ H = - \int \! d^d x \, \de (x_1) \, T_{11} (x) \, .
\eqno (7.1) $$
The minus sign in the definition (7.1) is a reflection of the rotation to
euclidean space, as defined with our conventions $H$ should have a positive
spectrum. Choosing the configuration specified by the points $x=(0,\bx), \,
y=(\half s , {\bf 0}), \, {\bar y} =(- \half s , {\bf 0})$ then from (3.1) we
may write
$$ \l T_{11} (x)  \, \O (y) \, \O ({\bar y}) \r = {d-1 \over d}\, a \,
{1\over (\bx^2 + \quar s^2 )^d} \, {1\over s^{2\eta-d}} \, .
\eqno (7.2) $$
It is then straightforward to evaluate the integral over $\bx$ to obtain
$$ \l \O (y) \, H \, \O ({\bar y}) \r = - 2{d-1\over d}S_d \, a \, {1\over
s^{2\eta + 1}} \, .
\eqno (7.3) $$
However since $H$ is the generator of translations along the 1 direction
$$ \l \O (y) \, H \, \O ({\bar y}) \r = - {\pr \over \pr s} \,
\l \O (y) \, \O ({\bar y}) \r = {2\eta N \over s^{2\eta + 1}} \, ,
\eqno (7.4) $$
given the expression (6.15) for the two point function of the operator $\O$.
Comparing (7.3) and (7.4) gives the same relation between $a$ and $N$ as
found earlier in (6.20).

For the three point function involving the energy momentum tensor and two
vector currents then from (3.11), using (3.13) and (3.14), we may similarly
obtain in this configuration
$$ \eqalign {
\l T_{11} (x)\, V_1(y) \, V_1({\bar y}) \r {}& =
{1\over (\bx^2 + \quar s^2 )^d \,  s^{d-2}} \biggl \{ 2 {d-1\over d}\, c\cr
& ~~~~~ + e \Bigl (
(d-1)^2 - (d-2)(d+1) \, {s^2 \bx^2 \over
(\bx^2 + \quar s^2 )^2} \Bigl ) \biggl \} \, , \cr
\l T_{11} (x)\, V_i(y) \, V_j({\bar y}) \r {}& =
{1\over (\bx^2 + \quar s^2 )^d \,  s^{d-2}} \biggl \{ -2 {d-1\over d}\, c
\, \de_{ij} \cr
& ~~~~~ + e \Bigl ( (d-3) \de_{ij} - (d-2)(d+1) \, {s^2 x_i x_j \over
(\bx^2 + \quar s^2 )^2} \Bigl ) \biggl \} \, , \cr}
\eqno (7.5) $$
where $i,j$ denote components orthogonal to the 1 direction. We may then
obtain
$$ \eqalign {
\l V_1(y) \, H\, V_1({\bar y}) \r = {}&
- 4{d-1\over d}S_d \, (c+e) \, {1\over s^{2d - 1}} \, , \cr
\l V_i(y) \, H\, V_j({\bar y}) \r = {}& \de_{ij} \,
4{d-1\over d}S_d \, (c+e) \, {1\over s^{2d - 1}} \, . \cr}
\eqno (7.6) $$
As before this may be directly related to the two point function of the
vector currents which as given in (2.23) requires
$$ \l V_1(y) \, V_1({\bar y}) \r = - C_V\,  {1\over s^{2d-2}} \, , \quad
\l V_i(y) \, V_j({\bar y}) \r =  \de_{ij} \, C_V \, {1\over s^{2d - 2}} \, .
\eqno (7.7) $$
Hence the same result (6.26) derived from the Ward identities may be obtained.

For the energy momentum tensor three point function the algebraic complications
are again more severe. It is actually more convenient to start from the
collinear configuration of section 4 with points at $(0,{\bf 0}), \,
y=(\half s,{\bf 0}),\,{\bar y}=(-\half s,{\bf 0})$ and consider the conformal
transformation
$$ x_\mu \to {(1+\quar s^2 b^2) x_\mu + (x^2  -\half s^2 b{\cdot x}
-\quar s^2 ) b_\mu \over 1+2b{\cdot x} + b^2 x^2} \, ,
\quad b_\mu=(0,{\bf b}) \, ,
\eqno (7.8) $$
so that $y\to y, \, {\bar y}\to {\bar y}$ and $(0,{\bf 0})\to
(0 , -\quar s^2 {\bf b}) = x$. Using this we may find by transformation
from (4.20) and (4.21)
$$ \eqalignno {
\l T_{11}(x)\, & T_{ij}(y)\, T_{k\ell} ({\bar y}) \r \cr
={} &{1\over (\bx^2 + \quar s^2)^d s^d} \biggl\{ \de \, \de_{ij} \de_{k\ell}
+\ep \bigl (\de_{ik}\de_{j\ell} + \de_{i\ell}\de_{jk}\bigl ){}
+ A \, x_i x_j x_k x_\ell {s^4\over (\bx^2 + \quar s^2)^4} \cr
& + \Bigl ( (\beta-\de)(\de_{ij} x_k x_\ell + \de_{k\ell} x_i x_j)
- (\gamma + \ep) \bigl (\de_{ik} x_j x_\ell + i \leftrightarrow j, k
\leftrightarrow \ell \bigl )\Bigl ) {s^2\over (\bx^2 + \quar s^2)^2}
\biggl \} \, ,\cr
\l T_{11}(x)\, & T_{i1}(y)\, T_{k1} ({\bar y}) \r \cr
={} &{1\over (\bx^2 + \quar s^2)^d s^d} \biggl\{  \Bigl ( \gamma
- (\gamma + \ep) {s^2\bx^2\over (\bx^2 + \quar s^2)^2} \Bigl)\de_{ik}
+ A \, x_i x_k {s^4\bx^2\over (\bx^2 + \quar s^2)^4} \cr
& {}~~~~~~~~~~~~~~~~~~~~~ + (\gamma +\ep - A) x_i x_k
{s^2\over (\bx^2 + \quar s^2)^2} \biggl \} \, , & (7.9) \cr
& A = \alpha -2\beta + 4\gamma + \de + 2\ep = (d+3)(d-2)(4a+2b-c) \, .\cr}
$$
Hence we obtain
$$ \eqalign {
\l T_{ij}(y)\, H\, T_{k\ell} ({\bar y}) \r
={} &- {2S_d \over s^{2d+1}}\, {1\over d} \biggl \{ \Bigl ({A \over d+2}
+2\beta +(d-2) \de \Bigl) \de_{ij} \de_{k\ell} \cr
& ~~~~~~~~~~~ +  \Bigl ({A\over d+2} -2\gamma + (d-2)\ep \Bigl )
\bigl (\de_{ik}\de_{j\ell} + \de_{i\ell} \de_{jk}\bigl ) \biggl \} \cr
= {}& \Bigl ( \half \bigl (\de_{ik}\de_{j\ell} + \de_{i\ell} \de_{jk}\bigl ){}
- {1\over d}\, \de_{ij} \de_{k\ell} \Bigl )\, {2d C_T \over s^{2d+1}}\, ,  \cr
\l T_{i1}(y)\, H\, T_{k1} ({\bar y}) \r
={} &- {2S_d \over s^{2d+1}} \, {1\over d} \Bigl \{2\gamma - (d-2)\ep
- {A\over d+2} \Bigl \}\, \de_{ik}
= - \half \de_{ik}\,{2dC_T \over s^{2d+1}}\, , \cr}
\eqno (7.10) $$
where we have used, from (4.22,23) and (4.25) $2\beta + (d-2)\de = -2(d-1)b +
(d-5)c$, $2\gamma - (d-2) \ep
= 2(d+3)(d-2) a + 2(d-3)b-3(d-1)c$ and the results for $A$ in (7.9) and $C_T$
in (6.42). The final result (7.10) is then as expected from the general
formula for the two point function of the energy momentum tensor in (2.23).
\bigskip
\leftline{\bigbf 8 Scale Anomalies}
\medskip
The discussion in this paper has so far been based on exact conformal
invariance. This is relevant for quantum field theories at critical points
where $\beta$ functions are zero and hence the operator contributions to
the trace of the energy momentum tensor vanish. However even at such
critical points for a curved space background and with other external fields
there are possible $c$ number contributions to the trace. For $d=4$ these
are local dimension four scalars and the trace may have the form
$$ g^{\mu\nu}\l T_{\mu\nu} \r = \half p\, J^2 - \quar \kappa \,
F^a{}_{\! \mu\nu} F^{a\mu\nu} - \beta_a F - \beta_b G + h \nab^2 R \, .
\eqno (8.1) $$
$F^a{}_{\! \mu\nu}$ is the field strength formed from the gauge field
$A^a{}_{\! \mu}$ while if $R_{\alpha\beta\gamma\de}$ is the Riemann
curvature formed from the metric $g_{\mu\nu}$, with $R_{\alpha\beta}$ the
Ricci tensor and $R$ the scalar curvature, then for general $d$ we define
$$ \eqalign {
F & =  R^{\alpha\beta\gamma\de} R_{\alpha\beta\gamma\de}- {4\over d-2}\,
R^{\alpha\beta} R_{\alpha \beta} + {2\over (d-2)(d-1)}\, R^2
= C^{\alpha\beta\gamma\de} C_{\alpha\beta\gamma\de} \, , \cr
G & =  R^{\alpha\beta\gamma\de} R_{\alpha\beta\gamma\de}- 4R^{\alpha\beta}
R_{\alpha \beta} +  R^2 = 6 \, R^{\alpha\beta}{}_{\! [\alpha\beta}
R^{\gamma\delta}{}_{\! \gamma\delta]} \, , \cr
C_{\alpha\beta\gamma\de} & = R_{\alpha\beta\gamma\de} - {2\over d-2} \bigl (
g_{\alpha[\gamma}R_{\delta]\beta} - g_{\beta[\gamma}R_{\delta]\alpha} \bigl )
{} + {2\over (d-1)(d-2)}\, g_{\alpha[\gamma}g_{\delta]\beta}  R \, , \cr}
\eqno (8.2) $$
where $C_{\alpha\beta\gamma\de}$ is the Weyl tensor, the traceless part of
the Riemann tensor.
In writing (8.1) we have supposed that the operator $\O$ which is coupled
to the source $J$ has a dimension $\eta \approx 2$, as is relevant for the
case of the operator $\phi^2$ in scalar field theories (if such a low
dimension operator
is present then generically the expansion (8.1) has several more possible
terms, we neglect such complications [27] here since scalar operators are only
considered as an illustrative exercise for more serious calculations
involving conserved vector fields and the energy momentum tensor). Many
authors have computed the trace of the energy momentum tensor on curved
space for free fields [23] and found
$$ \eqalign {
\beta_a = {}& - {1\over 64\pi^2}\, {1\over 10} \bigl ( \thir n_\phi
+ 2 n_\psi + 4 n_V \bigl ) \, , \cr
\beta_b = {}& {1\over 64\pi^2}\, {1\over 90} \bigl ( n_\phi
+ 11 n_\psi + 62 n_V \bigl ) \, , \cr}
\eqno (8.3) $$
with the notation of section 5 where now $n_V$ denotes the number of free
vector fields. In general $g^{\mu\nu}\l T_{\mu\nu} \r$ might be expected
to contain a term $\propto R^2$ but such a contribution can be shown to
be absent, at the critical point, due to integrability conditions for $W$
[27,28]. The coefficient $h$ in (8.1) is undetermined in general
since it may be modified at will
by the addition of an arbitrary local term $\propto \int d^4 x\sqrt g\,R^2$
to $W$. If present the trace identity for the two point function on flat
space becomes
$$ \l T_{\mu\mu}(x)\, T_{\si\rho}(0) \r = 2h (\pr_\si \pr_\rho -
\de_{\si\rho}\pr^2) \pr^2 \de^4(x) \, .
\eqno (8.4) $$
For simplicity we take $h=0$ henceforth.

The result for the trace anomaly in (8.1) is intimately connected with the
renormalisation group equation since, for $\mu$ an arbitrary renormalisation
mass scale, then by dimensional analysis for general $d$
$$\biggl ( \mu{\pr \over \pr \mu} + \int \! d^d x \, \Bigl ( 2 g^{\mu\nu}
{\de\over\de g^{\mu\nu}} + (d-\eta) J
\, {\de \over \de J} \Bigl )\biggl ) W = 0 \, .
\eqno (8.5) $$
Hence (8.1) implies, for $d=4$,
$$\eqalignno {
\mu{\pr \over \pr \mu} \l V^a {}_{\! \mu} (x) \, V^b {}_{\! \nu} (0) \r
= {}& - \kappa\, \de^{ab} S_{\mu\nu} \de^4(x) \, , & (8.6a)\cr
\mu{\pr \over \pr \mu} \l T_{\mu\nu} (x) \, T_{\si\rho}(0) \r
= {}& - 4\beta_a \, \Delta^T_{\mu\nu\si\rho} \de^4(x) \, , &(8.6b)\cr}
$$
where we define for arbitrary $d$
$$ \eqalign {
S_{\mu\nu} = {}& \pr_\mu\pr_\nu - \de_{\mu\nu} \pr^2 \, , \qquad \pr_\mu
S_{\mu\nu} = 0 \, \cr
\Delta^T_{\mu\nu\si\rho} = {}& \half \bigl ( S_{\mu\si} S_{\nu\rho} +
S_{\mu\rho} S_{\nu \si} \bigl ) {} - {1\over d-1}S_{\mu\nu}S_{\si\rho}\, ,
\quad \Delta^T_{\mu\mu\si \rho} = 0 \, . \cr}
\eqno (8.7) $$
Even for a conformal field theory a dependence on a mass scale $\mu$ arises
when the two point functions are carefully defined as distributions for $d=4$.
{}From (2.23) or (5.4) we may write for general $d$
$$\eqalign {
\l V^a {}_{\! \mu} (x) \, V^b {}_{\! \nu} (0) \r
= {}& -  \de^{ab}{C_V\over 2(d-2)(d-1)}\, S_{\mu\nu} {1\over x^{2d-4}} \, , \cr
\l T_{\mu\nu} (x) \, T_{\si\rho}(0) \r
= {}& {C_T \over 4(d-2)^2 d(d+1)} \Delta^T_{\mu\nu\si\rho} {1\over x^{2d-4}}
\, . \cr}
\eqno (8.8) $$
However $(x^2)^{-\lambda}$ is a singular function with poles at
$\lambda = \half d + n, \, n=0,1,\dots$
$$ {1\over (x^2)^\lambda} \sim {1\over d + 2n-2\lambda}\, {1\over 2^{2n}
n!}\, {\Gamma(\half d)\over \Gamma (\half d + n)}\, S_d (\pr^2)^n_{\vphantom d}
\de^d(x) \, .
\eqno (8.9) $$
Consequently $x^{-2d+4}$ although defined by analytic continuation in $d$ is
singular when $d=4$. For $d=4$ using differential regularisation
ensures a sensible distribution through writing
$$ \R {1\over x^4} = - {1\over 4} \pr^2 {1\over x^2}\bigl ( \ln \mu^2 x^2 +
a \bigl ) \, ,
\eqno (8.10) $$
with $a$ an arbitrary constant (which may be absorbed into $\mu$). With this
prescription
$$ \mu{\pr\over \pr \mu} \R {1\over x^4} = 2\pi^2 \de^4(x) \, .
\eqno (8.11) $$
Using this regularisation in (8.8) when $d=4$ then (8.6a,b) give
$$ \kappa = {\pi^2\over 6}\, C_V \, , \qquad \beta_a = - {\pi^2 \over 640} \,
C_T \, .
\eqno (8.12) $$
It is easy to check that the results of section 5, (5.5,6) and (5.16), are in
accord with (8.3).

For three point functions (8.1) leads to modified trace identities involving
extra pieces containing $\de$ functions for all three points coincident. Such
terms arise due to the need to subtract singularities of the schematic form
$$ {1\over \bigl ((x-y)^2\bigl )^{\de_{xy}} \bigl ((y-z)^2\bigl )^{\de_{yz}}
 \bigl ((z-x)^2\bigl )^{\de_{zx}}} \sim {1\over d+n
- (\de_{xy} + \de_{yz} +\de_{zx})}\, \pr^{2n} \de^d(x-y) \de^d(x-z) \, ,
\eqno (8.13) $$
for $n=0,1,\dots$, although detailed formulae are more complicated than
(8.9)\footnote{*}{For a particular case see ref. [29].}.
For illustrative purposes we consider
first the $\l T\O\O\r$ three point function when (8.1) implies
$$ \eqalign {
\l T_{\mu\mu} (x) \, \O (y) \, \O (z) \r
= {}& (d-\eta) \bigl ( \de^d(x-y)\, \l \O (x) \, \O (z) \r
+ \de^d(x-z)\, \l \O (x) \, \O (y) \r \bigl ) \cr
& + p \, \de^d(x-y) \de^d(x-z) \, , \cr}
\eqno (8.14) $$
replacing (6.13b). We show below that the additional term in (8.14)
involving $p$ is necessary for any $d$ when $\eta\approx \half d$.

To verify (8.14) we start from (6.17) which with (6.20) may be written
as
$$\hA_{\mu\nu}(s) = - {\eta\over (d-2)(d-1)S_d}\Bigl ( \pr_\mu \pr_\nu -
{1\over d}\, \de_{\mu\nu} \pr^2 \Bigl ) {1\over s^{d-2}} +
{d-\eta\over d}\, \de_{\mu\nu} \de^d(s) \, .
\eqno (8.15) $$
Based on this form the result provided by (3.1) and (3.2) for the complete
three point function may be re-expressed as
$$ \eqalign {
\l T_{\mu\nu} & (x) \, \O (y) \, \O (z) \r \cr
={}& {d\eta N\over (d-2)^2(d-1)S_d}
\Bigl ( \pr_\mu \pr_\nu - {1\over d}\, \de_{\mu\nu} \pr^2 \Bigl )
\biggl ( {1\over (x-y)^{d-2}(x-z)^{d-2} (y-z)^{2\eta-d+2}} \biggl )\cr
& - {2\eta N\over (d-2)^2 S_d} \biggl (
\Bigl ( \pr_\mu \pr_\nu - {1\over d}\, \de_{\mu\nu} \pr^2 \Bigl )
{1\over (x-y)^{d-2}}\, {1\over (x-z)^{d-2} (y-z)^{2\eta-d+2}} \cr
& ~~~~~~~~~~~~~~~~~+
\Bigl ( \pr_\mu \pr_\nu - {1\over d}\, \de_{\mu\nu} \pr^2 \Bigl )
{1\over (x-z)^{d-2}}\, {1\over (x-y)^{d-2} (y-z)^{2\eta-d+2}} \biggl ) \cr
& + {d-\eta \over d}\, \de_{\mu\nu} \bigl ( \de^d(x-y) + \de^d(x-z) \bigl )
N \R {1\over (y-z)^{2\eta}} \cr
& + {1\over d}\, p \, \de_{\mu\nu} \de^d(x-y) \de^d(x-z) \, , \cr}
\eqno (8.16) $$
which is consistent as $x\to y$ and also $x\to z$ with the regularised short
distance expansion represented by $\hA_{\mu\nu}(s)$ in (8.15) and agrees with
(3.1,2) for non coincident points. Each of the terms
on the r.h.s of (8.16) involving derivatives are well defined distributions
for $\eta\approx \half d$, the first since the derivative operators act on a
non singular expression for $\eta\approx \half d$ and
may be integrated by parts as in differential regularisation and the second
since, although there is a potential singularity when $\eta = \half d$,
the coefficient of this apparent singularity can only be proportional to
$\de_{\mu\nu} \de^d(x-y) \de^d(x-z)$ but such a factor cannot be present since
this term has been arranged to be traceless. In the piece corresponding to
the second term in $\hA_{\mu\nu}(s)$ in (8.15) we also introduce the
regularised expression, for $\eta\approx \half d$,
$$ \eqalign {
\R {1\over x^{2\eta}} = {}& {1\over x^{2\eta}} - {\mu^{2\eta -d}\over
d-2\eta}\, S_d \de^d(x) \cr
= {}& - {1\over d-2\eta} \pr^2 \Bigl ( {1\over 2\eta - 2}\,
{1\over x^{2\eta -2}} - {\mu^{2\eta -d}\over d-2}\, {1\over x^{d-2}} \Bigl )\cr
= {}& - {1\over 2(d-2)}\, \pr^2 {1\over x^{d-2}}\Bigl (\ln \mu^2x^2 + {2\over
d-2} \Bigl ) \ \ \hbox{for} \ \eta \to \half d \, , \cr}
\eqno (8.17) $$
which coincides with (8.10) for $d=4,\eta=2$ taking $a=1$. The last term
on the r.h.s. of (8.16) reflects the ambiguity of representing the whole
three point function as a well defined distribution, its coefficient has
been chosen to correspond with the extra term in the trace identity (8.14).
The identity (8.14) is then automatically satisfied if instead of (6.15) we
now take for the two point function the regularised form
$$ \l \O (x) \, \O (z) \r = N\, \R{1\over (x-z)^{2\eta}} \, .
\eqno (8.18) $$
The resulting expression given by (8.16) for
$\l T_{\mu\nu} (x) \, \O (y) \, \O (z) \r$ is hence a well defined
distribution on $\bR^{3d}$ for $\eta \approx \half d$ including the limiting
case $\eta = \half d$. If we now consider the divergence of
(8.16) standard calculations for $\eta \ne \half d$ give
$$ \eqalignno {
\pr_\mu \l T_{\mu\nu} (x) \, \O (y) \, \O (z) \r
={}& {\eta N\over d}\Bigl ( \pr_\nu \de^d(x-y) \, {1\over (x-z)^{2\eta}}
+ \pr_\nu \de^d(x-z) \, {1\over (x-y)^{2\eta}} \Bigl ) \cr
& - {d-\eta \over d}N\Bigl ( \de^d(x-y) \, \pr_\nu {1\over (x-z)^{2\eta}}
+  \de^d(x-z) \, \pr_\nu {1\over (x-y)^{2\eta}} \Bigl ) \cr
& + {d-\eta \over d}\, \pr_\nu \bigl ( \de^d(x-y) + \de^d(x-z) \bigl )
N \R {1\over (y-z)^{2\eta}} \cr
& + {1\over d}\, p \, \pr_\nu \bigl ( \de^d(x-y) \de^d(x-z)\bigl ) \cr
= {}& \pr_\nu \de^d(x-y) \, N\R {1\over (x-z)^{2\eta}}
+ \pr_\nu \de^d(x-z) \, N\R {1\over (x-y)^{2\eta}} \cr
& + {1\over d}\bigl ( p - \mu^{2\eta-d}S_d N \bigl )
\pr_\nu \bigl ( \de^d(x-y) \de^d(x-z)\bigl ) \, , & (8.19) \cr}
$$
using the definition of $\R$ in (8.17), where the final result allows taking
the limit $\eta \to \half d$ in each term without singularities.
Assuming the regularised form (8.18) for $\l\O \O \r$ then requiring the Ward
identity (6.13a) therefore determines finally $p = \mu^{2\eta-d}S_d N$ (the
same relation also follows from $\mu{\pr\over \pr\mu}\l \O(x) \O(0) \r =
p \de^d(x) $). In the above derivation
of (8.19) $\eta$ may be regarded as a convenient regularisation parameter, as
in analytic regularisation. Alternatively, if for instance $\eta=d-2$, the
subtraction introduced in (8.17) and subsequent discussion
is equivalent to using dimensional regularisation.

For the $\l TVV\r$ three point function then for $d=4$ (8.1) now leads to
a modified trace identity instead of (6.21)
$$ \l T_{\mu\mu} (x) \, V^a{}_{\! \si} (y) \, V^b{}_{\! \rho}(z) \r =
\de^{ab} \kappa \bigl ( \pr_\rho \de^4(x-y) \pr_\si \de^4 (x-z) -
\de_{\si\rho} \pr_\lambda \de^4(x-y) \pr_\lambda \de^4 (x-z) \bigl ) \, .
\eqno (8.20) $$
To derive this result we use dimensional regularisation since $d$ is the
only variable parameter in $\l TVV\r$ consistent with maintaining conformal
invariance. According to (8.13) $\l TVV\r$ has a potential singularity as a
distribution on $\bR^{3d}$ proportional to $\pr^{2n} \de^d(x-y) \de^d (x-z)$
for $d=2n+2$.
For $d = 4 -\vep$ a regularised form of the three point function is given by
$$ \eqalign {
\l T_{\mu\nu}(x)\, V_\si (y)\, V_\rho (z)\r = {}&
I_{\si\alpha}(x-y) I_{\rho\beta}(x-z) \, {t_{\mu\nu\si\rho}(X) \over
(x-y)^d (x-z)^d (y-z)^{d-2}} \cr
& - \bigl ( A_{\mu\nu\si\beta}(x-y) - \hA_{\mu\nu\si\beta}(x-y) \bigl ) \, C_V
{I_{\beta\rho}(y-z) \over (y-z)^{2d-2}} \cr
& - \bigl ( A_{\mu\nu\alpha\rho}(x-z) - \hA_{\mu\nu\alpha\rho}(x-z) \bigl )\,
C_V
{I_{\alpha\si}(z-y) \over (z-y)^{2d-2}} \cr
& + {\mu^{-\vep}S_4\over\vep}\, {C_V\over 12}\,D_{\mu\nu\si\rho}(x-y,x-z) \, .
\cr}
\eqno (8.21) $$
The second and third terms on the r.h.s. of (8.21) subtract the singular
contributions for any $d$ arising as $x\to y$ and $x\to z$ respectively and
replace them by the leading regularised coefficients in the short distance
expansion of $T_{\mu\nu}(x) V_\si(y)$ and $T_{\mu\nu}(x)V_\rho(z)$
as given by (6.24). The last term is necessary to subtract singular pieces
present as $\vep \to 0$ according to (8.13) which are then purely
local with support only for $x=y=z$. This satisfies the obvious symmetry
requirements $D_{\mu\nu\si\rho}(s,t) = D_{\nu\mu\si\rho}(s,t) =
D_{\mu\nu\rho\si}(t,s)$. From (5.4) and (8.8,9) the regularised
two point function using dimensional regularisation is easily seen to be
$$ \l V_{\mu}(x)\, V_\nu(0) \r = C_V {I_{\mu\nu}(x)\over x^{2d-2}} +
{\mu^{-\vep}S_4\over\vep}\, {C_V\over 12}\, S_{\mu\nu} \de^d (x) \, ,
\eqno (8.22) $$
with $S_{\mu\nu}$ as in (8.7). The expression (8.22) clearly agrees with
(8.6a), for $\vep\to 0$, given the relation between $\kappa$ and $C_V$ again.
The last term in (8.21) can now be determined
by the Ward identities reflecting conservation of $T_{\mu\nu}$ and $V_\si$.
In order to satisfy (6.21) it is necessary that this term should generate the
$\vep$ poles required by the regularised two point function as in (8.22)
since in section 5 we have verified that without this term the expression
given by (8.21) obeys the conservation Ward identities with the unregularised
form for the $\l VV\r$ two point functions.
Hence $D_{\mu\nu\si\rho}(x-y,x-z)$ is required to satisfy
$$ \eqalignno {
\pr_\mu D_{\mu\nu\si\rho}(x-y,x-z)&{} =
\pr_\nu \de^d(x-y) S_{\si\rho} \de^d(x-z) - \de_{\nu\si} \pr_\mu
\de^d(x-y) S_{\mu\rho} \de^d(x-z) \cr
&\, + \pr_\nu \de^d(x-z) S_{\si\rho} \de^d(x-y) - \de_{\nu\rho} \pr_\mu
\de^d(x-z) S_{\mu\si} \de^d(x-y) \, , \cr
& {\pr\over \pr y_\si} D_{\mu\nu\si\rho}(x-y,x-z) = 0 \, . & (8.23) \cr}
$$
This has a unique solution consistent with the symmetry requirements
$$ \eqalign {
D_{\mu\nu\si\rho}(s,t)&{} = \bigl ( \de_{\mu\si}\pr_\rho \de^d (s) \pr_\nu
\de^d (t) + \de_{\mu\rho}\pr_\nu \de^d (s) \pr_\si \de^d (t) - \de_{\si\rho}
\pr_\mu \de^d (s) \pr_\nu \de^d (t) + (\mu \leftrightarrow \nu)\bigl )  \cr
&\, - \de_{\mu\nu} \pr_\rho \de^d (s) \pr_\si \de^d (t)
+ \bigl ( \de_{\mu\nu}\de_{\si\rho} - \de_{\mu\si}\de_{\nu\rho}
- \de_{\mu\rho}\de_{\nu\si}\bigl ) \pr_\lambda \de^d (s) \pr_\lambda \de^d (t)
\, . \cr}
\eqno (8.24) $$
It is easy to see that
$$ D_{\mu\mu\si\rho}(s,t) = \vep \bigl ( \pr_\rho \de^d (s) \pr_\si \de^d (t)
- \de_{\si\rho} \pr_\lambda \de^d (s) \pr_\lambda \de^d (t) \bigl ) \, .
\eqno (8.25) $$
Hence applying this result in (8.21) shows that the regularised $\l TVV \r$
three point function is in accord with the modified trace identity (8.20)
for $d=4$ if $\kappa$ is given by (8.12) once more.

As before we finally consider the three point function for the energy momentum
tensor although the calculational details are more intricate. By virtue of
(8.1)
the trace identity is modified when $d=4$ from the result in (6.33) to
$$ \eqalign { \!\!\!\!\!
\l T_{\mu\mu} (x)  \, T_{\si\rho} (y) \, T_{\alpha \beta} (z) \r &{}
= 2  \bigl (\de^4(x-y)+\de^4(x-z)\bigl)\,
\l T_{\si\rho} (y) \, T_{\alpha \beta} (z) \r \cr
&\, - 4\bigl ( \beta_a\, \A^F_{\si\rho,\alpha\beta}(x-y,x-z)
+ \beta_b\, \A^G_{\si\rho,\alpha\beta}(x-y,x-z) \bigl ) \, ,\cr }
\eqno (8.26) $$
where we assume $h=0$ and we define, for general $d$,
$$ \eqalign {
{\de\over \de g^{\si\rho}(y)}\, {\de\over \de g^{\alpha\beta}(z)} \
F (x) \, \Bigl |_{g_{\mu\nu}=0} = 2\, {d-3\over d-2} \,
\A^F_{\si\rho\alpha\beta}(x-y,x-z) \, , \cr
{\de\over \de g^{\si\rho}(y)}\, {\de\over \de g^{\alpha\beta}(z)} \
G (x) \, \Bigl |_{g_{\mu\nu}=0} =  \A^G_{\si\rho\alpha\beta}(x-y,x-z) \, . \cr}
\eqno (8.27) $$
The $d$ dependent coefficients in the definition of $\A^F_{\si\rho\alpha\beta}$
are not essential but are chosen for later convenience. The particular form of
$ \A^F_{\si\rho,\alpha\beta}(s,t)$ is not required here but it is clearly
symmetric $\A^F_{\si\rho,\alpha\beta}(s,t) = \A^F_{\alpha\beta,\si\rho}(t,s)$
as is $\A^G_{\si\rho,\alpha\beta}(s,t)$. By explicit calculation, with
$\Delta^T_{\si\rho\alpha\beta}$ defined in (8.7),
$$ \eqalign {
\int \! d^d x & \, \A^F_{\si\rho,\alpha\beta}(x-y,x-z) =
\Delta^T_{\si\rho\alpha\beta} \de^d(y-z) \, , \cr
\A^G_{\si\rho,\alpha\beta}(x-y,x-z) = {}& {- 24}\bigl \{
\E^A_{\si\alpha\gamma\kappa,\rho\beta\delta\lambda}\pr_\kappa^{\vphantom g}
\pr_\lambda^{\vphantom g} \bigl( \pr_\gamma^{\vphantom g} \de^d (x-y)
\pr_\de^{\vphantom g} \de^d (x-z) \bigl ) {}
+ \si \leftrightarrow \rho \bigl \} \, , \cr}
\eqno (8.28) $$
where $\E^A_{\alpha\beta\gamma\delta,\mu\nu\si\rho}$ is the projector
for totally antisymmetric four index tensors, for any
$T_{\alpha\beta\gamma\delta}^{\vphantom g}$ then
$T_{[\alpha\beta\gamma\delta]}^{\vphantom g} =
\E^A_{\alpha\beta\gamma\delta,\mu\nu\si\rho}T_{\mu\nu\si\rho}^{\vphantom g}$
(if $d=4$ then we may write
$\E^A_{\alpha\beta\gamma\delta,\mu\nu\si\rho} = {1\over 24}
\ep_{\alpha\beta\gamma\delta}^{\vphantom g}\ep_{\mu\nu\si\rho}^{\vphantom g}$).

The anomalous additional terms in (8.26) arise due to extra singularities
present in the three point function when $d=4$. As in the $\l TVV\r$
case it is convenient to use dimensional regularisation and subtract the
singular poles in $\vep = d-4$. The possible counterterms for the $\l TTT\r$
three point function may be formed from
$$ \eqalign {
{\de\over \de g^{\mu\nu}(x)}\,{\de\over \de g^{\si\rho}(y)}\,
{\de\over \de g^{\alpha\beta}(z)} \,
\int \! d^dx \sqrt g \, F \, \Bigl |_{g_{\mu\nu}=0} = {}&{d-3\over d-2} \,
D^F_{\mu\nu\si\rho\alpha\beta}(x,y,z) \, , \cr
{\de\over \de g^{\mu\nu}(x)}\,{\de\over \de g^{\si\rho}(y)}\,
{\de\over \de g^{\alpha\beta}(z)} \,
\int \! d^dx \sqrt g \, G \, \Bigl |_{g_{\mu\nu}=0} = {}& \half \,
D^G_{\mu\nu\si\rho\alpha\beta}(x,y,z) \, , \cr}
\eqno (8.29) $$
where explicit forms for $D^F_{\mu\nu\si\rho\alpha\beta}$ and
$ D^G_{\mu\nu\si\rho\alpha\beta}$ are not essential at this stage, although
we may obtain
$$ \eqalign {
& D^G_{\mu\nu\si\rho\alpha\beta}(x,y,z) \cr
&  ~~~~ = {- 60}\bigl \{
\E^A_{\mu\si\alpha\gamma\kappa,\nu\rho\beta\delta\lambda} \,
\pr_\gamma^{\vphantom g} \pr_\delta^{\vphantom g} \de^d (x-y) \,
\pr_\kappa^{\vphantom g} \pr_\lambda^{\vphantom g} \de^d (x-z)  {}
+ \si \leftrightarrow \rho, \alpha\leftrightarrow \beta \bigl \} \, , \cr}
\eqno (8.30) $$
where $\E^A_{\mu\si\alpha\gamma\kappa,\nu\rho\beta\delta\lambda}$
corresponds to antisymmetrisation of five index tensors.

$D^F$ and $D^G$ are clearly symmetric local functions of thee points, from
their definitions in (8.29), and satisfy important identities. From Weyl
rescaling and invariance under diffeomorphisms
$$ {2\over \sqrt g}\, g^{\mu\nu} {\de\over \de g^{\mu\nu}}
\int \! d^d x\sqrt g \, F = \vep F \, , \quad
\nab^\mu{1\over \sqrt g} {\de\over \de g^{\mu\nu}}
\int \! d^d x\sqrt g \, F = 0 \, ,
\eqno (8.31) $$
we may derive, using (8.28),
$$ \eqalign {
& D^F_{\mu\mu\si\rho\alpha\beta}(x,y,z) \cr
& ~~ = {} \vep \, \A^F_{\si\rho,\alpha\beta}(x-y,x-z)
- 2 \bigl ( \de^d(x-y) + \de^d(x-z) \bigl )
\Delta^T_{\si\rho\alpha\beta} \de^d(y-z) \, , \cr
&  \pr_\mu^{\vphantom g} D^F_{\mu\nu\si\rho\alpha\beta}(x,y,z)\cr
& ~~ ={} -\pr_\nu \de^d(x-y)\,
\Delta^T_{\si\rho\alpha\beta} \de^d(x-z)  - \bigl \{ \pr_\si \bigl (
\de^d(x-y)\,
\Delta^T_{\rho\nu\alpha\beta} \de^d(y-z)\bigl ){} +
\si \leftrightarrow \rho \bigl \} \cr
& ~~~~~~ -\pr_\nu \de^d(x-z)\, \Delta^T_{\alpha\beta\si\rho} \de^d(x-y)
- \bigl \{ \pr_\alpha \bigl (\de^d(x-y)\,
\Delta^T_{\beta\nu\si\rho} \de^d(y-z)\bigl ){} +
\alpha \leftrightarrow \beta \bigl \} \, , \cr}
\eqno (8.32) $$
which imply also $\A^F_{\si\si,\alpha\beta}(s,t)=0, \, \pr^s_{\,\si}
\A^F_{\si\rho,\alpha\beta}(s,t)=0$. Similarly, or from the explicit form
(8.30) using $5\E^A_{\mu\si\alpha\gamma\kappa,\mu\rho\beta\delta\lambda}
= \vep \, \E^A_{\si\alpha\gamma\kappa,\rho\beta\delta\lambda}$,
$$  D^G_{\mu\mu\si\rho\alpha\beta}(x,y,z) = \vep \,
\A^G_{\si\rho,\alpha\beta}(x-y,x-z) \, , \quad
\pr_\mu^{\vphantom g} D^G_{\mu\nu\si\rho\alpha\beta}(x,y,z) = 0 \, .
\eqno (8.33) $$
As is well known for $d=4$ $\int d^4 x \sqrt g \, G$ is a topological
invariant, proportional to the Euler number. In consequence
$D^G_{\mu\nu\si\rho\alpha\beta}$ vanishes when $d=4$. However this involves
the vanishing of antisymmetric five index tensors which is valid strictly
only when $d=4$ and so $D^G_{\mu\nu\si\rho\alpha\beta}$ is a legitimate
and in fact necessary counterterm in the context of dimensional
regularisation when relations dependent on specific integer dimensions
should not imposed in a consistent treatment\footnote{*}
{Note that the first variation of $\int d^d x \sqrt g \, G$ defines
a tensor $H_{\mu\nu} =
-15 \, R^{\alpha\beta}{}_{\! [\alpha\beta}
R^{\gamma\delta}{}_{\! \gamma\delta} \, g_{\mu]\nu} =
2 R^{\alpha\beta\gamma}{}_{\! \mu} R_{\alpha\beta\gamma\nu} - 8
R^{\alpha\beta}R_{\alpha\mu\beta\nu} + 2 RR_{\mu\nu} - \half g_{\mu\nu}G$
which is a direct analogue of the usual Einstein tensor $R_{\mu\nu}
- \half g_{\mu\nu} R$. $ H_{\mu\nu} = H_{\nu\mu}$ is  zero for $d=4$ and
satisfies  $g^{\mu\nu}H_{\mu\nu} = \half\vep G , \ \nab^\mu H_{\mu\nu} = 0$.}.

Using the previous results we may now write a regularised form for the energy
momentum tensor three point function, analogous to (8.21), as
$$ \eqalign {
\l T_{\mu\nu}&(x)\, T_{\si\rho} (y)\, T_{\alpha\beta} (z)\r \cr
{}& = \I_{\si\rho,\si'\rho'}(x-y)\, \I_{\alpha\beta\alpha'\beta'}(x-z) \,
{t_{\si'\rho'\alpha'\beta'\mu\nu} (X) \over(x-y)^d (x-z)^d (y-z)^d} \cr
&\, - \bigl ( A_{\mu\nu\si\rho\gamma\delta}(x-y) -
\hA_{\mu\nu\si\rho\gamma\delta}(x-y)\bigl ) \, \half C_T \Biggl (
{I_{\gamma\delta\alpha\beta}(y-z) \over (y-z)^{2d}} +
{I_{\gamma\delta\alpha\beta}(x-z) \over (x-z)^{2d}} \Biggl ) \cr
&\, - \bigl ( A_{\mu\nu\alpha\beta\gamma\delta}(x-z) -
\hA_{\mu\nu\alpha\beta\gamma\delta}(x-z)\bigl ) \, \half C_T \Biggl (
{I_{\gamma\delta\si\rho}(z-y) \over (z-y)^{2d}} +
{I_{\gamma\delta\si\rho}(x-y) \over (x-y)^{2d}} \Biggl ) \cr
&\, - \bigl ( A_{\si\rho\alpha\beta\gamma\delta}(y-z) -
\hA_{\si\rho\alpha\beta\gamma\delta}(y-z)\bigl ) \, \half C_T \Biggl (
{I_{\gamma\delta\mu\nu}(z-x) \over (z-x)^{2d}} +
{I_{\gamma\delta\mu\nu}(y-x) \over (y-x)^{2d}} \Biggl ) \cr
& -{\mu^{-\vep}\over\vep}\, 4\bigl (\beta_a
D^F_{\mu\nu\si\rho\alpha\beta}(x,y,z)
+ \beta_b D^G_{\mu\nu\si\rho\alpha\beta}(x,y,z) \bigl ) \, . \cr}
\eqno (8.34) $$
By virtue of (8.32) and (8.33) this is in accord with the conservation Ward
identity in (6.33) and the modified trace identity (8.26) in the limit
$d\to 4$ if
$$ \l T_{\mu\nu}(x)\, T_{\si\rho} (0) \r = C_T\, {\I_{\mu\nu,\si\rho}(x)
\over x^{2d}} + {\mu^{-\vep}\over \vep}\, 4\beta_a \, \Delta^T_{\mu\nu\si\rho}
\de^d(x) \, .
\eqno (8.35) $$
This has the appropriate form for the dimensionally regularised two point
function
for the energy momentum tensor, based on (8.8,9), so long as the same relation
obtained earlier in (8.12) between $\beta_a$ and $C_T$ holds. The regularised
expression (8.34) leads to a renormalisation group equation for the three
point function
$$ \mu{\pr\over \pr\mu} \l T_{\mu\nu}(x)\, T_{\si\rho} (y)\,
T_{\alpha\beta} (z)\r = 4\beta^a \, D^F_{\mu\nu\si\rho\alpha\beta}(x,y,z)  \, ,
\eqno (8.36) $$
in the limit $d\to 4$ when the apparent $D^G_{\mu\nu\si\rho\alpha\beta}$
contribution disappears. Clearly $\beta_a$ arises from the scale dependence
introduced in the regularisation of the two and three point functions of the
energy momentum tensor even in the conformal limit whereas $\beta_b$ has a
rather different significance [30]. Using dimensional regularisation
the coefficients $\beta_a , \beta_b$ can both in principle be
determined by analysing the singular poles in $\vep$ of the three point
function according to an appropriate version of (8.13). Assuming the general
counterterm is just a linear combination of $D^F_{\mu\nu\si\rho\alpha\beta}$
and $D^G_{\mu\nu\si\rho\alpha\beta}$, as exhibited in (8.34), then $\beta_a$
and $\beta_b$ may be found as expressions linear in the three parameters
$a,b,c$
which are discussed in sections 3 and 6 and which specify the general three
point
function of the energy momentum tensor. Due to its complexity we have not
undertaken this calculation here. However
the result for $\beta_a$ may alternatively be found
given (8.12) and (6.42) when $d=4$. Using this and also the results for free
fields in section 5 with (8.3) gives
$$ \beta_a = -{\pi^4\over 64\times 30} (14a -2b -5c ) \, , \quad
\beta_b = {\pi^4\over 64\times 90} (9a -2b - 10c ) \, .
\eqno (8.37) $$
\bigskip
\leftline{\bigbf 9 Derivative Relations}
\medskip
In sections 2 and 3 we have endeavoured to derive expressions for conformally
invariant three point functions involving conserved vector currents and the
energy momentum tensor. In obtaining these results the conservation equations
such as in (2.22) had to be imposed as additional constraints. Here we show
that in some cases it is possible to obtain alternative representations in
which the conservation equations are automatic.

We consider first a general three point function involving a vector current
which from (2.25), with $d-1+\eta_- +q =0$, has the form
$$
\l V_\mu (x_1)\, \O_2^{i_2} (x_2) \, \O_3^{i_3} (x_3) \r =
{1\over x_{13}^{\, 2(d-1)}\, x_{23}^{\, 2\eta_2}}\,
I_{\mu \nu} (x_{13}) D_2^{\, i_2} {}_{\! j_2} (I(x_{23})) \,
t_\nu {}^{\! j_2 i_3} (X_{12}) \, ,
\eqno (9.1) $$
where from (2.27)
$$ \pr_\mu t_\mu {}^{\! i_2 i_3} (X) = 0 \, .
\eqno (9.2) $$
If this equation is trivially satisfied by writing
$$ t_\nu {}^{\! i_2 i_3} (X) = \pr_\rho
t_{\nu\rho} {}^{\! i_2 i_3} (X) \, , \quad
t_{\nu\rho} {}^{\! i_2 i_3} (X) = - t_{\rho\nu} {}^{\! i_2 i_3} (X) \, ,
\eqno (9.3) $$
then an alternative representation to (9.1) with manifest current
conservation is feasible. Since
$$ {\pr \over \pr X_{12\, \rho}} = x_{13}^{\, 2} I_{\si\rho}
(x_{13}^{\vphantom g} ) \, {\pr\over \pr x_{1\, \si}} \, ,
\eqno (9.4) $$
and with results such as
$$
\pr_{\mu} \Bigl ( {1\over x^{2(d-2)} } \, \I^A{}_{\!\!\!\! \mu \nu,\si \rho}
(x) \Bigl ){} = 0 \, \quad
\I^A{}_{\!\!\!\!\mu \nu,\si \rho} (x) = \half \bigl ( I_{\mu \si}(x)
I_{\nu \rho} (x) - I_{\mu \rho}(x) I_{\nu \si} (x) \bigl ) \, ,
\eqno (9.5) $$
where $\I^A_{\mu \nu,\si \rho}$ corresponds to an inversion
in the representation of $O(d)$ formed by
antisymmetric tensors, then we may write
$$  \eqalign {
\l V_\mu (x_1)\, & \O_2^{i_2} (x_2) \, \O_3^{i_3} (x_3) \r \cr
& = {\pr \over \pr x_{1 \, \nu}} \biggl (
{1\over x_{13}^{\, 2(d-2)}\, x_{23}^{\, 2\eta_2}}\,
\I^A{}_{\!\!\!\! \mu \nu, \si \rho} (x_{13}) D_2^{\, i_2} {}_{\! j_2}
(I(x_{23})) \, t_{\si\rho} {}^{\! j_2 i_3} (X_{12})  \biggl ) \, , \cr}
\eqno (9.6) $$
with $t_{\si\rho} {}^{\! i_2 i_3} (\lambda X) = \lambda^{-(d-2+\eta_-)}
t_{\si\rho} {}^{\! i_2 i_3} (X)$.

For a comparable treatment for the energy momentum tensor we first define
$\H^C$ as the space of tensors with the symmetries of the Weyl tensor
as defined in (8.2). Thus
$$ C_{\mu\si\rho\nu} \in \H^C \ \Rightarrow \
 C_{\mu\si\rho\nu} =  C_{[\mu\si][\rho\nu]} \, , \quad  C_{\mu[\si\rho\nu]}
= 0 \, , \quad  C_{\mu\si\rho\mu} = 0 \, ,
\eqno (9.7) $$
which implies $ C_{\mu\si\rho\nu} =  C_{\rho\nu\mu\si}$, so that $\H^C$ has
dimension ${1\over 12}d(d+1)(d+2)(d-3)$. It is easy to see that for any
$ C_{\mu\si\rho\nu}(x) \in \H^C$ implies $\pr_\si \pr_\rho C_{\mu\si\rho\nu}(x)
= h_{\mu\nu}(x)$ is a conserved symmetric traceless tensor. Hence starting
from (2.28) written as
$$
\l T_{\mu \nu} (x_1)\, \O_2^{i_2} (x_2) \, \O_3^{i_3} (x_3) \r =
{1\over x_{13}^{\, 2d \vphantom \eta} \, x_{23}^{\, 2\eta_2}}\,
\I_{\mu \nu,\si\rho}(x_{13}) D_2^{\, i_2} {}_{\! j_2}
(I(x_{23})) \, t_{\si \rho} {}^{\! j_2 i_3} (X_{12}) \, ,
\eqno (9.8) $$
where (2.30) now requires $\pr_\mu t_{\mu\nu} {}^{\! i_2 i_3} (X) = 0$,
we therefore assume that it is possible to take
$$ t_{\mu\nu} {}^{\! i_2 i_3} (X) = \pr_\alpha \pr_\beta
t_{\mu\alpha\beta\nu} {}^{\! i_2 i_3} (X) \, , \quad
t_{\mu\alpha\beta\nu} {}^{\! i_2 i_3} (X) \in \H^C \, .
\eqno (9.9) $$
Using (9.4) and results akin to (9.5) and in particular the symmetry
properties of tensors in $\H^C$, as given in (9.7), we may now show that
(9.8) is equivalent to
$$ \eqalign { \!\!\!\!\!\!
\l & T_{\mu \nu} (x_1) \, \O_2^{i_2} (x_2) \, \O_3^{i_3} (x_3) \r \cr
& = {\pr\over \pr x_{1\, \si}} \, {\pr\over\pr  x_{1\, \rho}} \biggl (
{1\over x_{13}^{\, 2(d-2) } \, x_{23}^{\, 2\eta_2}}\,
\I^C{}_{\!\!\!\! \mu\si\rho\nu,\mu'\si'\rho'\nu'}(x_{13}) D_2^{\, i_2} {}_{\!
j_2}
(I(x_{23})) \, t_{\mu'\si'\rho'\nu'} {}^{\! j_2 i_3} (X_{12})\biggl ) \, , \cr}
\eqno (9.10) $$
for $\I^C \in \H^C \times \H^C$ representing inversions on tensor fields
belonging
to the space $\H^C$, $\I^C{}_{\!\!\!\! \mu\si\rho\nu,\alpha\beta\gamma\delta}
\I^C{}_{\!\!\!\! \alpha\beta\gamma\delta,\mu'\si'\rho'\nu'} =
\E^C{}_{\!\!\!\!\mu\si\rho\nu,\mu'\si'\rho'\nu'}$ where $\E^C \in \H^C \times
\H^C$ acts as a projection operator onto $\H^C$. $\I^C$ and $\E^C$ are
symmetric and $\I^C$ may be written as
$\I^C{}_{\!\!\!\!\mu\si\rho\nu,\alpha\beta\gamma\delta}
= I_{\mu\mu'} I_{\si\si'} I_{\rho\rho'} I_{\nu\nu'}
\E^C{}_{\!\!\!\!\mu'\si'\rho'\nu',\alpha\beta\gamma\delta}$. If
$P_{\mu\si\rho\nu} = P_{[\mu\si][\rho\nu]}, \ P_{\si\rho} = \half (
P_{\mu\si\mu\rho} + P_{\mu\rho\mu\si} )$ then $\E^C$ is defined by
$$\eqalign {
\E^C{}_{\!\!\!\! \mu\si\rho\nu,\alpha\beta\gamma\delta }
P_{\alpha\beta\gamma\delta } & =
{\ts {1\over 3}} \bigl ( P_{\mu\si\rho\nu} + P_{\rho\nu\mu\si}
+ P_{\mu[\nu\rho]\si} - P_{\si[\nu\rho]\mu} \bigl ) \cr
&\ - {2\over d-2} \bigl ( \de_{\rho[\mu}P_{\si]\nu} - \de_{\nu[\mu}
P_{\si]\rho} \bigl ) {} + {2\over (d-2)(d-1)}\, \de_{\rho[\mu}\de_{\si]\nu}
\, P_{\lambda\lambda} \, . \cr}
\eqno (9.11) $$

The significance of such results as (9.6) and (9.10) is that in an expansion
of the effective action $W(g,A,J)$ then the contributions corresponding to
(9.6) involve just the abelian field strength $\pr_\mu A_\nu - \pr_\nu A_\mu$
and for (9.10) the Weyl tensor $C_{\alpha\beta\gamma\de}$. The latter follows
since in an expansion around flat space $g_{\mu\nu} = \de_{\mu\nu} +
h_{\mu\nu}$ then $C_{\mu\si\rho\nu} = (\Delta h)_{\mu\si\rho\nu} + {\rm O}
(h^2)$ where the differential operator $\Delta$ is defined by
$$ \int \! d^d x \, h_{\mu\nu}\, \pr_\si \pr_\rho C_{\mu\si\rho\nu} = \half
\int \! d^d x \, (\Delta h)_{\mu\si\rho\nu}\, C_{\mu\si\rho\nu} \ \
{\hbox{for}} \ \ C_{\mu\si\rho\nu}(x) \in \H^C \, .
\eqno (9.12) $$
Thus $\Delta_{\mu\si\rho\nu,\alpha\beta} = 2
\E^C{}_{\!\!\!\!\mu\si\rho\nu,\alpha\gamma\delta\beta}\pr_\gamma \pr_\delta $
or more explicitly from (9.11)
$$ \eqalign {
(\Delta h)_{\mu\si\rho\nu} = {}& - \pr_\rho\pr_{[\mu} h_{\si]\nu} + \pr_\nu
\pr_{[\mu} h_{\si]\rho} \cr
& - {1\over d-2} \Bigl ( \de_{\rho[\mu}(\L h)_{\si]\nu} -
\de_{\nu[\mu}(\L h)_{\si]\rho} \Bigl ) {}+ {1\over (d-1)(d-2)} \de_{\rho[\mu}
\de_{\si]\nu} (\L h)_{\lambda \lambda} \, , \cr
& (\L h)_{\si\rho} = - \pr^2 h_{\si\rho} + \pr_\si \pr_\lambda h_{\lambda\rho}
+ \pr_\rho \pr_\lambda h_{\lambda\si} - \pr_\si\pr_\rho h_{\lambda\lambda} \,
. \cr}
\eqno (9.13) $$
With these definitions it should be noted that
$$ \eqalign {
\Delta^T_{\mu\nu\si\rho}  = {}& 2 \, {d-2\over d-3} \,
\pr_\alpha \pr_\beta \Delta_{\mu\alpha\beta\nu,\si\rho} \, , \cr
\A^F_{\si\rho\alpha\beta}(x-y,x-z) = 4 \, {d-2\over d-3} \, {}&
\E^C_{\si\si'\rho'\rho,\alpha\alpha'\beta'\beta} \pr_{\si'}^{\vphantom g}
\pr_{\rho'}^{\vphantom g} \de^d(x-y) \pr_{\alpha'}^{\vphantom g}
\pr_{\beta'}^{\vphantom g} \de^d(x-z) \, , \cr}
\eqno (9.14) $$
with $\Delta^T$ defined by (8.7) and $\A^F$ by (8.27).

As an illustration of the applicability of these results we consider the
$\l TT \O\r $ three point function, for $\O$ a scalar field of dimension
$\eta$. The conformal invariant expression was constructed in (3.3,4) and
(3.6). It is easy to see that we may write
$$ {1\over X^{2d-\eta}} \, t_{\alpha\beta\gamma\delta} ( X) =
{d-3\over d-2} \, \Delta^T_{\alpha\beta\gamma\delta} {K\over X^{2d-\eta-4}} \,
, \eqno (9.15) $$
where the coefficients $a,b,c$ are determined in terms of $K$ and satisfy
(3.6). Hence we find
$$ \eqalign { \!\!
\l T_{\mu \nu}& (x_1) \, T_{\si \rho} (x_2) \, \O (x_3) \r =
{d-3\over d-2}\, {1\over x_{13}^{\, 2d \vphantom )}\,
x_{23}^{\, 2d\vphantom )}}
\,  \I_{\mu \nu,\alpha \beta}(x_{13}) \I_{\si \rho,\gamma\delta}(x_{23}) \,
\Delta^T_{\alpha\beta\gamma\delta} {K\over X_{12}^{2d-\eta-4}} \cr
& = \pr_{1\mu'} \pr_{1\nu'} \biggl ( {1\over x_{13}^{\, 2(d-2)}\,
x_{23}^{\, 2d\vphantom )}}
\,  \I^C{}_{\!\!\!\!\mu\mu'\nu' \nu,\alpha \beta \gamma\delta }(x_{13})
\I_{\si \rho,\alpha\delta}(x_{23}) \,
\pr_{X_{12} \beta}^{\vphantom )}\pr_{X_{12}\gamma}^{\vphantom )}
{4K\over X_{12}^{2d-\eta-4}} \biggl ) \cr
& = \pr_{1\mu'} \pr_{1\nu'} \pr_{2\si'} \pr_{2\rho'}
\biggl ( {1\over x_{13}^{\, \eta \vphantom d}\, x_{23}^{\, \eta \vphantom d}}
\, {4K \over x_{12}^{\, 2d-\eta-4}}
\,  \I^C{}_{\!\!\!\!\mu\mu'\nu' \nu,\alpha \beta \gamma\delta }(x_{13})
\I^C{}_{\!\!\!\!\si\si'\rho'\rho,\alpha\beta\gamma\delta}(x_{23}) \biggl ) \, .
\cr}
\eqno (9.16) $$
The conservation equation for the energy momentum tensor is automatically
satisfied as is appropriate since there are no Ward identities in this case.
It should be noted that for $\eta=0$ this three point function does not depend
on $x_3$ and is proportional to $\l T_{\mu \nu}(x_1) \, T_{\si \rho}(x_2)\r $.
\bigskip
\leftline{\bigbf 10 Conclusion}
\medskip
The results of this paper demonstrate that for general dimensions the form of
operator product expansions involving the energy momentum tensor, or the
conformal invariant expressions for three point functions, are rather more
complicated than in two dimensions where the essential formulae take the very
simple form given by (1.1,2). There are still many unanswered questions,
particularly in relation to positivity constraints. It is trivial that in
unitary theories the overall coefficient of the two point function $C_T$, and
also $C_V$, is positive which given (6.42) and (8.12) provides one condition
on the three parameters $a,b,c$ in the general three point function of the
energy momentum tensor and also requires $\beta_a < 0$ for the coefficient of
the $F$ term in the energy momentum tensor trace on a curved space background.
As remarked by Cappelli,
Friedan and Latorre [19] there is no presently known condition on $\beta_b$,
despite the results for free fields (8.3). It would be very nice to derive
further positivity conditions on $a,b,c$ which might imply positivity of
$\beta_b$. In general there are no such conditions on three point functions
but this is not true when the energy momentum tensor is involved since the
spectrum of the Hamiltonian $H$ is required to be positive. To proceed it may
be necessary to consider the operator product expansion in four point
functions [30]. In this context it should be pointed out that it is not
entirely clear if the operator product expansion in dimensions $d>2$ is
associative although this is probably essential for an algebraic operator
formulation such as is used in two dimensions. This question should
presumably be answerable at least in a perturbative context.

As an illustration of the potential use of some of our results we recapitulate
the well known derivation of the $c$-theorem [2,3,19]. In two dimensions for a
general massless field theory, or at distances small compared with any explicit
mass scale, the two point function for the energy momentum tensor takes the
regularised form
$$ \l T_{\mu\nu} (x)\, T_{\si\rho}(0) \r = S_{\mu\nu} S_{\si\rho}
\Omega (t) \, , \qquad t = \half \ln \mu x^2 \, ,
\eqno (10.1) $$
where $\mu$ is a renormalisation mass scale (note that in two dimensions from
(8.7) we may write $S_{\mu\nu} = - {\tilde \pr}_\mu {\tilde \pr}_\nu , \
{\tilde \pr}_\mu = \ep_{\mu\alpha} \pr_\alpha $
and hence $\Delta^T_{\mu\nu\si\rho} = 0$). In terms of (10.1) we
follow Zamoldchikov [3] and define
$$ \eqalign {
F(t) = {}& z^4  \l T_{zz} (x)\, T_{zz}(0) \r = {\ts {1\over 16}}
\Omega''''(t) - {\ts {3\over 4}} \Omega'''(t) + {\ts {11\over 4}} \Omega''(t)
- 3 \Omega '(t) \, , \cr
G(t) = {}& z^2 x^2  \l T_{zz} (x)\, T_{\mu\mu}(0) \r = -{\ts {1\over 4}}
\Omega''''(t) + {\ts {3\over 2}} \Omega'''(t) -2 \Omega''(t) \, , \cr
H(t) = {}& (x^2)^2_{\vphantom d}  \l T_{\mu\mu} (x)\, T_{\si\si}(0) \r =
\Omega''''(t) - 4 \Omega'''(t) + \Omega''(t) \, . \cr }
\eqno (10.2) $$
It is then easy to see that
$$ C'(t) = - {\ts {3\over 2}} H(t)  < 0 \ \ \hbox{for} \ \
C(t) = 2 F(t) - G(t) - {\ts {3\over 8}} H(t) \, ,
\eqno (10.3) $$
where if $T_{\mu\mu} = \beta^i \O_i $ then $H = G_{ij} \beta^i\beta^j $ with
$G_{ij}(t) = (x^2)^2_{\vphantom d}  \l \O_i (x)\, \O_j (0) \r $ the
Zamolodchikov
metric on the space of couplings. The flow of $C(t)$ is clearly monotonic
decreasing and is only stationary at the critical points where $\beta^i =0$
where, from (1.2), $C\to c$ the Virasoro central charge of the corresponding
conformal theory (as compared with our previous normalisation $T_{\mu\nu}\to
- 2\pi T_{\mu\nu}$ which is conventional for two dimensional conformal field
theories). An essential part of Zamolodchikov's argument was that the quantity
$C(t)$ is well defined in the conformal limit and hence the change $\Delta C$
in flowing between two conformal field theories is unambiguous, independent
of the path. It is possible to imagine then that it may be feasible to
define quantities in terms of the three point function, analogous to (10.2),
which have a well defined conformal limit and also nice properties under
renormalisation flow. It would probably be desirable to use the collinear
configuration since then, as described in section 4, the dependence on the
spatial variables $x,y,z$ factorises into a unique form in the conformal
limit. Presumably considering three point functions of the energy momentum
tensor in two dimensions gives nothing new, since the scale is determine by
the two point function, but for $d>2$ there are three independent coefficients
in general. For $d=4$ one linear combination gives $\beta_b$, as shown in
(8.37), which was suggested by Cardy [21] as a possible candidate for a
generalisation of the $c$-theorem to four dimensions.

In two dimensions the Virasoro central charge may be related to physically
measureable effects such as the universal finite size corrections to the
Casimir energy for a strip of width $L$ [32]. The derivation of this result
is not valid for $d>2$, due to the restricted nature of the conformal group in
this case, but can be used to provide an alternative generalisation of $c$
away from $d=2$ which differs from that provided by the coefficient of the
conformally invariant two point function for the energy momentum tensor. It
would be nice if the three independent coefficients in the energy momentum
tensor three point function could perhaps be related to such physically
observable
energies. Independent of such conjectures it should be feasible to obtain
expressions for the effective action $W$ on curved space backgrounds for
conformally invariant theories where at least to third order in the curvature
our results imply there are just three linearly independent forms whose
coefficients are related to the three parameters in the general conformally
invariant three point for the energy momentum tensor on flat space.
Assuming conformal invariance is therefore quite restrictive and should lead
to hopefully simpler results than that found recently for general theories
[33]. The corresponding two dimensional result takes the remarkably elegant
form
$$ W(g) = - {c\over 96\pi} \int \! d^2 x \sqrt g \, R {1\over \nab^2} R \, ,
\eqno (10.4) $$
which was found by integrating the two dimensional trace anomaly [34].
\bigskip
One of us (A.P.) would like the thank the European Community for a research
fellowship. H.O. would like to thank Professor S. Deser and Graham Shore for
very useful discussions. A.P. would also like to express his appreciation for
discussions with Professors A. Schwimmer and I. Antoniadis.
\bigskip
\bigskip
%\vfill\eject
\leftline{\bigbf Appendix}
\medskip
Here we collect various results needed in the derivation of the
conformal invariant form of the three point function for the
energy momentum tensor. In order to impose (3.16b) we use, with $h^1,h^2,h^3,
h^4,h^5$ defined in (3.2), (3.4) and (3.18),
$$ \eqalign {
\I_{\mu\nu,\mu'\nu'}h^5_{\mu'\nu'\si\rho\alpha\beta} ={}&
h^5_{\mu\nu\si\rho\alpha\beta} -2
h^4_{\alpha\beta\si\rho\mu\nu} -2
h^4_{\si\rho\alpha\beta\mu\nu} + 8
h^2_{\si\rho\alpha\beta}h^1_{\mu\nu}\, , \cr
\I_{\mu\nu,\mu'\nu'}h^4_{\mu'\nu'\si\rho\alpha\beta} ={}&
h^4_{\mu\nu\si\rho\alpha\beta} -4
h^2_{\mu\nu\si\rho}h^1_{\alpha\beta} -4
h^2_{\mu\nu\alpha\beta}h^1_{\si\rho} + 32
h^1_{\mu\nu}h^1_{\si\rho}h^1_{\alpha\beta}\, , \cr
\I_{\si\rho,\si'\rho'}h^4_{\mu\nu\si'\rho'\alpha\beta} ={}&
-h^4_{\mu\nu\si\rho\alpha\beta} +4
h^2_{\mu\nu\alpha\beta}h^1_{\si\rho} \, , \cr
\I_{\mu\nu,\mu'\nu'}h^3_{\mu'\nu'\si\rho} ={}&
h^3_{\mu\nu\si\rho} - 2 h^2_{\mu\nu\si\rho} + 8
h^1_{\mu\nu}h^1_{\si\rho} \, , \cr
\I_{\mu\nu,\mu'\nu'}h^2_{\mu'\nu'\si\rho} ={}&
-h^2_{\mu\nu\si\rho}  + 8 h^1_{\mu\nu}h^1_{\si\rho} \, , \cr
\I_{\mu\nu,\mu'\nu'}h^1_{\mu'\nu'} = {}& h^1_{\mu\nu} \, . \cr}
\eqno (A.1) $$
Note that these results are consistent with $\I^2 =1$.
For the conservation equations we use
$$\eqalign{
\Bigl ( \pr_\mu - d\, {X_\mu\over X^2}\Bigl ) h^1_{\mu\nu} = {}& 0 \, ,
\cr
\Bigl ( \pr_\mu - d\, {X_\mu\over X^2}\Bigl ) h^2_{\mu\nu\si\rho} = {}&
-2d\, {X_\nu\over X^2} h^1_{\si\rho} - {4\over d}\,
\pr_\nu h^1_{\si\rho} \, , \cr
\Bigl ( \pr_\mu - d\, {X_\mu\over X^2}\Bigl ) h^3_{\mu\nu\si\rho} = {}&
-2d\, {X_\nu\over X^2} h^1_{\si\rho} - d\, \pr_\nu h^1_{\si\rho}\, ,
\cr
\Bigl ( \pr_\mu - d\, {X_\mu\over X^2}\Bigl )
h^4_{\mu\nu\si\rho\alpha\beta} = {}&
-{d-2\over d}\, \pr_\nu h^2_{\si\rho\alpha\beta}
+4\, {X_\nu\over X^2}\bigl( h^2_{\si\rho\alpha\beta}
- 2(d+2)\, h^1_{\si\rho}h^1_{\alpha\beta}\bigl)\cr
&-2(d+2)\bigl(h^1_{\alpha\beta}\pr_\nu h^1_{\si\rho} +
h^1_{\si\rho}\pr_\nu h^1_{\alpha\beta}\bigl) \, , \cr
\Bigl ( \pr_\mu - d\, {X_\mu\over X^2}\Bigl )
h^4_{\alpha\beta\mu\nu\si\rho} = {}&
-{2\over d}\, \pr_\nu h^2_{\si\rho\alpha\beta}
+{X_\nu\over X^2}\bigl(2\, h^3_{\si\rho\alpha\beta}
-d\,h^2_{\si\rho\alpha\beta}\bigl)  \, , \cr
\Bigl ( \pr_\mu - d\, {X_\mu\over X^2}\Bigl )
h^5_{\mu\nu\si\rho\alpha\beta} = {}&
-d\, \pr_\nu h^2_{\si\rho\alpha\beta}
+{X_\nu\over X^2}\bigl(4\, h^3_{\si\rho\alpha\beta}
- 2d\, h^2_{\si\rho\alpha\beta}\bigl)  \, . \cr}
\eqno (A.2) $$
In addition we need
$$ \eqalign {
h^2_{\mu\nu\si\rho}\pr_\mu h^1_{\alpha\beta} = {}&
{X_\nu\over X^2}\bigl ( h^2_{\si\rho\alpha\beta}-4 h^1_{\si\rho}
h^1_{\alpha\beta}\bigl){} - {4\over d}\,h^1_{\si\rho}\pr_\nu
h^1_{\alpha\beta}\, , \cr
h^3_{\mu\nu\si\rho}\pr_\mu h^1_{\alpha\beta}+
h^3_{\mu\nu\alpha\beta}\pr_\mu h^1_{\si\rho}
= {}&\pr_\nu h^2_{\si\rho\alpha\beta} + 2\, {X_\nu\over X^2}
\bigl (h^2_{\si\rho\alpha\beta}-4 \, h^1_{\si\rho}
h^1_{\alpha\beta}\bigl){} \cr
& -2\, h^1_{\si\rho}\pr_\nu h^1_{\alpha\beta} - 2\,h^1_{\alpha\beta}\pr_\nu
h^1_{\si\rho}\, . \cr}
\eqno (A.3) $$
Also with ${\tilde h}_{\mu\nu\si\rho}$ as in (3.12) we find
$$ \eqalign {
\Bigl (\pr_\mu - d\,{X_\mu\over X^2}\Bigl ) {\tilde h}_{\mu\nu\si\rho} = {}& 0
\, , \cr
\Bigl (\pr_\si - d\,{X_\si\over X^2}\Bigl ) {\tilde h}_{\mu\nu\si\rho} = {}&
2d\, {X_\rho\over X^2} h^1_{\mu\nu} - 2\, \pr_\rho h^1_{\mu\nu}\, . \cr}
\eqno (A.4) $$

For applications in section 6 we also need formulae for the divergences of
the functions $H^i$ introduced in (6.30). It is convenient to define
$$ \eqalign {
D^1_{\si\rho\alpha\beta} (s) = {}& D^1_{\alpha\beta\si\rho} (s)
= \Bigl ( \pr_\si^{\vphantom h}
\pr_\rho^{\vphantom h} - {1\over d}\, \de_{\si\rho}^{\vphantom h}
\pr^2 \Bigl ) \, {1\over s^{d-2}} h^1_{\alpha\beta}(\hs) \, , \cr
D^2_{\nu\si\rho\alpha\beta} (s) = {}&
\Bigl ( \pr_\si^{\vphantom h}\pr_\rho^{\vphantom h}
- {1\over d}\, \de_{\si\rho}^{\vphantom h}\pr^2 \Bigl )
\Bigl ( \de_{\nu\alpha}^{\vphantom h}\pr_\beta^{\vphantom h}
+ \de_{\nu\beta}^{\vphantom h}\pr_\alpha^{\vphantom h}
- {2\over d}\, \de_{\alpha\beta}^{\vphantom h}\pr_\nu^{\vphantom h}\Bigl )
{1\over s^{d-2}} \, , \cr
D^3_{\si\rho\alpha\beta} (s) = {}& \Bigl ( \de_{\si\alpha}^{\vphantom h}
\pr_\rho^{\vphantom h} \pr_\beta^{\vphantom h} + (\si\leftrightarrow \rho,
\alpha \leftrightarrow \beta)
- {4\over d} \de_{\si \rho}^{\vphantom h} \pr_\alpha^{\vphantom h}
\pr_\beta^{\vphantom h}
- {4\over d} \de_{\alpha \beta}^{\vphantom h} \pr_\si^{\vphantom h}
\pr_\rho^{\vphantom h}
+ {4\over d^2}\de_{\si\rho}^{\vphantom h}\de_{\alpha \beta}^{\vphantom h}
\pr^2 \Bigl ) {} {1\over s^{d-2}} \, . \cr}
\eqno (A.5) $$
We may then find
$$ \eqalign {
\pr_\mu^{\vphantom h}H^1_{\alpha\beta\mu\nu\si\rho}(s) = {}&
- {2\over d}\, \pr_\nu^{\vphantom h} D^1_{\si\rho\alpha\beta}(s)
- {1\over d(d-2)}\, D^2_{\nu\alpha\beta\si\rho} (s) \, , \cr
\pr_\mu^{\vphantom h}H^2_{\mu\nu\si\rho\alpha\beta}(s) = {}&
- {2(d-1)\over d(d-2)}\, \pr_\nu^{\vphantom h} D^3_{\si\rho\alpha\beta}(s)
\, , \cr
\pr_\mu^{\vphantom h}H^2_{\alpha\beta\mu\nu\si\rho}(s) = {}&
{2\over d}(d-2)\, \pr_\nu^{\vphantom h} D^1_{\si\rho\alpha\beta}(s)
- {2\over d-2}\, D^2_{\nu\alpha\beta\si\rho} (s) \, , \cr
\pr_\mu^{\vphantom h}H^3_{\mu\nu\si\rho\alpha\beta}(s) = {}&
- {1\over d}(d-1)(d-2)\, h^3_{\si\rho\alpha\beta}\pr_\nu^{\vphantom h} \,
S_d \de^d(s) \, , \cr
\pr_\mu^{\vphantom h}H^3_{\alpha\beta\mu\nu\si\rho}(s) = {}&
D^2_{\nu\alpha\beta\si\rho} (s) \, , \cr
\pr_\mu^{\vphantom h}H^4_{\mu\nu\si\rho\alpha\beta}(s) = {}&
2D^2_{\nu\alpha\beta\si\rho} (s) + 2D^2_{\nu\si\rho\alpha\beta} (s)
- {2\over d}\, \pr_\nu^{\vphantom h} D^3_{\si\rho\alpha\beta}(s) \, , \cr
\pr_\mu^{\vphantom h}H^4_{\alpha\beta\mu\nu\si\rho}(s) = {}&
{1\over d}(d-2)\, \pr_\nu^{\vphantom h} D^3_{\si\rho\alpha\beta}(s) \cr
& ~~~ - (d-2)\Bigl (h^3_{\si\rho\nu\alpha}\pr_\beta^{\vphantom h}
+ h^3_{\si\rho\nu\beta}\pr_\alpha^{\vphantom h} - {2\over d}\,
\de_{\alpha\beta}^{\vphantom h}h^3_{\si\rho\nu\mu}\pr_\mu^{\vphantom h}\Bigl )
S_d\de^d(s) \, . \cr}
\eqno (A.6) $$
\vfill\eject
\noindent {\bigbf References}
\medskip
\parskip=0cm
\parindent=20pt
\item {1.} P. Ginsparg, {\it in} ``Champs, Cordes et Ph\'enom\`enes
Critiques", (E. Br\'ezin and J. Zinn-Justin eds.) North Holland, Amsterdam
1989.
\vskip 6pt
\item {2.} J. Cardy, {\it in} ``Champs, Cordes et Ph\'enom\`enes Critiques",
(E. Br\'ezin and J. Zinn-Justin eds.) North Holland, Amsterdam 1989.
\vskip 6pt
\item {3.} A.B. Zamolodchikov, {\it JETP Lett.} {\bf 43} (1986), 43; {\it
Sov. J. Nucl. Phys.} {\bf 46} (1988), 1090;\hfill\break
A.W.W. Ludwig and J.L. Cardy, {\it Nucl. Phys.} {\bf B285} [FS19] (1987), 687.
\vskip 6pt
\item {4.} G. Mack and A. Salam, Ann. Phys. (N.Y.) {\bf 53} (1969), 174.
\vskip 6pt
\item {5.} C. Callan, S. Coleman and R. Jackiw, {\it Ann. Phys. (N.Y.)}
{\bf 59} (1970), 42;\hfill\break
D.J. Gross and J. Wess, {\it Phys. Rev.} {\bf D2} (1970) 753;\hfill\break
S. Coleman and R. Jackiw, {\it Ann. Phys. (N.Y.)} {\bf 67} (1971), 552.
\hfill\break
R. Jackiw, {\it in} ``Lectures on Current Algebra and its
Applications'', by S.B. Treiman, R. Jackiw and D.J. Gross, Princeton
University Press, Princeton 1972.
\vskip 6pt
\item {6.} A.M. Polyakov, {\it JETP Lett.} {\bf 12} (1970), 381; {\it Soviet
Phys. JETP} {\bf 30} (1970), 150;\hfill\break
A.A. Migdal, {\it Phys. Lett.} {\bf 37B} (1971) 98,
386;\hfill\break
G. Parisi and L. Peliti, {\it Lett. al Nuovo Cim.} {\bf 2} (1971), 627;
\hfill\break
S. Ferrara, R. Gatto, A.F. Grillo and G. Parisi, {\it Lett. al
Nuovo Cim.} {\bf 4} (1972), 115; {\it Phys. Lett.} {\bf 38B} (1972), 333;
\hfill\break
S. Ferrara, A.F. Grillo and R.Gatto, {\it Ann. Phys. (N.Y.)} {\bf 76} (1973),
161.
\vskip 6pt
\item {7.} E. Schreier, {\it Phys. Rev.} {\bf D3} (1971), 980.
\vskip 6pt
\item {8.} G. Mack and K. Symanzik, {\it Comm. Math. Phys.} {\bf 27} (1972),
247; \hfill\break
G. Mack and I.T. Todorov, {\it Phys. Rev.} {\bf D8} (1973), 1764;\hfill\break
M. L\"uscher and G. Mack, {\it Comm. Math. Phys.} {\bf 41} (1975), 203.
\vskip 6pt
\item {9.} K. Koller, {\it Comm. Math. Phys.} {\bf 40} (1975), 15.
\vskip 6pt
\item {10.} G. Mack, {\it Comm. Math. Phys.} {\bf 53} (1977), 155.
\vskip 6pt
\item {11.} G. Mack, {\it Comm. Math. Phys.} {\bf 55} (1977), 1.
\vskip 6pt
\item {12.} E.S. Fradkin and M.Ya. Palchik, {\it Nuovo Cim.} {\bf 34} (1976),
438; {\it Phys. Rep.} {\bf 44} (1978), 249.
\vskip 6pt
\item {13.} J. Cardy, {\it Nucl. Phys.} {\bf B290} (1987), 355.
\vskip 6pt
\item {14.} J. Polchinski, {\it Nucl. Phys.} {\bf B303} (1988), 226.
\vskip 6pt
\item {15.} A. Cappelli and A. Coste, {\it Nucl. Phys.} {\bf B314} (1989),
707.
\vskip 6pt
\item {16.} E.S. Fradkin and M.Ya. Palchik, {\it Int. J. Mod. Phys.} {\bf A5}
(1990), 3463.
\vskip 6pt
\item {17.} D.J. Gross, {\it in} ``Methods in Field Theory'',
Les Houches Lectures 1975, (R. Balian and J. Zinn-Justin eds.)
North-Holland, Amsterdam 1976;\hfill\break
P. Olesen, {\it Nucl. Phys.} {\bf B104} (1976), 125.
\vskip 6pt
\item {18.} I. Jack and H. Osborn, {\it Nucl. Phys.} {\bf B343} (1990), 647.
\vskip 6pt
\item {19.} A. Cappelli, D. Friedan and J.I. Latorre, {\it Nucl. Phys.}
{\bf B352} (1991), 616;\hfill\break
D.Z. Freedman, J.I. Latorre and X. Vilas\'{\i}s, {\it Mod. Phys.
Lett.} {\bf A6} (1991), 531;\hfill\break
A. Cappelli, J.I. Latorre and X. Vilas\'{\i}s-Cardona, {\it Nucl. Phys.}
{\bf B376} (1992), 510.
\vskip 6pt
\item {20.} G.M. Shore, {\it Phys. Lett.} {\bf B253} (1991), 380;
{\bf 256} (1991), 407.
\vskip 6pt
\item {21.} J.L. Cardy, {\it Phys. Lett.} {\bf B215} (1988), 749.
\vskip 6pt
\item {22.} H. Osborn, {\it Phys. Lett.} {\bf B222} (1989), 97.
\vskip 6pt
\item {23.} N.D. Birrell and P.C.W. Davies, ``Quantum Fields in Curved Space",
Cambridge University Press, Cambridge (1982).
\vskip 6pt
\item {24.} D.Z. Freedman, K. Johnson and J.I. Latorre, {\it Nucl. Phys.}
{\bf B371} (1992), 329;\hfill\break
D.Z. Freedman, G. Grignani, K. Johnson and N. Rius, {\it Ann. Phys. (N.Y.)}
{\bf 216} (1992), 75;\hfill\break
J.I. Latorre, C. Manuel and X. Vilas\'{\i}s-Cardona, Barcelona preprint,
UB-ECM-PF 93/4, hep-th/9303044.
\vskip 6pt
\item {25.} R.J. Crewther, {\it Phys. Rev. Lett.} {\bf 28} (1972), 1421.
\vskip 6pt
\item {26.} Ya.S. Stanev, {\it Bulg. J. Phys.} {\bf 15} (1988), 93.
\vskip 6pt
\item {27.} H. Osborn, {\it Nucl. Phys.} {\bf B363} (1991), 486.
\vskip 6pt
\item {28.} L. Bonora, P. Cotta-Ramusino and C. Reina, {\it Phys. Lett.}
{\bf B126} (1983), 305;\hfill\break
P. Pasti, M. Bregola and L. Bonora, {\it Classical and Quantum
Gravity} {\bf 3} (1986), 635.
\vskip 6pt
\item {29.} I. Jack and H. Osborn, {\it Nucl. Phys.} {\bf B234} (1984), 331.
\vskip 6pt
\item {30.} S. Deser and A. Schwimmer,
{\it Phys. Lett.} {\bf B309} (1993), 279.
\vskip 6pt
\item {31.} K-H. Rehren and B. Schroer, {\it Nucl. Phys.} {\bf B295[FS1]}
(1988), 229.
\vskip 6pt
\item {32.} H.W.J. Bl\"ote, J.L. Cardy and M.P. Nightingale, {\it Phys. Rev.
Lett.} {\bf 56} (1986), 742;\hfill\break
I. Affleck, {\it Phys. Rev. Lett.} {\bf 56} (1986), 746.
\vskip 6pt
\item {33.} A.O. Barvinsky, Yu. V. Gusev, V.V. Zhytnikov and G.A. Vilkovisky,
``Covariant Perturbation Theory (IV). Third Order in the Curvature",
University of Manitoba preprint.
\vskip 6pt
\item {34.} A.M. Polyakov, {\it Phys. Lett.} {\bf B103} (1981), 207.
\bye